\documentclass[useAMS,usegraphicx,usenatbib,a4paper]{mn2e}
\usepackage[dvips, bookmarks, pdfborder={0 0 0.1}]{hyperref}
\usepackage{mathrsfs}
\usepackage{amssymb,amsmath}
\usepackage{placeins}
\usepackage{multirow}
\usepackage{aas_macros}
\usepackage{color}
\usepackage{mn2e-bst} 

\usepackage[normalem]{ulem}

\graphicspath{{figures/}}

\newcommand{\res}{\mathscr{R}}
\newcommand{\SC}{\Sigma_{\rm crit}}
\newcommand{\DS}{\Delta\Sigma}

\newcommand{\msunh}{h^{-1}\mathrm{M}_\odot }
\newcommand{\msunhh}{h^{-2}\mathrm{M}_\odot }
\newcommand{\lsunhh}{h^{-2}\mathrm{L}_\odot }
\newcommand{\VC}{\boldsymbol{\chi}}
\renewcommand{\vec}[1]{\boldsymbol{#1}}
\newcommand{\myplot}[1]{\includegraphics[width=8cm,height=8cm]{#1}}
\newcommand{\myplottwo}[2]{\myplot{#1}\myplot{#2}}

\newcommand{\mytab}{\begin{table}[htb]}
\newcommand{\myfig}{\begin{figure}[htbp]}
\newcommand{\mybibstyle}{mymn}

\begin{document}
%\linenumbers %for GAMA review
\title[GAMA halo mass from weak lensing]{Galaxy and Mass Assembly (GAMA): The halo mass of galaxy groups from maximum-likelihood weak lensing}
\author[J. Han et al.]
{Jiaxin Han,$^{1}$\thanks{jiaxin.han@durham.ac.uk}
Vincent R. Eke,$^{1}$
Carlos S. Frenk,$^{1}$
Rachel Mandelbaum,$^{2}$
\newauthor
Peder Norberg,$^{1}$
Michael D. Schneider,$^{3,4}$
John A. Peacock,$^5$
Yipeng Jing,$^6$
\newauthor
Ivan Baldry,$^7$
Joss Bland-Hawthorn,$^8$
Sarah Brough,$^9$
Michael J. I. Brown, $^{10}$
\newauthor
Jochen Liske,$^{11}$
Jon Loveday,$^{12}$
and Aaron S. G. Robotham$^{13}$\\
$^1$Institute of Computational Cosmology, Department of Physics,
University of Durham, South Road, Durham, DH1 3LE\\
$^2$McWilliams Center for Cosmology, Department of Physics, Carnegie Mellon University, Pittsburgh, PA 15213, USA\\
$^3$Lawrence Livermore National Laboratory, PO Box 808 L-210, Livermore, CA 94551-0808, USA\\
$^4$Department of Physics, University of California, One Shields Avenue, Davis, CA 95616, USA\\  
$^5$Institute for Astronomy, University of Edinburgh, Royal Observatory, Edinburgh EH9 3HJ, UK\\
$^6$Center for Astronomy and Astrophysics, Physics Department, Shanghai Jiao Tong University, Shanghai 200240, China\\
$^7$Astrophysics Research Institute, Liverpool John Moores University, IC2, Liverpool Science Park, 146 Brownlow Hill,\\
Liverpool, L3 5RF\\
$^8$Sydney Institute for Astronomy, School of Physics A28, University of Sydney, NSW 2006, Australia\\
$^9$Australian Astronomical Observatory, PO Box 915, North Ryde, NSW 1670, Australia\\
$^{10}$School of Physics, Monash University, Clayton, Victoria 3800, Australia\\
$^{11}$European Southern Observatory, Karl Schwarzschild-Str. 2, 85748 Garching, Germany\\
$^{12}$Astronomy Centre, University of Sussex, Falmer, Brighton BN1 9QH\\
$^{13}$ICRAR M468, UWA, 35 Stirling Highway, Crawley, WA 6009 
}

\maketitle

\begin{abstract}
  We present a maximum-likelihood weak lensing analysis of the mass
  distribution in optically selected spectroscopic Galaxy Groups
  (G$^3$Cv5) in the Galaxy And Mass Assembly (GAMA) survey, using
  background Sloan Digital Sky Survey (SDSS) photometric galaxies. The
  scaling of halo mass, $M_h$, with various group observables is
  investigated. Our main results are: 1) the measured relations
  of halo mass with group luminosity, virial volume and central
  galaxy stellar mass, $M_\star$, agree very well with predictions
  from mock group catalogues constructed from a GALFORM
  semi-analytical galaxy formation model implemented in the Millennium
  $\Lambda$CDM N-body simulation; 2) the measured relations of halo
  mass with velocity dispersion and projected half-abundance radius
  show weak tension with mock predictions, hinting at problems in the
  mock galaxy dynamics and their small scale distribution; 3) the
  median $M_h|M_\star$ measured from weak lensing depends more
  sensitively on the lognormal dispersion in $M_\star$ at fixed $M_h$ than it
  does on the median $M_\star|M_h$. Our measurements suggest an
  intrinsic dispersion of $\sigma_{\log(M_\star)}\sim 0.15$; 4)
  Comparing our mass estimates with those in the catalogue, we find
  that the G$^3$Cv5 mass can give biased results when used to select
  subsets of the group sample. Of the various new halo mass estimators
  that we calibrate using our weak lensing measurements, group
  luminosity is the best single-proxy estimator of group mass.
\end{abstract}

\section{Introduction}
Even though the nature of dark matter will ultimately be determined by
observations of its particle properties, its gravitational effect has
so far been the cleanest way to map its distribution in the
universe. Weak gravitational lensing is one of the main techniques for
mapping dark matter on large and intermediate
scales~\citep[e.g.][]{Bartelmann01}. As its name suggests, weak
lensing is the production of weak distortions (shear) in the shapes of
background, or source, galaxies by foreground
masses. Usually one has no prior knowledge of the intrinsic shape of a
source galaxy, resulting in uncertainties much larger than the
gravitational shear signal, so the extraction of shape distortions has
to be done in a statistical way, for example by measuring the shear-shear correlation 
function on large scales~\citep[e.g.,][]{CFHTCosmicShear}, or by stacking a large
number of source galaxies around many lenses on smaller scales. Early applications of stacked lensing to low mass groups have been carried out by \citet{Hoekstra01} and \citet{Parker05} who measured the average mass-to-light ratio of groups in the Canadian Network for Observational Cosmology Field Galaxy Redshift Survey (CNOC2). Stacked lensing measurements have also been made using
galaxies and groups in many current large surveys, including the 
SDSS~\citep{Mandelbaum06b,Mandelbaum06a,Johnston07,Sheldon09},
CFHT Lensing Survey~\citep{CFHTVelander,CFHTHudson}, COSMOS~\citep{COSMOS12} and Deep Lens Survey~\citep{DLS}.
These studies estimate the average density profile of the dark matter haloes of the
lenses, and derive scaling relations between halo mass and other observational properties. 

Even though stacked lensing analyses can give a
non-parametric estimate of the matter density profile around lenses with
similar properties, the interpretation of the stacked signal can be
difficult. This is because the stacked profile is an average over
all the contributing haloes of unknown mass distribution, and this
average typically has a complicated weighting determined by the error of
each shape measurement, the number of pairs within each radial bin,
and the redshifts of lenses and sources. To account 
somewhat for these averaging effects, one usually
parametrizes the distribution of halo masses and the clustering of
haloes using the framework of halo occupation distribution (HOD)
models \citep[e.g.][]{HODrev,Mandelbaum05a,COSMOS12}, and fits for the HOD
parameters given the stacked profiles.%, in analogy to the HOD modelling of galaxy clustering \citep[e.g.][]{Zheng07,Brown08}. 

In this work we carry out a weak lensing analysis of galaxy groups
from the Galaxy And Mass Assembly~\citep[GAMA, ][]{GAMA} survey. GAMA
is an ongoing spectroscopic survey of moderate sky
coverage. %, which will be $360~\rm{deg}^2$ when completed.
As large scale surveys go, it has deep spectroscopy as well as
uniform, yet high, completeness ($>98$\%) down to $r_{\rm
  AB}=19.8$. This makes possible the construction of a large and
accurate galaxy group catalogue~\citep[G$^3$Cv5,][]{G3C}, able to reach
lower halo masses than other existing catalogues of the local
universe. In addition, the survey region of GAMA was selected to
overlap several companion surveys at different wavelengths, ranging
from radio to x-ray. These complementary data provide a detailed
picture of the properties of GAMA galaxies. The variation of galaxy
properties with environment, defined by the mass distribution probed
by weak lensing, can be investigated using gravitational shear
measurements of background galaxies taken from the photometric SDSS
data in the same region. Fortunately, the redshift distribution of
GAMA groups peaks at $z\sim 0.2$, where the lensing efficiency of the
SDSS galaxy sample also peaks. These lens and source samples are
described in more detail in Section~\ref{sec_data}. Since our default lens sample is subject to a survey flux limit and a group multiplicity selection, most of the measured mass-observable relations in this work are subject to some selection effects and should not be taken as general relations for a volume-limited sample. In order to draw some general conclusions on galaxy formation, however, we only compare our measurements with mock galaxy catalogues that incorporate the same selection function. These mock catalogues are also described in Section~\ref{sec_data}.

As the galaxy number density of our source sample ($\sim
1~\mathrm{arcmin}^{-2}$) is much lower than some dedicated lensing
surveys (e.g., $\sim 20~\mathrm{arcmin}^{-2}$ in CFHTLS and $\sim 70~\mathrm{arcmin}^{-2}$ in COSMOS), and because
the lens sample is restricted by the small GAMA sky coverage ($\sim 150 \mathrm{deg}^2$ in this work) compared
with SDSS for instance, we do not have any obvious advantage in
signal-to-noise over existing measurements. Hence efficient
utilization of the lensing signal is crucial to our analysis. To this
end, we go beyond the popular stacked analyses, and perform a maximum-likelihood
analysis on the shapes of individual background galaxies, broadly following
the method in \cite{Hudson98}
\citep[see also][]{Schneider97, Hoekstra03, Hoekstra04}. The key difference between
our approach and stacked lensing is that we fit the shapes of each
source galaxy explicitly, while stacked lensing only estimates or fits
the average tangential shear for sub-samples of sources binned
in radius, and around lenses binned according to mass proxies. 
Our method requires no binning in the source sample, and can be applied with or without binning in the lens sample.
Not binning the data avoids information losses, leading to good measurements with our small
sample. Another advantage of our method is that it is free from
the averaging ambiguity associated with stacking, since the mass of
each lens is explicitly modelled. With this method, the large number
of available observational properties associated with GAMA groups can now all be
linked with the underlying halo masses, to provide valuable
constraints on galaxy formation models. We will also show that our maximum-likelihood weak lensing method is an ideal tool for model selection, to pick up the tightest mass-observable relation observationally. We describe our method in
Section~\ref{sec_method}, and its practical application in
Section~\ref{sec_app}.  

As a first application of our maximum-likelihood weak lensing (MLWL)
method, we extract the scaling relations of halo mass to various group
observables, including velocity dispersion, luminosity, radius,
virial volume and stellar mass of the group's central galaxy. With MLWL we give
both non-parametric measurements of these relations by binning only the
lens sample according to observable, and parametric fits by modelling the
mass-observable relation as a power-law with no binning at all. The G$^3$Cv5 comes with estimated
halo masses calibrated using mock catalogues. These mass
estimates are also examined with MLWL, to see if they differ
from our measurements. Starting from MLWL we also
construct several new mass estimators, which we compare with predictions from
a semi-analytical galaxy formation model and
previous measurements. These results are described and discussed in
Sections~\ref{sec_result} and \ref{sec_estimators}, with all the fits summarized in Table~\ref{table_par}. 

Weak lensing measurements can be compared with predictions from galaxy formation models to gain insight into the various physical processes in the model. In this comparison, it is crucial that one properly accounts for the observational selection effects. \citet{Hilbert10} first compared the weak lensing measured mass-richness relation with the prediction from semi-analytic galaxy formation models. They construct mock clusters by picking cluster haloes from simulation snapshots, and applying observational selection functions to the member galaxies of the mock clusters. In this work, we improve the treatment of selection effects in two aspects. First, a light-cone galaxy catalogue~\citep{Merson13} is constructed from a semi-analytic galaxy formation model, to account fully for the selection function of the galaxy survey. Second, identical group finding algorithms~\citep{G3C} are applied to both the real and mock galaxy catalogues, to account fully for the selection effect introduced by group finding. We also have compared many more mass-observable relations. All the relations in Table~\ref{table_par} are subject to sample selection, and we only compare them with mock catalogues constructed with the same selection function as the real data. The only exception is in the comparison of our stellar mass-halo mass relation with those from other works, where we make an additional measurement for a volume-limited central galaxy sample.
%The reader who is only interested in the results can directly skip to Sections~\ref{sec_result} and \ref{sec_estimators}.

To summarize the structure of the paper, we describe our lens and source samples in Section~\ref{sec_data} along with the mock catalogues to which we compare our measurements; the general MLWL method is described in Section~\ref{sec_method}, with its application to our samples described in Section~\ref{sec_app}; the results are presented and discussed in Sections~\ref{sec_result} and \ref{sec_estimators}; finally, we conclude in Section~\ref{sec_conclusion}.

The units throughout this paper, wherever not explicitly specified,
are ${\rm km\,s}^{-1}$ for velocity, $h^{-1}\rm{Mpc}$ for length,
$\msunh$ for halo mass, $\msunhh$ for galaxy stellar mass, and
$\lsunhh$ for luminosity, where $H_0=100h$~km s$^{-1}$ Mpc$^{-1}$. The
$\log()$ function throughout is the common (base 10) logarithm, while
the natural logarithm is $\ln()$. Unless explicitly stated, the lens
sample covers groups with three or more members. The relevant
cosmological parameters, which only appear in the distance
calculations of our measurements, are $\Omega_M=0.3$ and
$\Omega_\Lambda=0.7$. \footnote{The mock catalogues with which we
  compare are constructed from the $\Lambda$CDM Millennium simulation
  which has a different cosmology ($\Omega_M=0.25$,
  $\Omega_\Lambda=0.75$). However, our lensing measurements are very
  insensitive to cosmology. Switching to Millennium/WMAP9/Planck
  cosmologies only introduces a $\sim 1$ percent difference into the fitted
  parameters.} % and $h=0.7$.

\section{Data Samples}\label{sec_data}

The lens and source samples used in this work are described in
detail in Sections~\ref{sec_lens} and~\ref{sec_source}
respectively. Section~\ref{sec_mock} contains a description of the mock
GAMA group catalogues, to which we compare our measurements.

\subsection{Lens Catalogue: GAMA Galaxy Group Catalogue (G$^3$Cv5)}\label{sec_lens}
We use the fifth version of the GAMA Galaxy Group Catalogue
\citep[hereafter G$^3$Cv5]{G3C}\footnote{We updated the version number to the internal version number of the group catalogue as in the GAMA database. However, the catalogue refers to the same one as in \citet{G3C}, and the G3Cv1 quoted in the previous version of this paper.} in the three equatorial GAMA regions
($12\times4~\rm{deg}^2$ each) as our lens sample. The galaxy groups
were identified in the 3-year GAMA I data using a modified
Friends-of-Friends (FoF) algorithm \citep{2PIGG} and calibrated against a set of
mock catalogues constructed from the GALFORM \citep{Bower06}
semi-analytical model, following the method described in
\cite{Merson13}. The GAMA I data used here are 
uniformly limited to $r_{\rm AB}=19.4$ across the three regions.
Group properties are found to be robust to the effects of
interlopers and are median unbiased. The G$^3$Cv5 catalogue contains
$\sim 12 200$ groups with two or more members and includes $\sim 50\%$
of all the GAMA galaxies down to a magnitude limit of $r_{\rm AB}\leq 19.4$.

Applying the G$^3$Cv5 group finding algorithm to mock GAMA surveys shows
that approximately half of the two-member groups contain galaxies from 
different dark matter haloes. % i.e.\ they are not bijectively matched. 
These groups would have particularly 
unreliable properties, so we exclude all binary groups from this study,
reducing the sample to $\sim 4500$ groups. In addition, we exclude groups 
for which the measured velocity dispersion is smaller than the assumed
velocity measurement errors or for which the stellar mass of the central 
galaxy has not been estimated \citep[mainly due to missing photometry in the
GAMA I reprocessed multi-wavelength imaging; ][]{Hill11, Kelvin12, GAMASM}.
This removes a further 164 groups. The central galaxy of the group is
defined in the iterative way recommended by \cite{G3C}, where the
galaxy furthest from the galaxy luminosity-weighted projected centre
is rejected and this process repeated until the 
brighter of the final two galaxies is chosen. This is the preferred choice of centre according to \citet{G3C} who find the iterative centre to be less affected by interlopers than the Brightest Cluster Galaxy (BCG) or the luminosity-weighted centre. We use these iterative central galaxies to define the centres of our groups. 
This central galaxy is identical to the BCG for
$\sim90$\% of the groups, and it makes little difference in our measurement
if we choose the BCG as group centre instead. 
Stellar masses for group central galaxies were inferred using a stellar
population synthesis model, adopting a Chabrier IMF \citep{GAMASM}. 

The redshift distribution of our group sample, i.e.\ lens catalogue,
is shown in the lower panel of Fig.~\ref{sample_dist}, peaking at
$z\sim0.2$ and extending to $z\sim0.5$.

\subsubsection{$\mathrm{G}^3\mathrm{Cv5}$ Mass Estimators}\label{sec_datamass}
For each GAMA group, after measuring the group velocity dispersion
with the gapper estimator \citep{Gapper} and correcting for a
velocity measurement error, the dynamical mass of the group is
estimated via 
\begin{equation}\label{eq_dynmass}
M_{\rm dyn}\, = \, A_{\rm dyn} \, \sigma_v^2 \, R_{50}.
\end{equation}
$R_{50}$ is the projected half-abundance radius containing 50 percent
of the group members \citep{G3C}. We adopt this definition of group
radius throughout this paper. The prefactor $A_{\rm dyn}\sim 10$ was
calibrated as a function of redshift and multiplicity in the mock
catalogues by \cite{G3C}.
%\citep[see][for more details of mass definitions]{Lilian},
%\footnote{by comparing $M_{\rm dyn}$ with $M_{DHalo}$, where the latter is
%  the sum of the 
%  constituent subhalo masses. In terms of a virial definition,
%  $M_{DHalo}$ is close to $M_{200b}$, the virial mass enclosing an
%  average density that is 200 times the background matter
%  density(\cite{G3C, Lilian}).
%
The mass definition used in the calibration process is not exactly the
commonly used $M_{200b}$, but closely related to it as mentioned in
\citet{G3C}\citep[see][for 
  more details of the exact mass definition used and how it compares
  to $M_{200b}$]{Lilian}. 
The other G$^3$Cv5 mass estimator, the luminosity 
mass, %\footnote{$M_{\rm lum}$ is not available in G$^3$Cv5, but its description is given in \cite{G3C}} 
 comes from rescaling the total group luminosity
\begin{equation}\label{eq_lummass}
M_{\rm lum} \, = \, A_{\rm lum} \, L_{\rm grp}.
\end{equation}
$L_{\rm grp}$ is the total $r$-band luminosity of the group, corrected
for the fraction of light in galaxies below the survey flux limit
using the GAMA luminosity function \citep{G3C}. Throughout this paper
we refer to the $r$-band $L_{\rm grp}$ as the group luminosity. Most of
the GAMA groups contain members fainter than $M^*=-20.44+5\log
h$ \citep{Blanton03,Loveday12}, and the group luminosity is dominated by
galaxies around $M^*$, so the correction factor is below $3$ for about
$90$ per cent of the groups and $\sim 2$ at $z=0.2$, the median group
redshift. The prefactor $A_{\rm lum}$ is calibrated using $M_{\rm dyn}$ for 
the observed groups as a function of redshift and
multiplicity. Consequently, $M_{\rm lum}$ is median unbiased with
respect to $M_{\rm dyn}$. 

\begin{figure}
\includegraphics[width=0.5\textwidth]{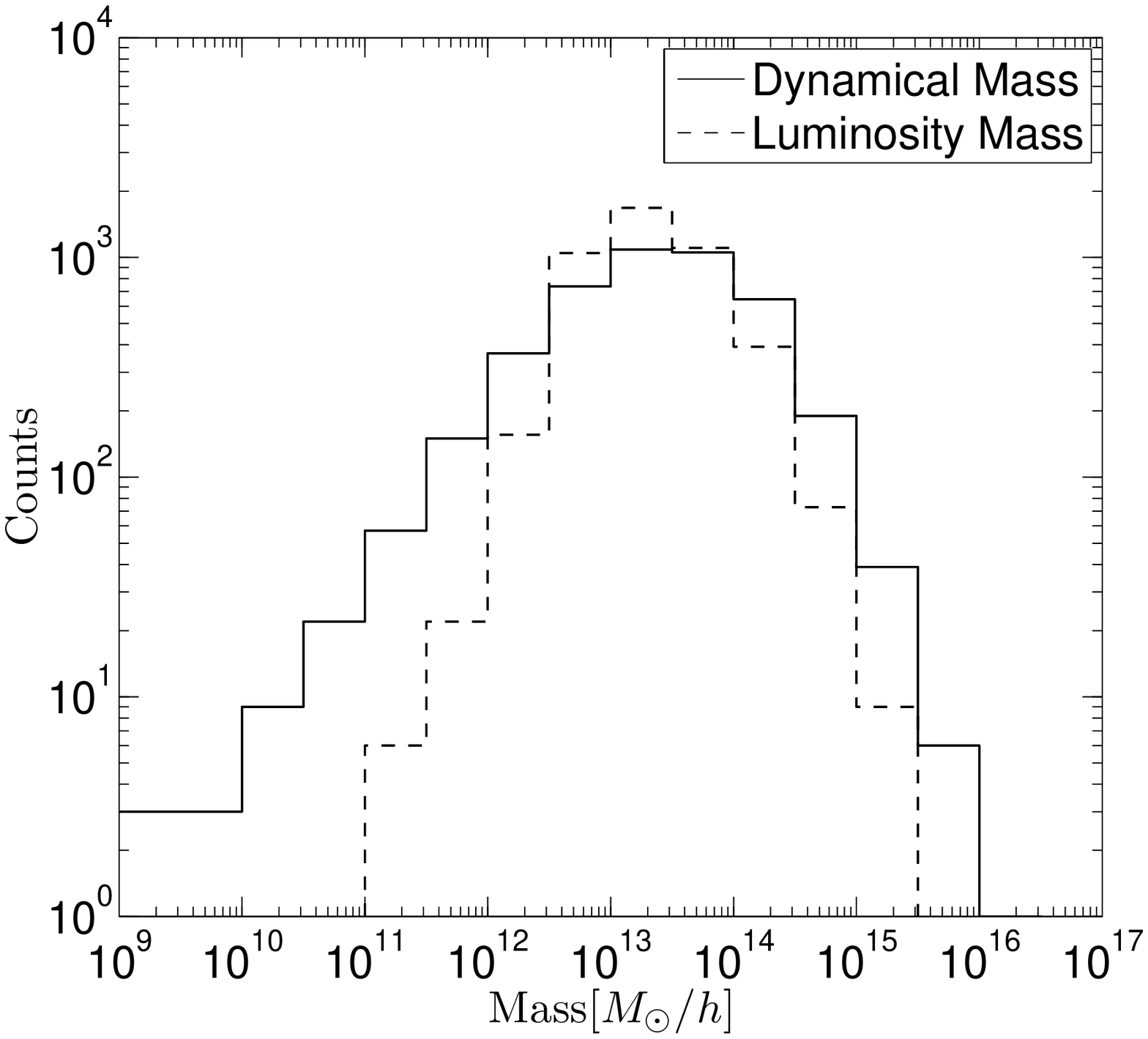}
\includegraphics[width=0.5\textwidth]{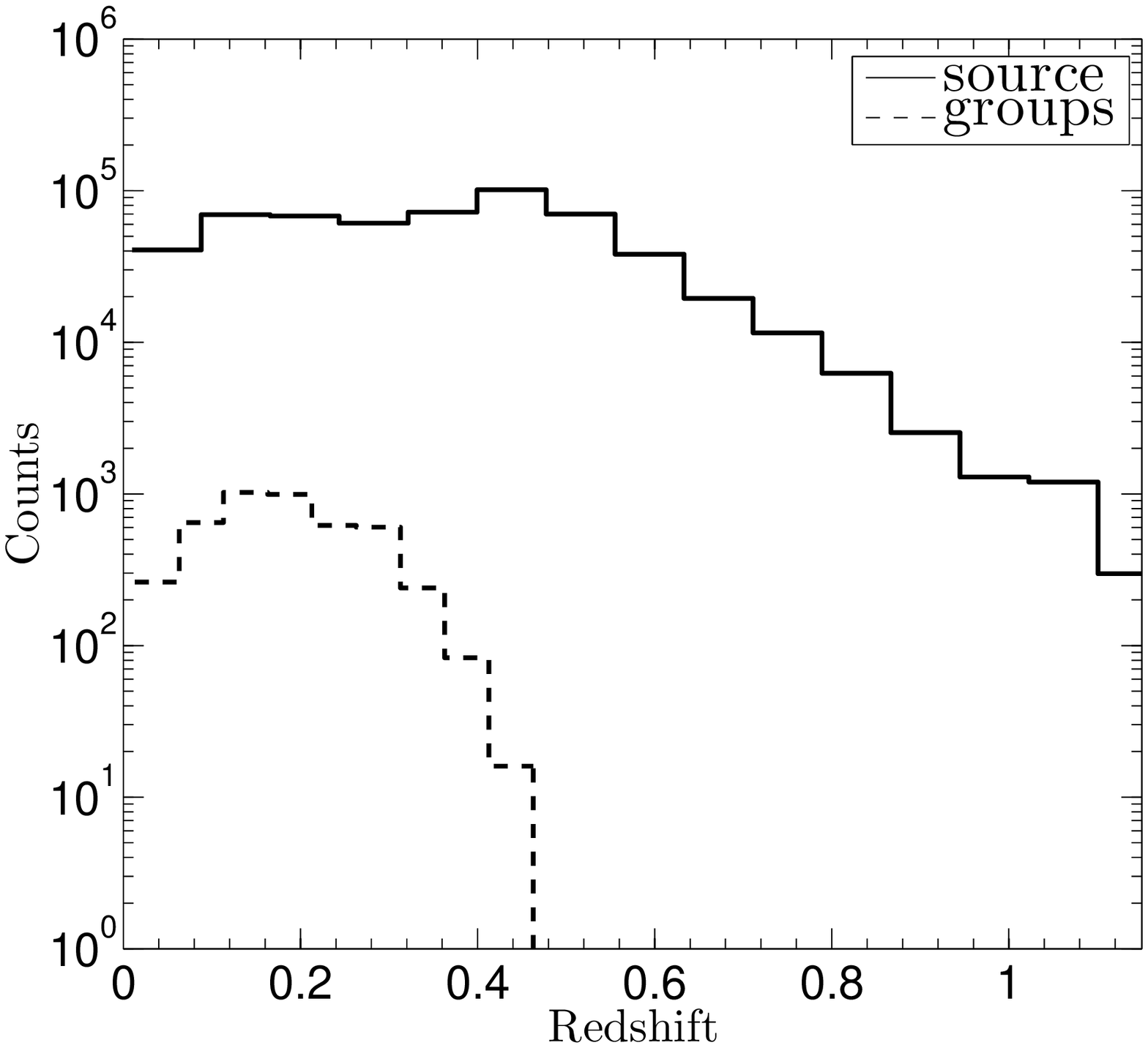}
\caption{The mass and redshift distributions of our lens and source
  samples. Upper panel: distribution of the G$^3$Cv5 dynamical and
  luminosity mass estimates for our group sample. Lower panel: redshift
  distribution of our GAMA group sample and the SDSS source
  galaxies. Note that only groups with at least three members are
  considered in this work.}\label{sample_dist} 
\end{figure}

As shown in the top panel of Fig.~\ref{sample_dist}, the GAMA groups
mainly reside in haloes of $10^{13}-10^{14}\msunh$. The dynamical mass
has a broader distribution than the luminosity mass, reflecting the
larger dispersion in the former estimate, particularly for the groups
with low membership. We find that the same luminosity mass calibration
method applied to the mock groups suggests that halo mass should be
more tightly correlated with luminosity mass than with dynamical mass.

\subsection{Source Catalogue: SDSS shape measurements}\label{sec_source}
We use as source galaxies those from the SDSS DR7 within and around
the three GAMA regions. The approach we follow is to measure a per-galaxy shape distortion, and then relate those to the shear applied to the ensemble of galaxies. The shapes of these galaxies are measured
by \citet{Mandelbaum05} and \citet{Reyes12} using the re-Gaussianisation technique, which we briefly describe in
Appendix~\ref{app_shape} for completeness. As recommended by \citet{Mandelbaum05} and
\citet{Reyes12}, we keep only those galaxies with extinction-corrected
$r$-band model magnitudes brighter than 21.8, $r$-band extinction
below 0.2, and galaxy resolution\footnote{The galaxy resolution is a
  measurement of how extended the galaxy is compared to the width of
  the PSF; see \citet{Mandelbaum05} for the exact definition.} above
$1/3$ in both the $r$ and $i$ bands. The photometric redshifts of
these galaxies are estimated using the template fitting algorithm
ZEBRA\footnote{\url{http://www.astro.ethz.ch/research/Projects/ZEBRA}}
\citep{ZEBRA}, the application of which in weak lensing is thoroughly
discussed by \citet{Nakajima12}. We further require that the ZEBRA
photo-$z$ determinations are successful using a non-starburst
template, considering the typically large photo-$z$ errors for
starburst galaxies. The final sample consists of $\sim 5.6\times 10^5$
galaxies, corresponding to a number density of $\sim
1~\rm{arcmin}^{-2}$. The lower panel of Fig.~\ref{sample_dist} shows
the distribution of galaxy photo-$z$ values, which peaks around
$z=0.5$ and extends to $z>1.0$.  The use of photo-$z$s for source
galaxies could introduce a bias and boost the error in our lensing
mass measurement. We address these issues with the help of Monte-Carlo
simulations as detailed in Section~\ref{sec_prior}, \ref{sec_sys} and
\ref{sec_result_sys}. %????

\subsubsection{Galaxy ellipticity}
For a purely elliptical galaxy image following a 2D Gaussian
brightness profile, its shape can be simply quantified by the axis
ratio and the direction of the major axis. Equivalently, we can
measure an ellipticity, $\VC=(\chi_1,\chi_2)$, defined as
\begin{equation}\label{eq_ellip}
\chi_1+i\chi_2=\frac{1-q^2}{1+q^2} e^{i2\phi},
\end{equation}
where $q$ is the minor to major axis ratio. $\phi$ is the position angle of the major axis, defined in a reference frame where the positive $x$ and $y$ axes point to the east and the north on the sky respectively. Note that the ellipticity is not a vector, since a rotation of the reference frame by $\psi$ transforms $\VC$ as
\begin{equation}
\left( \begin{array}{c} \chi_1' \\ \chi_2' \end{array} \right) = R(-2\psi)  \left( \begin{array}{c} \chi_1 \\ \chi_2 \end{array} \right)
\end{equation}, where 
\begin{equation}\label{eq_rotate}
R(\theta)=\begin{pmatrix} \cos\theta & \sin\theta \\ -\sin\theta & \cos\theta \end{pmatrix} 
\end{equation}
 is the rotation matrix.
However, we still write it in a vector form to simplify equations involving dot products of ellipticities and shears later, where $\VC\cdot\VC'=\chi_1\chi_1'+\chi_2\chi_2'$.  

A real galaxy image is typically the convolution of a non-Gaussian intrinsic image with a non-Gaussian point spread function (PSF). In Appendix~\ref{app_shape}, we describe how we measure the ellipticities for real galaxies. How the ellipticity relates to the lens distribution will be described in Section~\ref{sec_theo}.

\subsection{Mock Group Catalogues: Millennium light-cones}\label{sec_mock}
The GAMA group finder has been run on a set of nine light-cone mock
galaxy catalogues by \cite{G3C} to produce mock group catalogues that
allow us to compare the model predictions with observations and
investigate sample variance. The mock galaxy catalogues were created following the approach
developed by \cite{Merson13}, briefly summarised here.

Firstly a GALFORM semi-analytical galaxy formation model 
\citep[][in this case]{Bower06}
is run on merger trees extracted from the Millennium simulation
\citep{Millennium} to create the galaxy distribution within each
simulation snapshot. Using the individual snapshots (with replications if necessary), a
galaxy lightcone is generated by sampling the galaxies according to
their redshift and distance away from the observer. An interpolation
on galaxy position, velocity and k-correction is applied between
snapshots to avoid any abrupt transitions or features at snaphot
boundaries. All other galaxy properties are fixed to the earlier
snapshot. Finally the GAMA survey selection function is applied to the
galaxies in the lightcone to produce a mock GAMA survey. When
applying the survey selection function we force the mock luminosity function to
reproduce perfectly the observed luminosity function, by abundance matching. This
changes the $r$-band magnitudes of the original GALFORM predicted
magnitudes by less than 0.15 mags typically. 

This process is repeated for each of the nine different GAMA lightcone
mocks, all extracted from the same Millennium simulation, with some
limited attempt at reducing any overlap between each of them.
Futher details of their construction and limitations are given in
\citet{G3C} and \cite{Merson13}. 

The G$^3$Cv5 grouping algorithm was run on these GAMA mock surveys,
yielding the so-called mock group catalogues. Each mock group has the
same set of observational properties (and measured in exactly the same
way) as the real GAMA groups. For the purpose of this study, we also
associate a true halo mass, $M_h$, with each group by selecting the mass
of the dark matter halo hosting the iteratively-determined central
galaxy of the mock group\footnote{We 
  note that this matching differs from what was done in \citep{G3C},
  as discussed further in Section~\ref{sec_diagnose}.}.

\section{Method: Maximum Likelihood Weak Lensing}\label{sec_method}
One might wonder whether the Maximum Likelihood Weak Lensing (MLWL) technique is simply stacked weak lensing  extended to the unbinned limit, that is, the case in which one has at most a single lens-source pair inside each bin. While stacked weak lensing usually works pair by pair for all the
lens-source pairs, the MLWL method used here operates source by source. With
$N_s$ source galaxies one has $2N_s$ independent
observables since each source galaxy has two ellipticity components (see Equation~\eqref{eq_ellip}).
 Coupled with $N_\ell$ lens galaxies, this gives $N_\ell N_s$
tangential shear measurements. However, these $N_\ell N_s$ pairs are generally not
independent, and the covariance matrix of these measurements has at most
$2N_s$ non-zero eigenvalues. Hence the matrix is not invertible when $N_\ell
N_s>2N_s$ and, in this case, it is not possible to write down the
joint probability distribution function (PDF) of these variables. In
other words, attempts to extend stacked lensing to the unbinned limit will fail when the total number of
radial bins from all the mass bins exceeds
the intrinsic number of degrees of freedom in the source sample
($2N_s$). However, there is still a PDF associated with the dataset in
the unbinned case when one works in the eigenspace, i.e. directly with the shear of each source galaxy rather than
the tangential shear of each lens-source pair, as we do here. %We note that in practice stacked weak lensing does not enter the $N_\ell N_s>2N_s$ regime, but the above discussion points out the formal difference between stacked weak lensing and MLWL. %Thus, MLWL is not simply stacked lensing in the unbinned limit.

We start by describing the ellipticity of each source galaxy with a 2
dimensional Gaussian distribution, which enables us to write down a
likelihood function combining all the source galaxies. The model
dependence enters through the expected ellipticity of each
source. This is achieved by modelling the predicted ellipticity field
as a superposition of the shear field from all the foreground
lenses. Each lens is modelled as a circularly symmetric mass
distribution with a single parameter. In principle, the mass parameter
of all the lens haloes can be estimated simultaneously by optimizing
the joint likelihood of all the source ellipticities. However, this
means a parameter space with a dimensionality of the number of lenses,
$\sim 4500$ for our sample, which is not easily manageable. Besides,
the huge number of parameters also means extremely low signal to noise
for the estimate of each individual parameter. Instead of fitting the
mass of each halo explicitly, we reduce the dimensionality by
predicting their masses from group observables (e.g., group
luminosity), and only fit the parameters of a mass-observable
relation. We also take a second approach by dividing the sample into a
small number of bins according to observables, and fit a single
mass to all the haloes within each bin. In the following subsections
we give a detailed description of our implementation. 

\subsection{Predicting the ellipticity of source galaxies}\label{sec_theo}
The shear field, generated by foreground lenses, transforms the intrinsic ellipticities of source galaxies into observed ellipticities. To linear order, with our definition the ellipticity transforms 
under a small applied shear $|\vec{\gamma}|\ll 1$ as~\citep[e.g.,][]{BJ02}
\begin{equation}\label{eq_shear}
\VC=\VC_0+2\vec{\gamma}-(2\vec{\gamma}\cdot\VC_0)\VC_0.
\end{equation} If the intrinsic ellipticity, $\VC_0$, of each galaxy is randomly 
oriented with no correlation between its two components, then the expectation value of the observed ellipticity can be found from Equation~\eqref{eq_shear} as
\begin{equation}
\mathrm{E}[\VC]=2\res \vec{\gamma},
\end{equation} with 
\begin{equation}\label{eq_responsivity}
 \res=1-\sigma_{SN}^2
\end{equation}
 being the shear responsivity~\citep{BJ02}. Here
 $\sigma_{SN}^2=\mathrm{E}[\chi_{0,1}^2]=\mathrm{E}[\chi_{0,2}^2]$ is
 the intrinsic shape noise. In the presence of measurement errors, the
 responsivity (Equation~\ref{eq_responsivity}) is still valid, and can
 be equivalently derived from the more general Equation~(5.33) in
 \citet{BJ02} with a constant weight function.  
 
 It also follows from Equation~\eqref{eq_shear} that when
 $|\vec{\gamma}|\ll 1$, the predicted ellipticity $\hat{\VC}={\rm
   E}[\VC]$ due to the shear contribution from different lenses adds
 up linearly for the same source: 
\begin{eqnarray}
\left( \begin{array}{c} \hat{\chi}_1 \\ \hat{\chi_2} \end{array} \right) = 2\res\sum_\ell R(2\phi_\ell+\pi)  \left( \begin{array}{c}  \gamma_{t,\ell} \\  \gamma_{\times,\ell} \end{array} \right)
\label{eq_pred}
\end{eqnarray}
% \hat{\chi}_1&=&2\res\sum_\ell [-\cos(2\phi_\ell) \gamma_{t,\ell} + \sin(2\phi_\ell) \gamma_{\times,\ell}]\mbox{~~and}\label{eq_pred}\\
% \hat{\chi}_2&=&2\res\sum_\ell [-\sin(2\phi_\ell) \gamma_{t,\ell} - \cos(2\phi_\ell) \gamma_{\times,\ell}],\label{eq_pred2}
where $R(\theta)$ is the rotation matrix defined in Equation~\eqref{eq_rotate}, $\gamma_{t,\ell}$ and $\gamma_{\times,\ell}$ are the
tangential and cross shear produced by lens $\ell$ at the position
of the source, and $\phi_\ell$ is the position angle of the lens in the
reference frame of the source. Note the tangential reference frame is rotated by $\phi_\ell+\pi/2$ from the local source frame. The summation runs over all 
contributing lenses (i.e., foreground haloes).

\subsection{Lens models}
 We model the mass distribution of
each group as a spherical NFW \citep{NFW96,NFW97} halo, with a single
parameter, $M$, 
defined as the virial mass such that the average matter density inside
the virial radius equals 200 times the mean density of the
universe. The concentration parameter is fixed as in
\cite{Duffy08},
\begin{equation}
c=10.14\left(\frac{M}{2\times 10^{12}\msunh}\right)^{-0.081}(1+z)^{-1.01}.
\end{equation} %This equation is derived from simulations with WMAP5 cosmology.
 The data used in this study are consistent with these
assumptions, but do not provide sufficient leverage to fit the
internal halo profile while also determining the dependence of total
mass on other group observables; hence the restricted lens model
adopted here. %As a reference, shifting the concentration by 20\% typically results in less than 10\% shift in our fitted mass in the opposite direction.
We denote the virial radius following this definition as $R_{200b}$.
We also considered a central point source component
representing the stellar mass of the central galaxy, but found it had
almost no impact on the results and can be safely ignored for this
analysis. In this work we have also neglected the contribution to the
lensing signal from line-of-sight structures, the impact of lens group
asymmetries, and contamination from radial alignments of group member
galaxies. Previous studies \citep{Marian10,Schneider12,Schneider13}
suggest that these effects are likely to be present at a level of no
more than a few per cent. The contributions from these uncertainties, 
as well as the contributions from any lenses absent from our sample 
to the observed source shapes, are effectively considered as part of 
the intrinsic shape noise.

These circularly symmetric lenses induce only tangential shear in the source images,
 \begin{align}
\gamma_{\times,\ell}&=0 \mbox{~~and}\label{eq_crossshear}\\
\gamma_{t,\ell}&=\frac{\DS}{\SC}\label{eq_tshear}
\end{align}\citep[see e.g.][]{Schneider05}.
Here $\DS=\Sigma(<r) - \Sigma(r) $ is the difference between the mean
physical surface overdensity, $\Sigma$, within a radius $r$ and that at
$r$. Analytical expressions for $\DS(r)$ for NFW haloes can be found in
\citet{NFWlens}. The critical physical surface density is defined as 
\begin{equation}
\SC=\frac{c^2}{4\pi G}\frac{D_s}{D_{\ell s}D_{\ell}},
\end{equation}
where $D_\ell$, $D_s$ and $D_{\ell s}$ are the angular diameter distances to
the lens, the source and that between the two respectively. In
calculating these distances we adopt the fitting formula of 
\cite{DLfit}, which is accurate to within 1 percent, for relevant
cosmologies.

\subsection{Likelihood function}
 Following \citet{Hudson98}, we assume the noise in the observed $\VC$ follows a bi-variate Gaussian
 distribution. The probability of observing each source galaxy with
 shape $\VC=(\chi_1, \chi_2)$ is given as: 
\begin{equation}
p(\VC|\hat{\vec{\chi}})=\frac{1}{2\pi\sigma^2}e^{-\left[\frac{(\chi_1-\hat{\chi}_1)^2}{2\sigma^2}+\frac{(\chi_2-\hat{\chi}_2)^2}{2\sigma^2}\right]},
\end{equation}
where the single-component variance, $\sigma^2=\sigma_{\chi}^2+\sigma_{SN}^2$, is the sum of the measurement
noise, $\sigma_\chi$, and intrinsic shape noise,  $\sigma_{SN}$. For our
source sample, $\sigma_{SN}\simeq 0.4$ \citep{Mandelbaum05} provides the
dominant contribution to the total noise.\footnote{\citet{Reyes12}
  found the measurement noise in \citet{Mandelbaum05} was
  underestimated, leading to an overestimate of the shape noise used
  here. \citet{Reyes12} estimates the shape noise to be $\sigma_{SN}\simeq 0.35-0.37$. Adopting this shape noise would
  lead to a $\sim 2-4$ per cent increase in the responsivity. As a result, the derived masses would be lowered by a similar factor.}~ The predicted ellipticity,
$\hat{\vec{\chi}}$, is given by Equations~\eqref{eq_pred},
%\eqref{eq_pred2}, 
\eqref{eq_crossshear} and \eqref{eq_tshear}.

The likelihood function of our full lens-source sample is
written as,
\begin{equation}\label{eq_like}
L=\prod_i p_i,
\end{equation}
where $i$ runs over all the source galaxies. In principle, an
imperfect PSF correction could break the statistical independence of
the likelihoods for individual source galaxies so the combined
likelihood is no longer a simple product as above. Even in this case,
however, the large shape noise will make Equation~\eqref{eq_like} true to
good approximation.

\subsection{Likelihood ratio}\label{sec_likerat}
The log-likelihood function can be written as
\begin{equation}\label{eq_likerat}
\ln (L/L_0)=\sum_i \left[\frac{(2\vec{\chi}-\hat{\vec{\chi}})\cdot
    \hat{\vec{\chi}}}{2\sigma^2}\right]_i ,
\end{equation}
where $\ln L_0$ is a constant quantifying the log-likelihood
of the observed shapes given a model that predicts no gravitational shear.
This $L_0$ is independent of the model parameters, varying only with the data.

The observed ellipticity is the sum of a true ellipticity, produced by
the gravitational shear from the entire mass distribution, and a noise
component, i.e. $\vec{\chi}=\vec{\chi}_T+\vec{\epsilon}$, while the
predicted ellipticity can be written in terms of a difference from the true
shear-induced ellipticity, through
$\hat{\vec{\chi}}=\vec{\chi}_T-\vec{\chi}_\Delta$.
Note $\vec{\chi}_\Delta=0$ would correspond to a perfect model. With this
decomposition, the likelihood ratio reduces to 
\begin{align}
2\ln(L/L_0)&=\sum_i \left(\frac{\VC_T^2-\VC_\Delta^2 +2\vec{\epsilon}\cdot \hat{\VC}}{\sigma^2}\right)_i\\
&=\sum_i \left(\frac{S}{N}\right)_i^2-\left(\frac{\Delta S}{N}\right)_i^2 + \left(\frac{2 C_{MN}}{N^2}\right)_i,\label{eq_sigtonoi}
\end{align}
with the first term representing the signal-to-noise ratio of
the data, the second term deriving from discrepancies between the
model and actual gravitational shears, and the last term being the
model-noise cross-correlation. 

Intrinsic alignments of background galaxies and systematic biases in
the measured ellipticities can both produce regions of the survey in which 
$\langle\vec{\epsilon}\rangle\neq 0$. In addition, the predicted ellipticities near to
survey boundaries will typically have a preferential alignment,
because no contribution is taken into account in the model from lenses
just outside the survey region. Consequently, the model-noise
cross-correlation term in equation~\eqref{eq_sigtonoi} may well be
non-zero near to the edges of the survey region, meaning that the most
likely model will be biased by this cross-correlation. In
stacked weak lensing, 
this cross-correlation term shows up as systematic shear
\citep[see e.g.][]{Hirata04,Mandelbaum05}. It can equivalently be
understood as the non-zero residual of tangential shear averaged
inside boundary-crossing annuli around sources. This systematic shear
can be avoided by excluding the contribution to the source galaxy
predicted ellipticity from any lens more distant than the
nearest survey boundary, as we will do in Section~\ref{sec_cut}. 

In the absence of the cross-correlation term, the
interpretation of the likelihood ratio is now clear: it is simply the
difference between the signal-to-noise ratio in the data and that from
the model deficiency. Since $L_0$ is model independent, maximizing the
likelihood function is equivalent to 
minimizing the model deficiency. This makes MLWL an ideal tool to search for the tightest mass-observable relations, or the least-scatter halo mass estimators, as we will do in Section~\ref{sec_estimator}. 

According to Equation~\eqref{eq_sigtonoi}, one needs to avoid regions in the dataset
where $|\Delta S|\gtrsim S$ is expected, to avoid degrading the overall
signal-to-noise or producing biased fits. It also becomes clear from
Equation~\eqref{eq_sigtonoi} that stacking or binning reduces the
signal-to-noise by averaging the model within each bin, thus contributing to the
model deficiency term.

More generally, one can define a test statistic
\begin{equation}\label{eq_TS}
TS=2\ln(L/L_0),
\end{equation}
where $L$ is the maximum likelihood value of a full model while $L_0$
is that of the same model with some parameters fixed (the null
model). According to Wilks's theorem \citep{Wilks}, if the data follow
from the null model, then $TS$ follows a $\chi^2$ distribution with $n$
degrees of freedom, where $n$ is the difference between the number of free
parameters in the full and null models. For a measured $TS$, the
probability that such a large $TS$ is compatible with noise, or the
$p$-value, is simply $p=P(\chi^2_n>TS)$. This can be converted to a
Gaussian significance of $\Phi^{-1}(1-p/2)~\sigma$, where the function
$\Phi^{-1}$ is the quantile function of the Gaussian distribution. The covariance 
among parameters can also be estimated from the Hessian matrix of the likelihood ratio with 
respect to the parameters. 
Expanding the likelihood ratio locally around the maximum-likelihood value to leading order,
we have
\begin{equation}
\delta TS= \delta {\bf c}^T H \delta {\bf c},
\end{equation} where $\mathbf{c}$ is a column vector of model parameters, and $H=\nabla_c\ln(L)$ is the Hessian matrix. $\delta TS$ is now the likelihood ratio with respect to a null model at $\mathbf{c}+\delta\mathbf{c}$. Setting $\delta TS=1$, the $p=0.317$ critical value of the $\chi^2$ distribution, one obtains the covariance matrix of parameters as ${\Sigma}_{\rm cov}={H}^{-1}$. Alternatively, the errors on the parameters can be estimated using random catalogues, which are particularly useful when accounting for additional systematics. This will be described in Sections~\ref{sec_prior} and~\ref{sec_sys}.

\subsection{Prior distribution of nuisance parameters}\label{sec_prior}
So far we have assumed that the only uncertainties come from
the measured shapes of the source galaxies. However, the likelihood function
depends implicitly on the redshift of each source galaxy and
the inferred centre and predicted mass for each lens.
The uncertainties in these implicit parameters can be
accounted for by marginalizing over their prior distributions. With
any observed source redshift, $z_o$, observed lens centre, $\mathbf{X}_o$,
and predicted lens mass, $M_o$, the probability distributions of the
actual values can be written as $P(z|z_o)$, $P(\mathbf{X}|\mathbf{X}_o)$ and
$P(M|M_o)$. 
%Since any mass-observable relation is in general a
%correlation rather than a deterministic relation, $P(M|M_o)$ is
%generally not a $\delta$-function. 
Given knowledge about these additional
stochastic processes, the likelihood function can be more generally
written as
%\begin{figure*}
\begin{align}\label{eq_likefull}
L=\int &\prod_{s=1}^m p(\VC_s|z_s,\mathbf{X}_1,\mathbf{X}_2,...\mathbf{X}_n,M_1,M_2,...M_n) \\\nonumber
&dP(z_s|z_{o,s})\prod_{\ell =1}^n dP(\mathbf{X}_\ell|\mathbf{X}_{o,\ell}) dP(M_\ell|M_{o,\ell}),
\end{align}
%\end{figure*}
where the subscripts $s$ and $\ell$ represent different sources
and lenses in the sample. The above equation arises from assuming independent
prior distributions, $P(z_s|z_{o,s})$, $P(\mathbf{X}_\ell|\mathbf{X}_{o,\ell})$ and
$P(M_\ell|M_{o,\ell})$, for the different sources and lenses, although it is
straightforward to generalise to correlated distributions. 

Equation~\eqref{eq_likefull} is too computationally intensive to
solve directly. Instead, we take an indirect approach in our
likelihood optimization and continue to use Equation~\eqref{eq_like}
as our likelihood function. Ignoring the stochastic processes
mentioned above should lead to both a biased parameter estimation and
underestimated parameter uncertainties. In addition, Wilks's theorem would no
longer hold to 
interpret $TS$. To determine these biases and errors, we
will apply the same fitting process to a set of random catalogues in
which these additional stochastic processes have been introduced and the actual
parameter values are known. Comparing the distribution of the fitted
parameters with the input values, we can measure the bias and errors
introduced by this method. These can then be used to correct the
results inferred from the real observations, and calibrate where
significance levels lie within the $TS$ distribution.

To construct the random catalogues, the prior
distributions are chosen as follows:
\begin{eqnarray}
P(z|z_o)&=&\mathcal{N}(z_o,\sigma_z),\label{eq_pdf_z}\\
\mathrm{d}P(\mathbf{X}=\{r,\theta\}|\mathbf{X}_o)&=&\frac{r}{\sigma_\mathbf{X}^2}\exp\left(-\frac{r^2}{2\sigma_\mathbf{X}^2}\right)\mathrm{d}r\frac{\mathrm{d} \theta}{2\pi},\label{eq_PDF_X}\\
P(\log M|M_o)&=&\mathcal{N}(\log M_o,\sigma_{\log M}),\label{eq_Mprior}
\end{eqnarray}
where $\mathcal{N}(\mu,\sigma)$ is a Gaussian distribution with mean
$\mu$ and standard deviation $\sigma$, $r$ and $\theta$ are the
$2-$dimensional separation and position angle of $\mathbf{X}$ with respect to
$\mathbf{X}_o$, and $\sigma_{\log M}$ represents the width of the true halo
mass distribution for the appropriate mass estimator. For each source
galaxy, the photo-$z$ standard deviation is obtained by 
symmetrizing the ZEBRA one-sigma confidence limits, $z_\ell,z_u$, through
$\sigma_z=\sqrt{[(z_{u}-z_{o})^2+(z_{\ell }-z_{o})^2]/2}$. These estimated
photo-$z$ uncertainties are similar to those obtained by
\citet{Nakajima12}, who compared their estimated galaxy photo-$z$
values to 
spectroscopic redshifts. %Note that while $P(z_o|z)$ can be biased, $P(z|z_o)$ is unbiased, and fortunately we need the latter.

\begin{figure}
\myplot{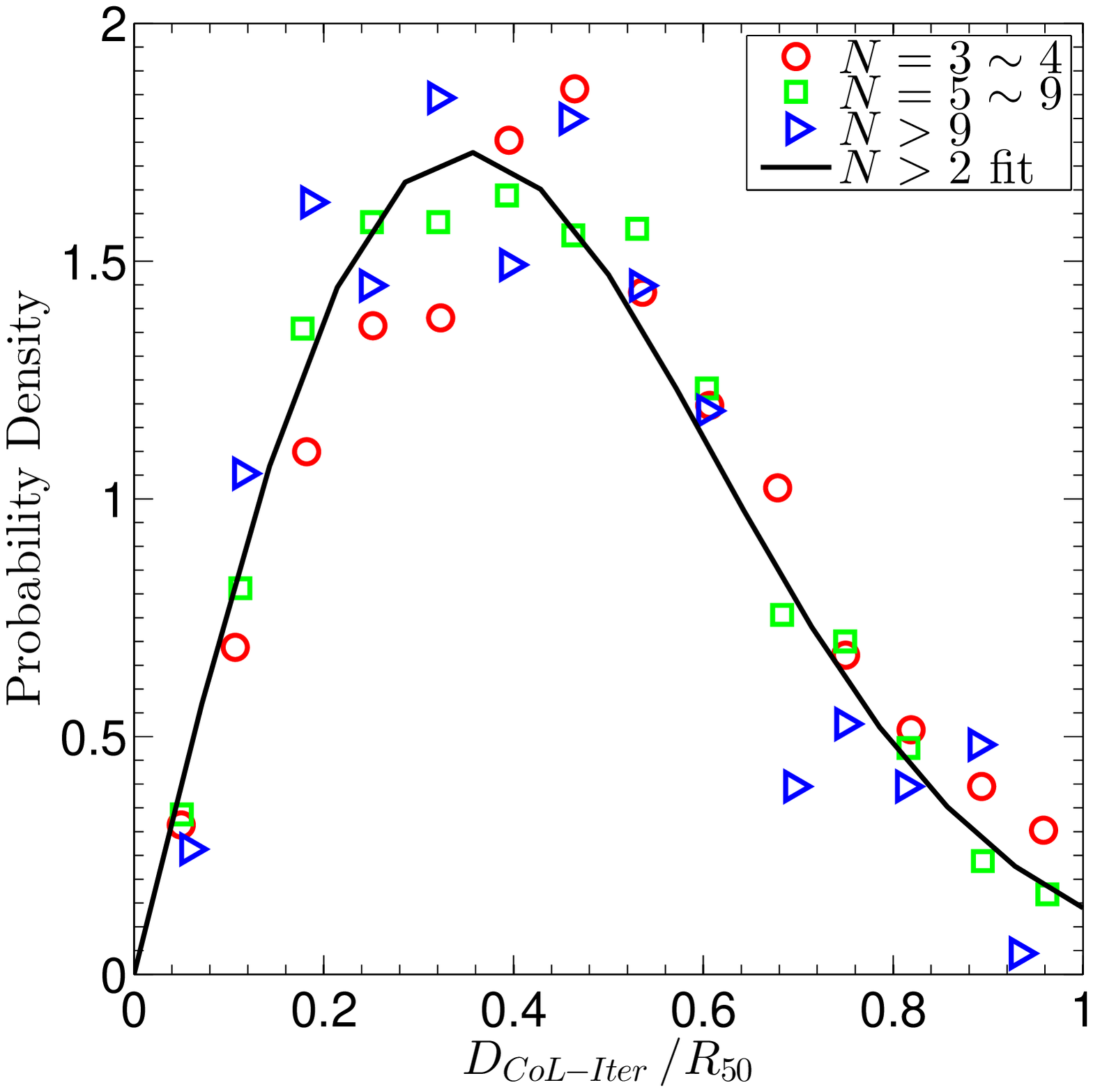}
\caption{Distribution of the offset between the centre of light and
  the iterative centre for G$^3$Cv5 groups. The separation between the two
  centres, $D_{CoL-Iter}$, is normalised by the comoving size, $R_{50}$,
  of each group. Different symbols represent
  different multiplicity ranges and the black solid line is a
  Rayleigh distribution fitted to the whole sample.}\label{fig_offset} 
\end{figure}

We choose to centre haloes on the
iteratively-determined centres of light, which in most cases coincide
with the locations of the BCGs. Although the existence of an offset
between BCG and the real halo centre is well recognised \citep[see,e.g.,][and
references therein]{Skibba11,Zitrin12,George12}, a reliable 
and general quantification of the offsets is not yet available. In
order to estimate the prior distribution of offsets between projected mass
and iteratively-determined centres of haloes, we start from investigating the 
displacements between different observational proxies of the halo centre. As shown in Figure~\ref{fig_offset}, the displacement
between the iteratively-determined and overall light centres in the real G$^3$Cv5,
when expressed in units of the group radius, $R_{50}\sim
0.2R_{200b}$, is almost independent of group observed mass or multiplicity. Its
distribution can be well fitted with a 2D-Gaussian with mean 0 and $\sigma_{\tilde{\bf X}}=0.35$ in each dimension, where $ \tilde{\mathbf{X}}=\mathbf{X}/R_{50}$. In Figure~\ref{fig_offset} only the distribution of $|\tilde{\bf X}|=D_{CoL-Iter}/R_{50}$ is plotted, which is a Rayleigh distribution (see Equation~\ref{eq_PDF_X}). Assuming the offset distributions of these two centres from the projected mass centre are independent and identical, this implies that $\sigma_{\tilde{\bf X}}=0.35/\sqrt{2}\simeq 0.2$. Hence we model the offset between observed and true halo centres with a Gaussian distribution having $\sigma_\mathbf{X}=0.2R_{50}$. The median, $\sigma_\mathbf{X}$, in our sample is $\sim 0.03\,h^{-1}\rm{Mpc}$, comparable with the estimation of $\sim 0.01\,h^{-1}\rm{Mpc}$ by \cite{Zitrin12} for their SDSS cluster sample, and with the BCG offset of $\sim 0.02\rm{Mpc}$ estimated in \citet{George12}.
 
To estimate the distribution of the true halo mass at a given
observationally-inferred mass, we make use of the mock G$^3$C
catalogues. In these, the true halo mass is well described by a log-normal
distribution for any given dynamical (luminosity) mass measurement,
with a dispersion $\sigma_{\log M}=0.6-0.8$ ($\sigma_{\log
M}=0.5-0.7$), increasing (decreasing) with mass. For a 
measured luminosity mass in the range of $10^{13}- 10^{15}\msunh$,
which comprises the majority of our lenses, the dispersion stays
roughly constant at $\sigma_{\log M}=0.5$, as is evident in the
upper-right panel of Fig.~\ref{fig_scaling}. In what follows, we adopt
$\sigma_{\log M}=0.5$ by default, but use $\sigma_{\log M}=0.7$ 
for the dynamical mass estimator. 

Note that if the full likelihood function (Equation~\ref{eq_likefull}) is used, then in principle any parameters that are part of the distribution function can be fitted. For example, one would be able to measure the mean logarithmic halo mass, or the median mass parameter for the lognormal prior distribution in halo mass (Equation~\ref{eq_Mprior}). As stated above, in this work we choose an alternative approach by fitting a simple likelihood function to the data ignoring the priors, and then calibrate the fitted parameters using random catalogues accounting for the priors. This is equivalent to the full likelihood fitting and, in the remainder of the paper, we will quote our calibrated best-fit mass-observable relations as the median mass-observable relation assuming a lognormal mass dispersion as in Equation~\eqref{eq_Mprior}.

\section{Data analysis}\label{sec_app}
\subsection{Data cuts}\label{sec_cut}
In order to minimize systematic uncertainties, a series of
cuts are applied to select the lens-source pairs for the likelihood
analysis, as described below. 
\begin{description}
\item[\textbf{Closed-circle cut.}]  The model-noise correlation 
  discussed in Section~\ref{sec_likerat} is most significant on
  intermediate ($\sim 10$~Mpc) to large scales for our sample, but in
  this paper we are considering the small-scale matter distribution in and
  around haloes. Thus, instead of making any correction, we completely
  avoid this boundary effect by imposing a closed-circle
  cut. Specifically, when predicting the gravitational shear at the
  location of galaxy $s$,
  $\hat{\vec\chi}_s$, our model only includes contributions from
  lenses that are 
  closer to the source than is the nearest survey boundary.

\item[\textbf{Virial cut.}] We model each lensing group with a single
  NFW halo. However, on large scales the two-halo term, i.e., the
  matter distribution contributed by nearby haloes, dominates over the
  single halo term. Since our sample does not cover arbitrarily low
  mass haloes, our modelled matter distribution will be missing the
  contribution from low mass haloes that are not
  modelled. \citet{Hayashi08} showed that a sharp transition in the
  halo-mass cross-correlation, $\xi_{hm}$, happens at a scale where
  $\xi_{hm}\sim 6$, corresponding to $\sim 3 R_{200b}$ for the halo
  mass and redshift ranges relevant here. Within this distance, the
  mass distribution is well described by a single halo profile. To
  avoid the correlation-dominated regime, we limit the analysis to
  within $2R_{200b}$ of each lens, where the virial radius $R_{200b}$
  of the group is estimated from its luminosity mass. This cut
  essentially decouples the lens modelling of haloes from each other,
  except for close by ones. As we further discuss in Appendix~\ref{app_2halo},
  the haloes that are not modelled should introduce a bias of less
  than $3\%$ in our mass estimates when our data cuts are applied.
%In the language of halo models, the correlation term
% describes the contribution from other surrounding haloes, and is
% partly accounted for by our simultaneous modelling of all the haloes
% in the catalogue.  We have estimated that within this radial range,
% the contribution from the linear correlation term is typically $<10\%$. 
%\footnote{Though we will see that the G3C luminosity mass
% overestimates the true mass by $\sim 2$, corresponding to a factor of
% $1.3$ in $R_{200b}$, our radial selection is still well within the
% halo term dominated regime.}  
%question: why such large improvement when 5->2? completeness effect(
% ignored lenses)? no, just the small groups contribute too much on
% large scale, hence bias the fits. 

\item[\textbf{Centre-offset cut.}] 
If the iteratively-determined lens centre is offset from the projected
centre of mass, then this will lead to an underestimate of the halo mass
with the largest errors being made in the central density profile. To
reduce any bias associated with this effect, lens-source pairs are
not used where the source is within a projected distance
$r_p=0.3R_{50}$ of a lens. This 
corresponds to a median physical radius of $\sim 0.04 \,h^{-1}\rm{Mpc}$. %, or a median angular seperation of $\sim 20$ arcsec.

\item[\textbf{Obscuration cut.}] It can be difficult to measure the
shapes of background galaxies that lie close, in
projection, to the bright central galaxies of foreground groups
\citep[e.g.][]{Hirata04,Mandelbaum05}. \cite{Budzynski12} found that
the obscuration 
radius of SDSS DR7 galaxies is about 5 arcsec, so we exclude any
source galaxy that lies within 10 arcsec of a foreground halo
centre in our sample. Adopting an even larger
cut of 20 arcsec makes no significant difference to our results. 

\item[\textbf{Redshift cut}] As a result of the large uncertainty in
  photometric redshift measurements, some foreground galaxies
  containing no real shear signal could be mistakenly identified as
  background source galaxies. To prevent excessive contamination from
  foreground galaxies, we only use source galaxies that have a
  photometric redshift at least $\Delta z=0.3$ higher than the
  spectroscopic redshift of the lens. This choice of redshift buffer
  is more conservative than the $\Delta z=0.1$ in
  \citet{Mandelbaum06b}, where they chose to correct for the remaining
  contamination with a boost factor. As there is a low lensing
  efficiency for small lens-source redshift separations, we estimate
  that our choice of the redshift buffer only reduces the
  signal-to-noise of the resulting measurements by $\sim25\%$ compared
  to no buffering, while effectively removing all foreground galaxy
  contamination (see Appendix~\ref{app_boost}).  This large redshift
  cut also reduces systematics from the photo-z prior distributions,
  since $\Sigma_{\rm crit}$ is less sensitive to source redshift for
  larger redshift separations between lens and source.

\item[\textbf{Multiplicity cut}.] As stated in Section~\ref{sec_data},
  we only keep groups with three or more members as lenses by default. Note that this creates a multiplicity-limited sample whose mass-observable relations could generally differ from those in a volume-limited sample. To compare our results with thoeretical models properly, we will use mock catalogues with the same selection criteria as the observational sample. We discuss this selection effect further when we compare our measured mass-light relation and stellar mass-halo mass relation with other measurements in Sections~\ref{sec_ML}, \ref{sec_SMHM}, and Appendix \ref{app_Nabs}.

%\item lens-selection cut? (exclude sources dominated by signal from
%  excluded lenses? no! effect so small, safely ignore.) 
\end{description}

\subsection{Assessing systematics: Bias, Error and Significance}\label{sec_sys}
Random catalogues are used to calibrate the MLWL method for systematic
errors introduced by the cuts applied to the data, described above,
and the ignorance of prior distributions highlighted in
Section~\ref{sec_prior}. These catalogues are Monte Carlo realizations
that also provide estimates of the statistical
significance of our results. We will call them \emph{random}
catalogues, to differentiate from the galaxy mock catalogues
described in Section~\ref{sec_mock}, which are produced using a
physically motivated galaxy formation model. 

The random catalogues are based on the subsets of G$^3$Cv5 groups and
SDSS source galaxies that we are considering. For any lens, an
observed mass is assigned from the parametrized mass-observable
relation that we try to calibrate.  True lens centres and masses are
chosen for each group from the distributions given in
Equations~\eqref{eq_PDF_X} and \eqref{eq_Mprior}, while the source
galaxies are assigned true redshifts via Equation~\eqref{eq_pdf_z} and
``intrinsic'' ellipticities according to a bivariate Gaussian with
$\sigma=\sqrt{\sigma_{\chi}^2+\sigma_{SN}^2}$. This realisation of the
source galaxy population is lensed by the set of inferred true lenses,
ignoring the two-halo term, which is estimated to produce under 3\%
bias in our results (see Appendix~\ref{app_2halo}). Thousands of 
random catalogues were created, with all observational properties
except source shapes being identical with those in the real data.

%For each source galaxy, the distortion produced by all the lenses are
%linearly added on top its intrinsic ellipticity in its local reference
%frame, to produce an observed ellipticity. We repeat this process 1000
%times to generate 1000 independent catalogues. We adopt a constant
%mass dispersion of $\sigma_{\log M}=0.5$ by default except for
%dynamical mass estimates, in which case $\sigma_{\log M}=0.7$ is
%suggested by the mock catalogues. Now each mock realization has
%exactly the same observational properties except source shapes, while
%their true properties follow our adopted prior distributions. 

The same analysis procedure is applied to each random
catalogue, adopting the same data cuts as were used with the real
data, yielding distributions of the
fitted parameters from the random catalogues. In all cases
studied, the distributions can be reasonably well described by a
normal distribution. For each parameter $x$ with mean, $\mu_x$, and standard
deviation, $\sigma_x$, in the random catalogues, we derive its bias
$\Delta_x=x_0-\mu_x$, where $x_0$ is the model input value. These derived biases
and uncertainties are then applied to the parameters inferred from the
real data, and we will quote our final measurements as 
\begin{equation}\label{eq_param}
x=\hat{x}+\Delta_x \pm \sigma_x,
\end{equation}
where $\hat{x}$ is the best fit parameter value from the likelihood
analysis on the real dataset. We leave the bias term explicitly in the
result, because it depends on our assumptions of the prior
distributions.

As explained in Section~\ref{sec_likerat}, in order to translate $TS$ to a significance value, one needs to know the distribution of $TS$ under the condition that the data are described by the null model. In the standard likelihood analysis, this is given by Wilks's theorem, to be a $\chi^2_n(TS)$ distribution. However, our $TS$ is estimated from a simple likelihood model which is only a crude description of the data distribution, because the prior distributions are ignored in the simple model. This invalidates Wilks's theorem. The distribution of our estimated $TS$ is therefore expected to differ from $\chi^2_n(TS)$, and we calibrate this distribution using our random catalogues. As in \cite{Han2012}, we find that the $TS$ distribution in the presence of systematics no longer follows a $\chi^2_n(TS)$ distribution. However, when scaled as $TS'=TS/b$, where $b$ is a consant to be determined, $TS'$ can be very well described by a $\chi^2_n(TS')$ distribution with the same number of degrees of freedom. In what follows, instead of interpreting $TS$ by reference to a $\chi^2_n(TS/b)$ distribution, we will make use of $TS'=TS/b$ which behaves as a standard $\chi^2_n$ variable. This $TS'$ serves as a ``corrected'' $TS$ and can be used to obtain significance levels using conventional $\chi^2_n$ distributions. By fitting the $TS$ distribution in the random catalogues with a $\chi^2_n(TS/b)$ distribution, we can derive the scale factor $b$ and use it to correct the observed $TS$ in the real measurements. 

\section{Results and Discussion}\label{sec_result}
In this section we describe in detail our measured mass-observable
relations, and compare these results with those in previous
studies. As mentioned in the beginning of Section~\ref{sec_method}, we
will be taking two complementary approaches in our fitting: 1) 
splitting the whole sample into several bins according to some
observable and fitting a single mass to all the haloes within each
bin; 2) predicting the mass of each halo from a mass-observable
relation and fitting for the parameters of this relation
globally. While the former approach is able to give a non-parametric
description of the mass-observable relations, the latter uses
all the lenses more efficiently and yields analytical descriptions of
the relations. We summarize these parametrized fits in
Table~\ref{table_par}, along with an estimated bias and uncertainty
inferred from random catalogues, in the form of
Equation~(\ref{eq_param}). We also list the correlation coefficients,
estimated from the likelihood surface between parameters from the
fitting, as well as the $TS$ with respect to a null model with no
gravitational shear.
%, as a reference for the goodness of fit of the halo mass model.
A visualization of the halo density profiles
through stacked lensing is provided in Appendix~\ref{app_stack}. Since
we will be showing the best fits together with systematic corrections
throughout the following sections, we first provide an overview of
those systematic corrections. 

\subsection{Overview of the systematic corrections}\label{sec_result_sys}
Systematic corrections to the fits are derived following the procedure
detailed in Section~\ref{sec_sys} by applying our likelihood fitting
to random catalogues that account for the prior distributions
described in Section~\ref{sec_prior}. After incorporating these
additional stochastic processes in the random catalogues, the derived
errors are generally larger than those estimated from direct
likelihood fitting of the real data. The systematic biases barely
affect the power-law slopes in our models, and are generally smaller
than the parameter uncertainties.

We use the random catalogues to check for effects of both the data cuts and the prior distributions. To check the effect of the data cuts alone, we first create two sets of random catalogues with no prior distributions. In generating these catalogues, one set has the same data cuts as in Section~\ref{sec_cut} when predicting the shear field, while the other set has no data cuts at all. In both cases, the true parameters and their errors are accurately recovered from direct likelihood fitting on the
random catalogues. This confirms that our adopted data cuts do not bias the
results, and that our MLWL method is working well when no systematics
from prior distributions are present. To calibrate the effect of the priors, we generate one set of random catalogues for each mass-observable relation that we wish to calibrate. Direct likelihood fitting is performed on these catalogues with no priors in the model, and the bias and error of the fitted parameters are extracted from their distributions after the fitting.

To see the separate effects of photo-$z$,
centre-offset and mass-dispersion on the fitting, we generate one
additional set of mocks including each effect individually for the last model 
in Table~\ref{table_par}, $M=AM_{\rm lum}$,
and assess their biases. This
yields $\Delta_{\log(A)}=0.02,0.08,-0.17$ for photo-$z$, centre-offset
and mass-dispersion biases respectively, revealing that one tends to
underestimate the mass when ignoring photo-$z$ bias and centre-offset
effects, while overestimating the mass by assuming there is no dispersion in the
mass-observable relation. Our estimated photo-$z$ bias translates into
$-5$ per cent in mass, in good agreement with that obtained by
\cite{Nakajima12} at our median group redshift $z=0.2$ \citep[ assuming $M\propto \DS^{1.5}$; see][]{Mandelbaum10}. \citet{Mandelbaum05a} found that the
stacked lensing-estimated mass lies in between the mean and median
value of the actual mass of the sample in the presence of a mass
scatter. This is consistent with our result that for a log-normal mass
scatter, the estimated mass is higher than the median (or the mean in
$\log M_h$). Note that the overall bias $\Delta_{\log(A)}=-0.09$ is
roughly the summation of the three, but is dominated by the mass
dispersion bias. Assuming a mass dispersion of 0.7 dex would lead to
$\Delta_{\log(A)}=-0.32$. In Table~\ref{table_par} we adopt a mass
dispersion of 0.5 dex by default, and use 0.7 dex for the dynamical
mass estimators (marked by $\dag$ in the table), leading to larger biases in their parameters.  

\begin{figure*}
\myplottwo{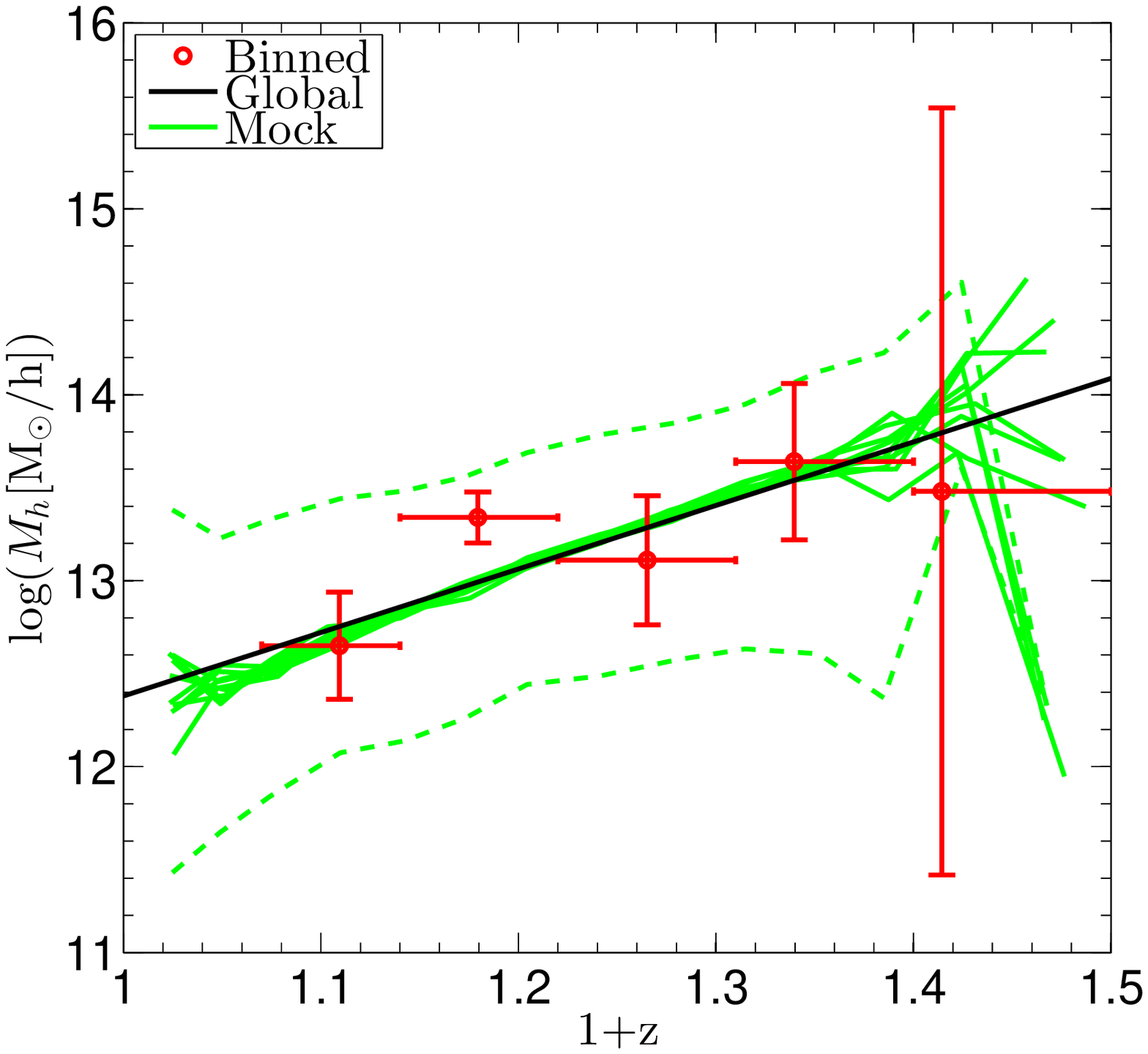}{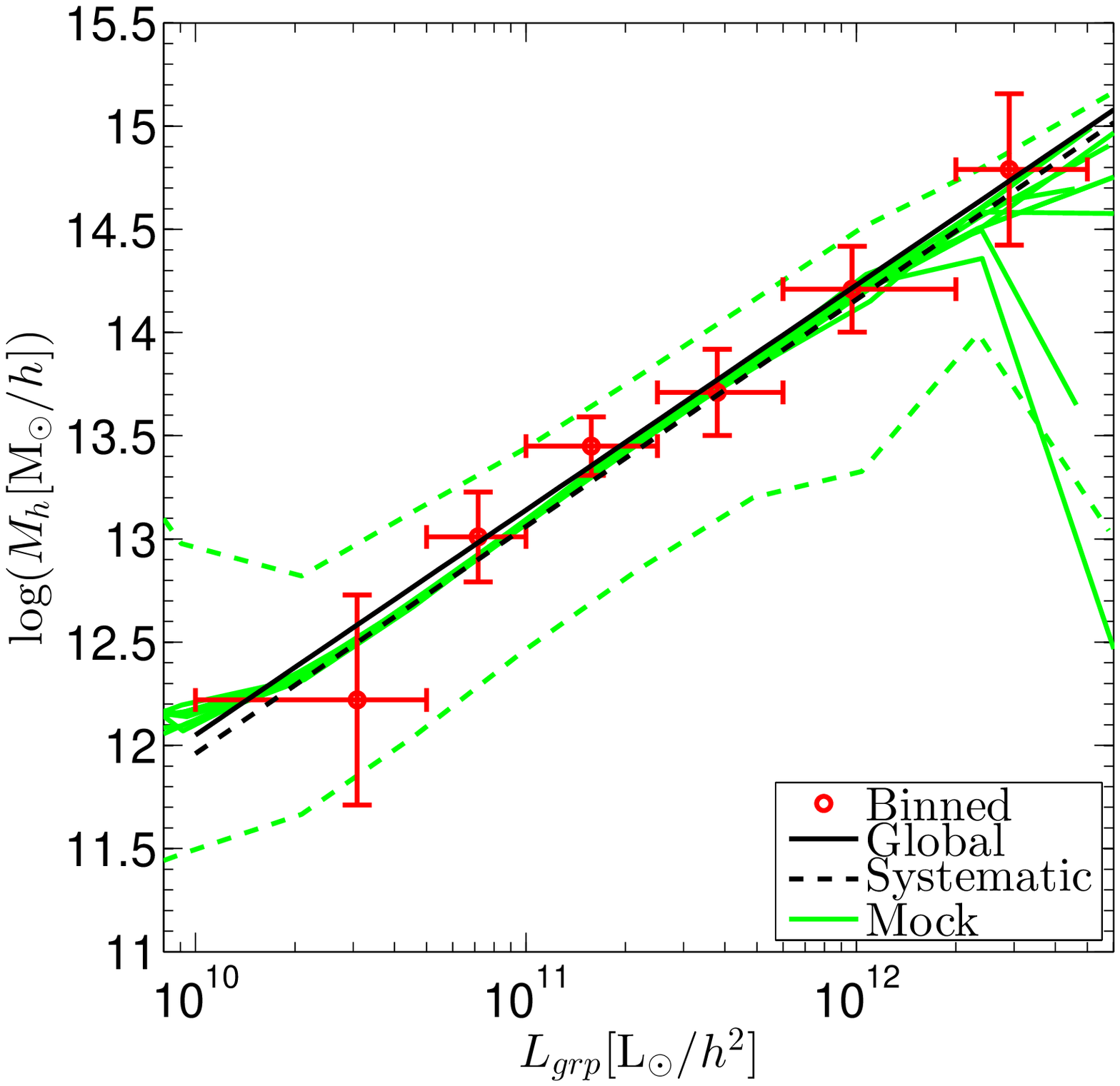}
\myplottwo{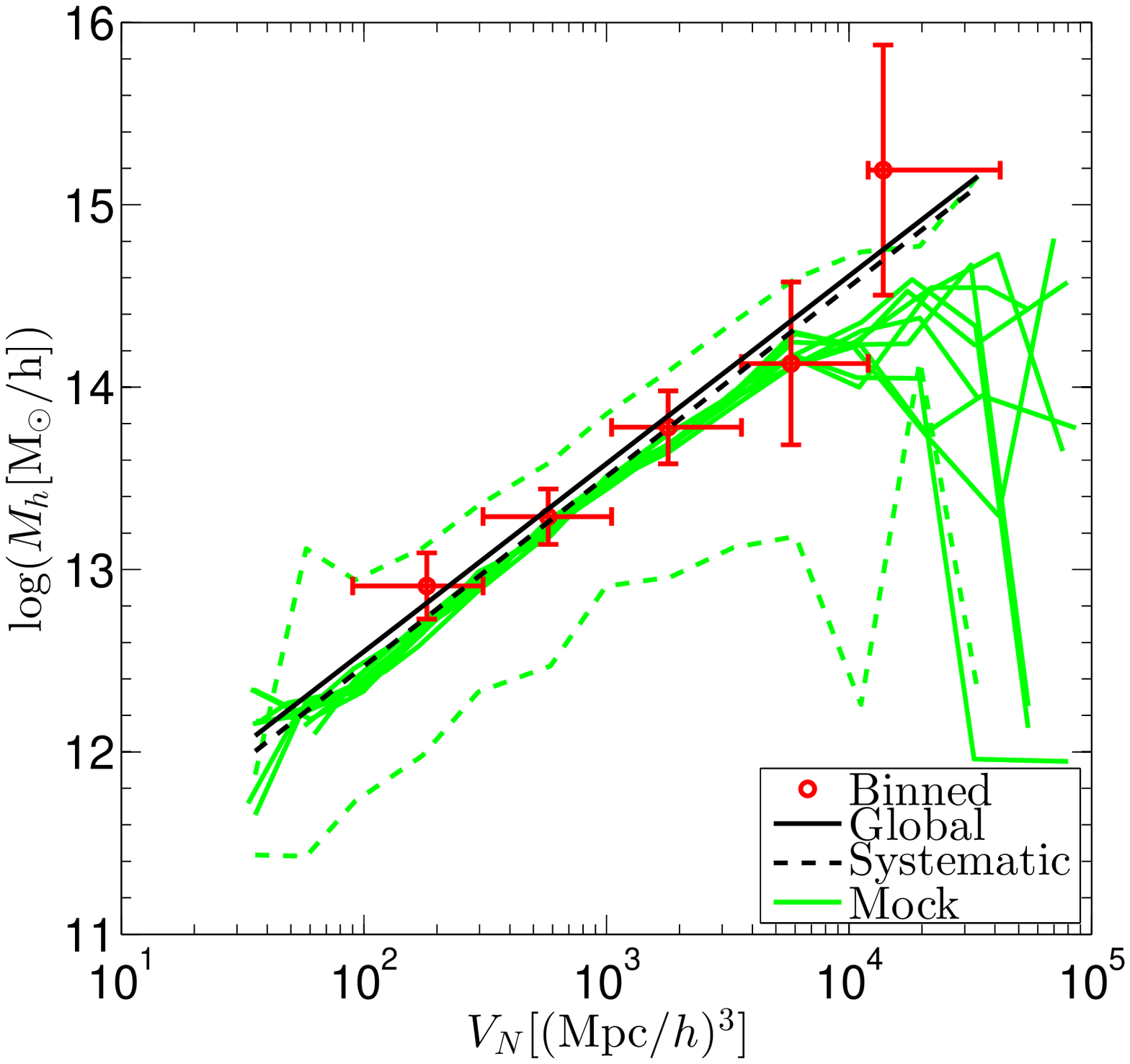}{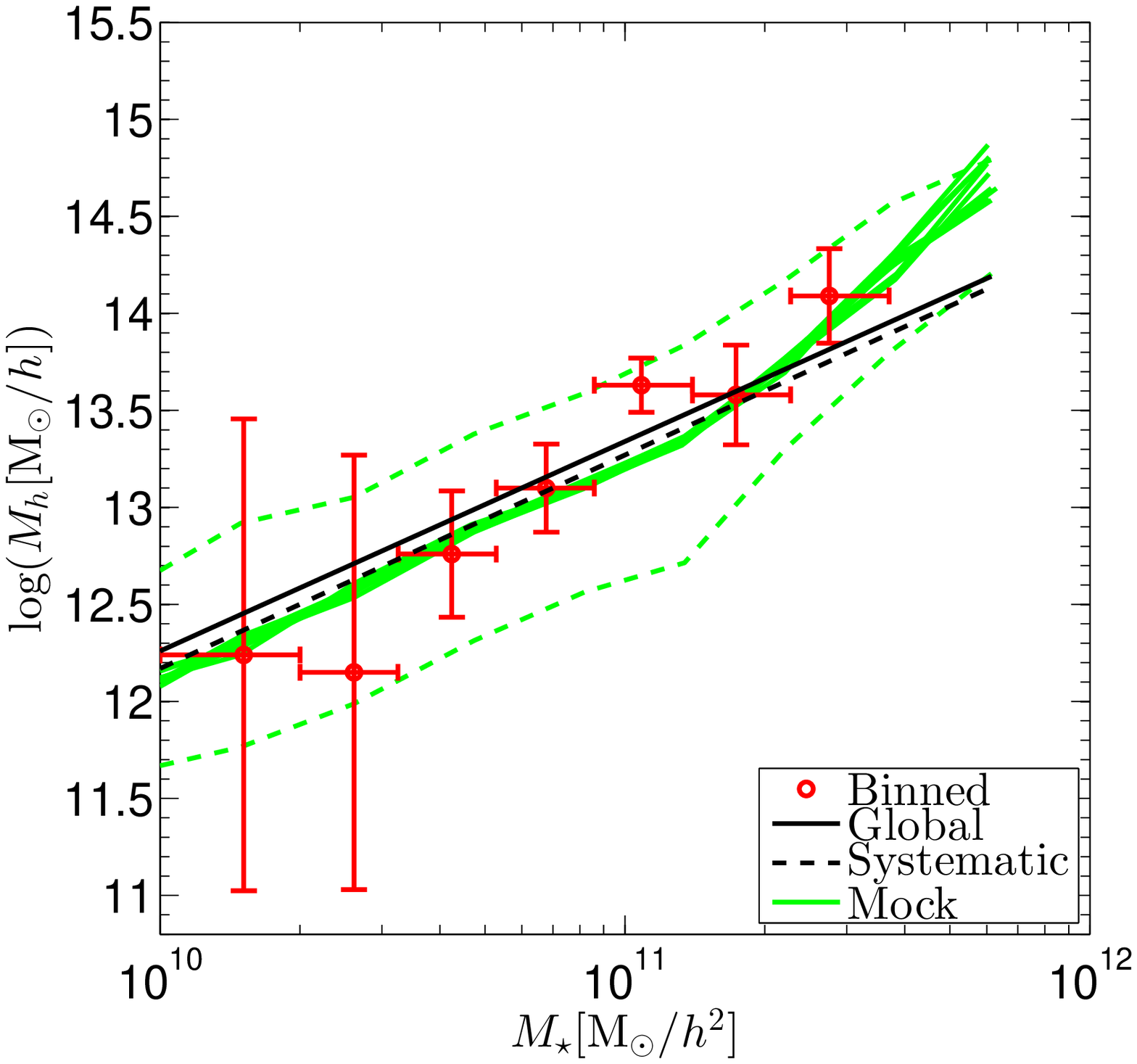}
\caption{Halo mass scaling with observed group properties. The
  upper-left panel contains the dependence of group mass on redshift,
  showing the selection in our sample. The other three panels
  show the scaling of halo mass with $r$-band group luminosity, $L_{\rm grp}$,
  multiplicity volume, $V_N$, and stellar
  mass of the central galaxy, $M_{\star}$. In each panel, the data
  points with errorbars show the halo mass measured using MLWL
  within each observable bin. The vertical errorbars are estimated
  from the weak lensing likelihood, while the horizontal errorbars simply
  mark the span of each bin. Black solid lines are the globally
  fitted power-law scalings from the MLWL method. The black
  dashed lines show the global fits after systematic correction (except in
  the redshift dependence panel). Note we have not applied systematic corrections to the binned measurements.
   Green solid lines depict the median relation extracted from 9 mock GAMA catalogues, while the green
  dashed lines mark the typical 16th and 84th (i.e., $\pm 1\sigma$) percentiles of the halo mass distribution in one mock
  catalogue.}\label{fig_scaling} 
\end{figure*}

\subsection{Mass-Observable Scaling Relations}\label{sec_scaling}
\begin{figure*}
\myplottwo{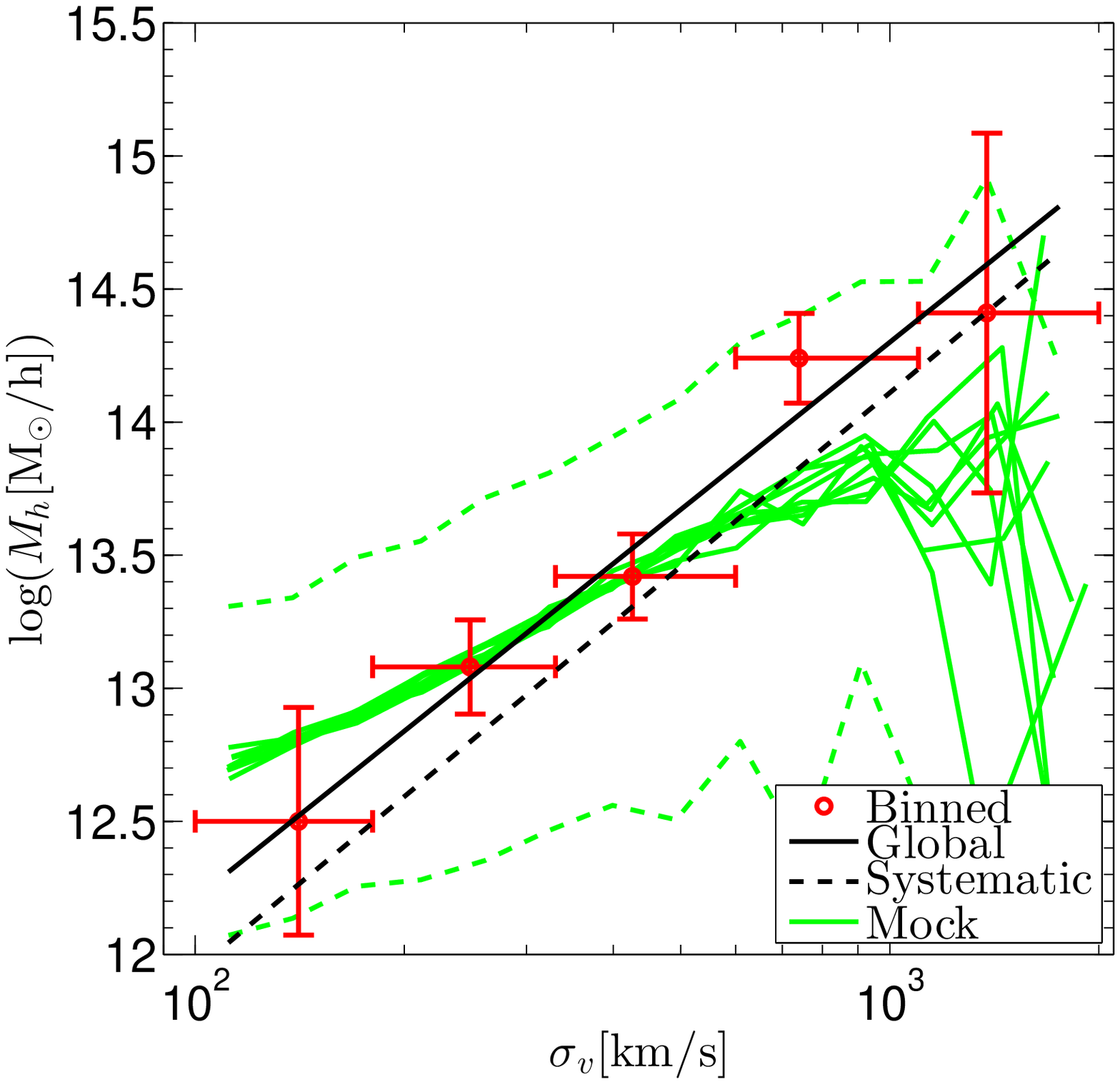}{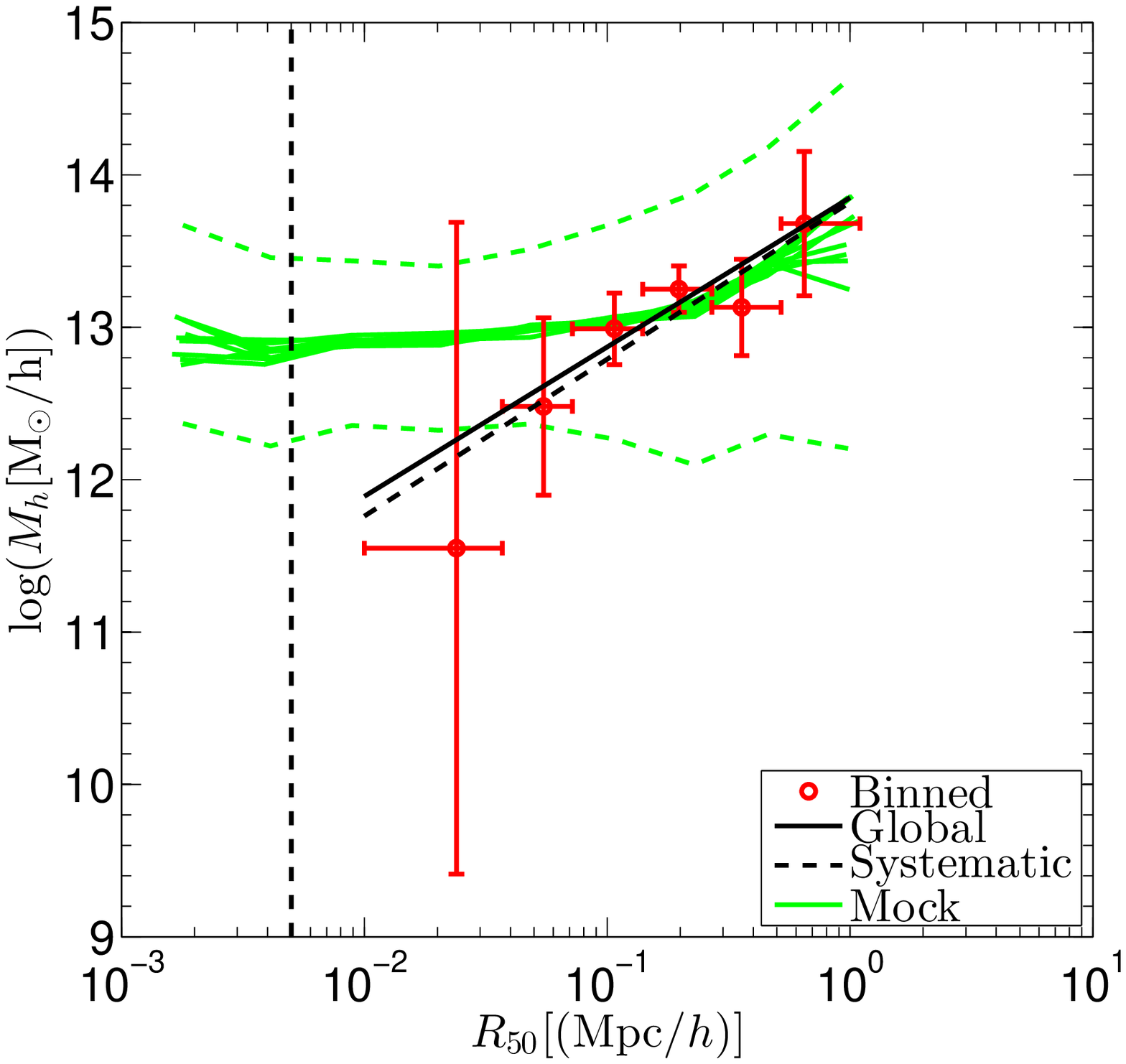}
\caption{Halo mass scaling with group velocity dispersion, $\sigma_v$,
  and observed radius, $R_{50}$. $R_{50}$ is defined to be the
  projected radius containing $50\%$ of the observed galaxies in the
  group. Line styles are the same as in Fig.~\ref{fig_scaling}. The
  vertical black dashed line in the $M_h(R_{50})$ panel marks the
  spatial resolution of the Millennium simulation underpinning the
  mocks.}\label{fig_scaling2} 
\end{figure*}

Before parametrizing the mass-observable relations with particular
functional forms, we can measure them in a non-parametric way by
splitting the lens sample according to a single mass proxy. We fit a
single mass parameter to each subsample of lenses. These measurements
will provide a consistency check with the parametrizations adopted in
our global fits. Unlike stacked lensing, we still do maximum
likelihood fitting over every source galaxy after splitting the lens
sample. As a result of our virial cut, the different lenses are mostly
decoupled from each other, and it makes little difference whether we
fit the bins jointly or independently. For the same reason, we
expect the systematic correction to each binned measurement to be the
same as that obtained for the global fit. In addition, because we
present the binned measurements mostly to reassure that our
parametrizations of the mass-observable relations are reasonable, and
because it is computationally expensive to estimate the systematic
corrections, we make no attempt to derive the corrections for the
binned measurements.

In Figs~\ref{fig_scaling} 
and \ref{fig_scaling2} we explore the
scaling relations for various mass observables: group velocity dispersion, 
$\sigma_v$; group total luminosity, $L_{\rm grp}$; stellar mass of the
iterative central galaxy, $M_*$; the multiplicity volume, $V_N$; and the
observed radius, $R_{50}$. Note that the total luminosity has been
corrected for unobserved galaxies in the group by integrating the
GAMA galaxy luminosity function.

A similar correction can be done for the observed multiplicity of
groups, $N$, to derive an absolute multiplicity $N_{\rm abs}$. Equivalently, we
choose to introduce a multiplicity volume as 
\begin{equation}\label{eq_VolMult}
V_N=\frac{N}{n(z)},
\end{equation}
where $n(z)$ is the average number density of observed galaxies at
redshift $z$. This volume translates into $N_{\rm abs}$ when multiplied by
the expected number density of galaxies down to the desired absolute
magnitude limit. Even though group redshift is not a physical mass
proxy, we still include it in Fig.~\ref{fig_scaling}
to show the group sample selection varies with redshift.

\begin{figure*}
\myplottwo{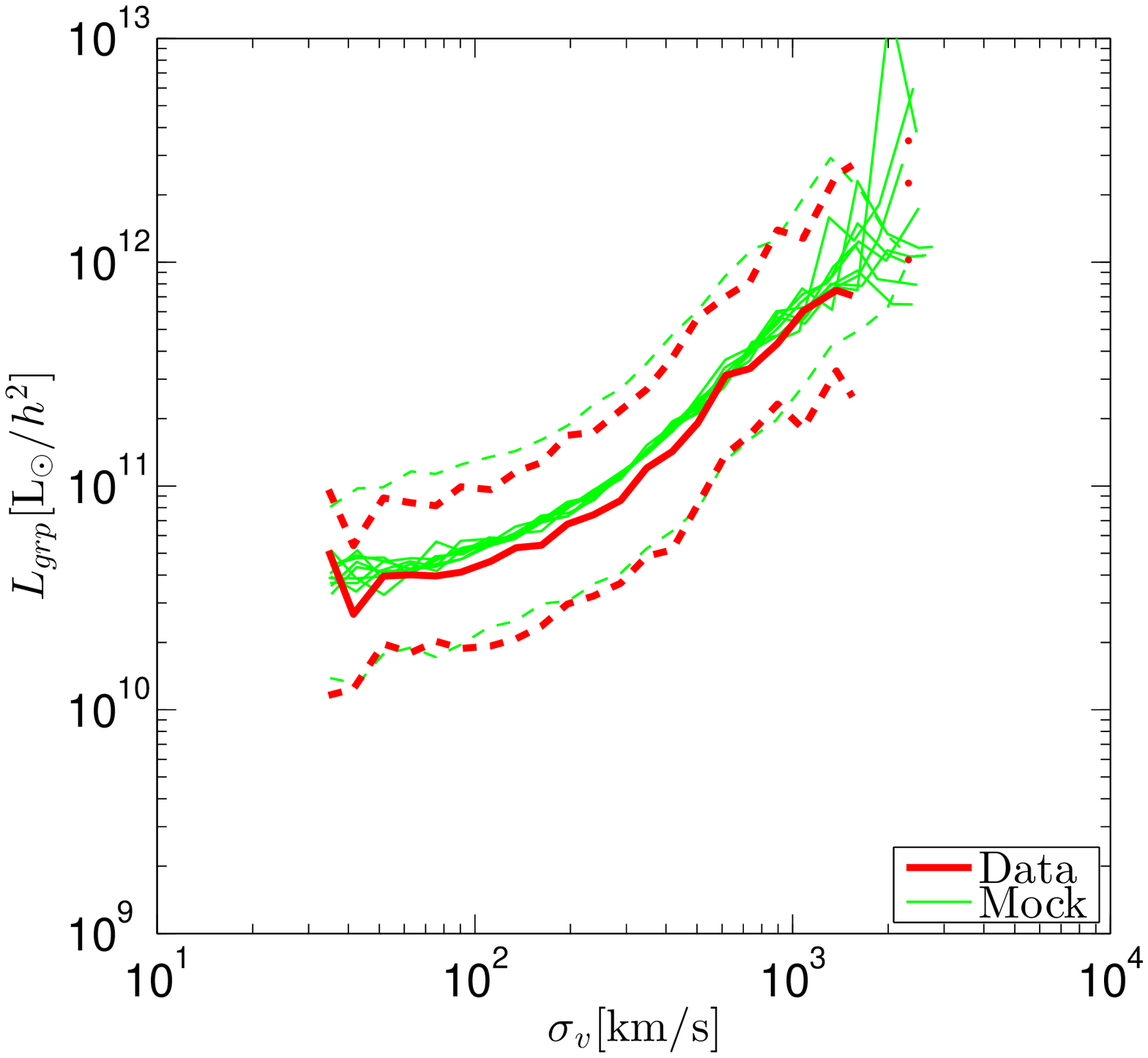}{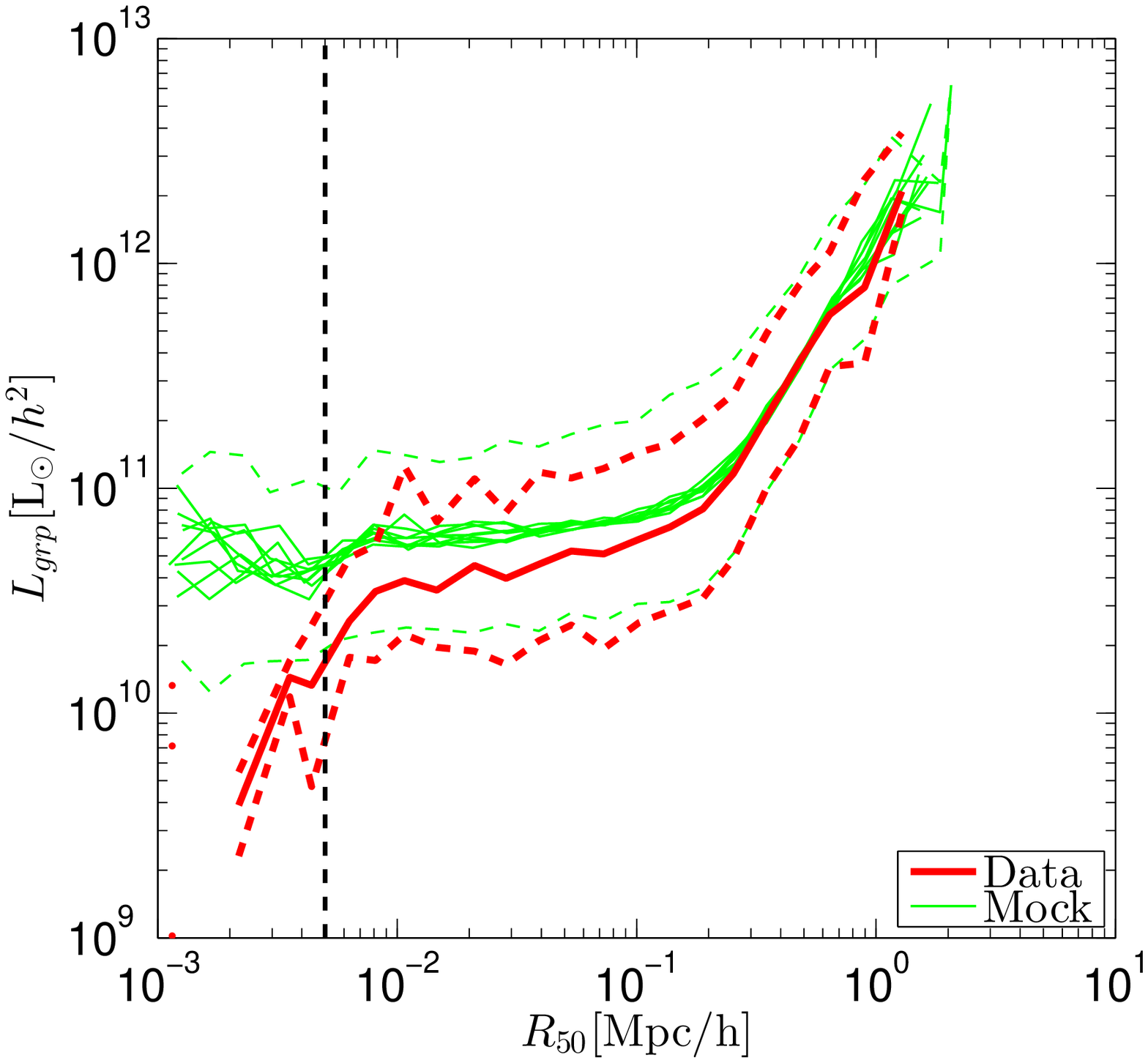}
\caption{The scaling of group luminosity with velocity dispersion, 
  $\sigma_v$, and size, $R_{50}$. In both panels, the solid lines show
  the median group luminosity at fixed $\sigma_v$ or $R_{50}$, while
  the dashed lines plot the 16th and 84th (i.e., $\pm 1\sigma$) percentiles. The red thick lines
  represent the results from the GAMA group catalogue, while the green thin lines
  are those from the mocks. Note that for the mocks we have plotted one
  median line for each mock catalogue, but only one set of $\pm
  1\sigma$ percentiles. In the $L_{\rm grp}(R_{50})$ panel, the vertical
  black dashed line marks the spatial resolution of the Millennium
  simulation underpinning the mocks.}\label{fig_scaling3} 
\end{figure*} 

All of these scaling relations can be well described by power-law functions.
Hence we provide global fits of these relations by modelling the mass of each group
as a power-law function of a single observable of the group, and
maximizing the likelihood of the entire sample. This produces the
parametrized relations in Table~\ref{table_par}, which are shown
with black lines in Figs~\ref{fig_scaling}
and~\ref{fig_scaling2}. As expected they all agree with the binned 
measurements. For each fit (except that for the redshift selection),
we also plot a dashed line showing the relation after applying 
the systematic correction listed in Table~\ref{table_par}. 

In Figs~\ref{fig_scaling} and \ref{fig_scaling2} we also compare our
measurements with predictions from mocks. From each of 9 mocks, we
extract the 16th, 50th and 84th percentiles of the halo mass
distribution as functions of the group properties. 
The measured redshift evolution of the halo mass
agrees well with that determined from the mocks, reflecting the identical
redshift selection in the real and mock catalogues. Both
the measured $M_h(L_{\rm grp})$ and $M_h(V_N)$ relations agree remarkably well
with those in the mocks. The former
may not be too surprising since the semi-analytical model underlying the
mocks is tuned to reproduce the galaxy luminosity function. In
addition, a further adjustment has been made to the galaxy
luminosities in the mocks, attributing any residual difference in the
redshift-dependent luminosity distribution to survey
photometry. However, these adjustments are guided by
the global luminosity function, while our measurement probes the
halo occupation distribution (HOD) of the galaxies, revealing that the
mocks have correctly populated haloes with (total) light. According to
our definition, $V_N$ is a measurement of the average clustering of
galaxies out to group boundaries. If one assumes that galaxies trace the
distribution of dark matter with a constant bias, then
$V_N\sim (4\pi/3)\Delta R_{200b}^3$ where $\Delta=200$ according to
our virial convention. Thus, one expects $M\propto V_N$, which
is what we see in Fig.~\ref{fig_scaling}, where the global fit
gives $M_h\propto V_N^{1.03+0.01\pm0.23}$. This agreement with the mocks
indicates that we have correctly modelled the average spatial
distribution of galaxies inside and outside groups. The measured
$M_h(M_\star)$ relation also agrees quite well with the mock predictions. 

The globally fitted $M_h(\sigma_v)$ relation is slightly steeper than
that seen in the mocks. This difference is too large to be accounted
for by the anticipated systematic error in the slope, and
persists when we use only $N>5$ groups. Examination of the
binned measurement shows that the difference originates from the
lower measured mass at low $\sigma_v$ and higher measured mass at high
$\sigma_v$. In both of these regimes, the uncertainties are much
larger than those at intermediate velocity dispersion
due to either intrinsically small halo mass or the
small number of stacked groups at high mass. Overall, the measurement
is marginally consistent with the mock predictions according to the
error bars. The small tension may originate from a velocity bias of
satellite subhaloes in the dark matter-only Millennium simulation.
%compared to simulations where baryonic physics are also
%incorporated. 
For example, \citet{Munari13} find that while galaxies
trace dark matter closely in SPH simulations, the velocity dispersions
of subhaloes in simulations with cooling are generally lower than those
in dark matter-only or adiabatic simulations, due to the longer
survival times for low velocity subhaloes in the former. The situation
is further complicated by the existence of ``orphan'' galaxies in
semi-analytical models, i.e. galaxies whose associated dark matter
substructure has become unresolved in the simulation. In this case,
the galaxy position is chosen to be that of the most
bound particle from the previously associated subhalo. This
results in a velocity distribution that follows that of the dark matter
particles. However, as found in \citet{Munari13}, there is a halo mass
dependent velocity dispersion bias between the subhaloes and 
dark matter particles. This could have given rise to the different
$M-\sigma_v$ slope we see in the mock. 
%Note that it is hard to explain the
%discrepancy with the velocity measurement error that is present in
%the real data but currently absent in the mocks. Adding noise in the
%mock catalogue tends to smear out the mass-observable correlation,
%making the relation flatter. In the real data, the velocity
%measurement error has been taken into account when estimating the
%group velocity dispersion. It makes little difference to our result if
%we use the raw velocity dispersion, for which such a correction is not
%applied. %\textcolor{red}{ToBeDone: What about velocity err/estimator
         %err? ($N/(N-1)$ factor). has some effect but not clear}
         %%adding noise in mock VelDisp does not help to lower the low
%%VelDisp mass, because in the mock the mass stays almost constant at
%%low mass end.  
A similar discrepancy is observed for the $M_h(R_{50})$ relation as well,
but only at small $R_{50}$, where the predicted halo mass is
almost constant while the measured mass keeps decreasing with
$R_{50}$. Despite this difference, the prediction is still marginally consistent with the measurements within the error ranges. 

If the $M_h(\sigma_v)$ and $M_h(R_{50})$ scaling relations are indeed
different in the data and in the mocks, then one might expect
different scaling of $L_{\rm grp}$ with $\sigma_v$ or $R_{50}$ as well,
since we have seen that $L_{\rm grp}$ is a good halo mass proxy in
both real and mock data. We compare these light-observable relations
in Fig.~\ref{fig_scaling3}. While there is little to distinguish
between the real and mock $L_{\rm grp}(\sigma_v)$ relations,
there is an obvious difference in the $L_{\rm grp}(R_{50})$ scaling
between the two, similar to the difference observed in the
$M_h(R_{50})$ relation. Note that this difference is most pronounced
near the spatial resolution $\epsilon=5 h^{-1}\rm{kpc}$ of the
Millennium simulation underpinning the mock catalogues, but is still
observable out to $R_{50}\sim 40\epsilon$. Similar results have been
found when measuring the galaxy correlation function, where
the autocorrelation function of red galaxies on small scales in the
GAMA mocks significantly exceeds that in the real data
\citep{Daniel}. This discrepancy ties in with studies of the radial
distribution of satellite galaxies, which find an overprediction of
model red satellites \citep{Budzynski12,Guo13,Wenting}. Our result is also in line with \citet{G3C} who find an overprediction of the number of compact groups in the mocks. 

The discrepancy between mock and data compact group luminosities could also be due to a selection effect caused by imperfections in the SDSS photometry. As the GAMA input catalogue was constructed from the SDSS photometric galaxies, selection effects in the latter could propagate to the GAMA catalogue. It is known that near bright galaxies, the flux level of the background sky could be overestimated in the SDSS\citep{SDSSDR6}, leading to an underestimate of the flux of neighbouring galaxies. As a result, faint galaxies in the vicinity of bright ones could be missing from the flux-limited GAMA galaxy catalogue, which in turn could remove bright and compact groups from the group catalogue. Note this type of selection is not implemented in the current GAMA mock catalogues, which could result in an excess of bright groups at small $R_{50}$ compared to observations.
%Because the GAMA input catalogue was constructed from SDSS photometric galaxies, bright and compact groups could be missing in the group catalogue if the SDSS deblending fails to seperate their members as individual galaxies. 

The model's treatment of orphan galaxies, which dominate the
galaxy population in the inner halo, may also be responsible for the differences between model and data $M_h(\sigma_v)$ and $M_h(R_{50})$ relations. Changes in how the positions
of these galaxies are modelled, and in the dynamical friction time 
estimation can both affect the satellite abundance and hence the size
distribution of sample groups. For 
example, \citet{Jiang08} found that the dynamical friction time scale
inferred by \citet{Bower06} is overestimated for major mergers,
resulting in an excess of orphan galaxies in the
model. Lastly, when constructing the light-cone mocks,
it is necessary to interpolate the position and velocity of galaxies between
simulation snapshots in order to place galaxies in an
observer's past light cone. Even though \citet{Merson13} have 
tried several different interpolation methods, we do not
exclude the possibility that those interpolations could distort the
spatial and velocity distributions of galaxies, contributing to our
observed discrepancies.

\subsection{Comparison of group mass to light ratio with 2PIGG measurement}\label{sec_ML}
\begin{figure*}
\myplottwo{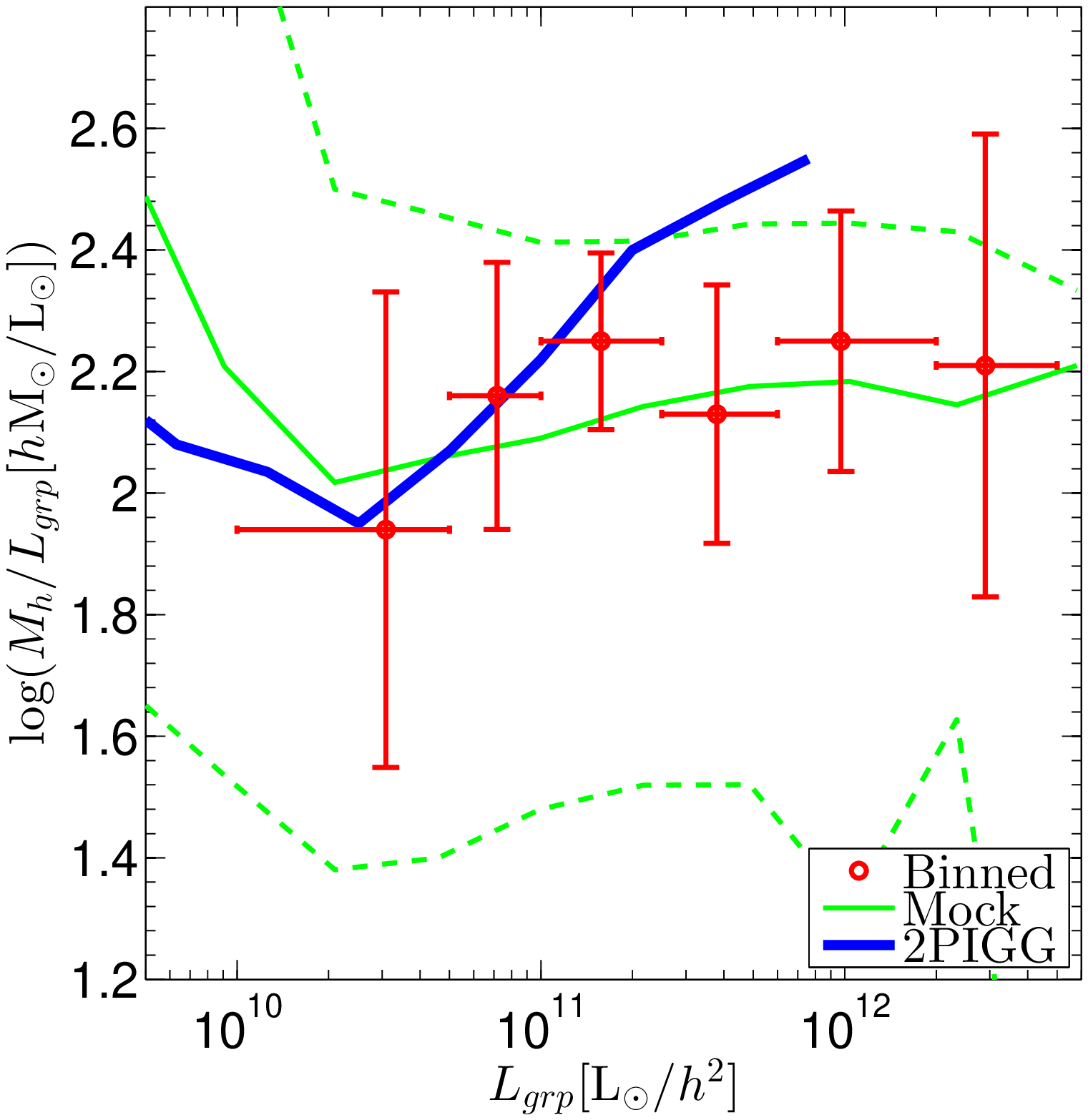}{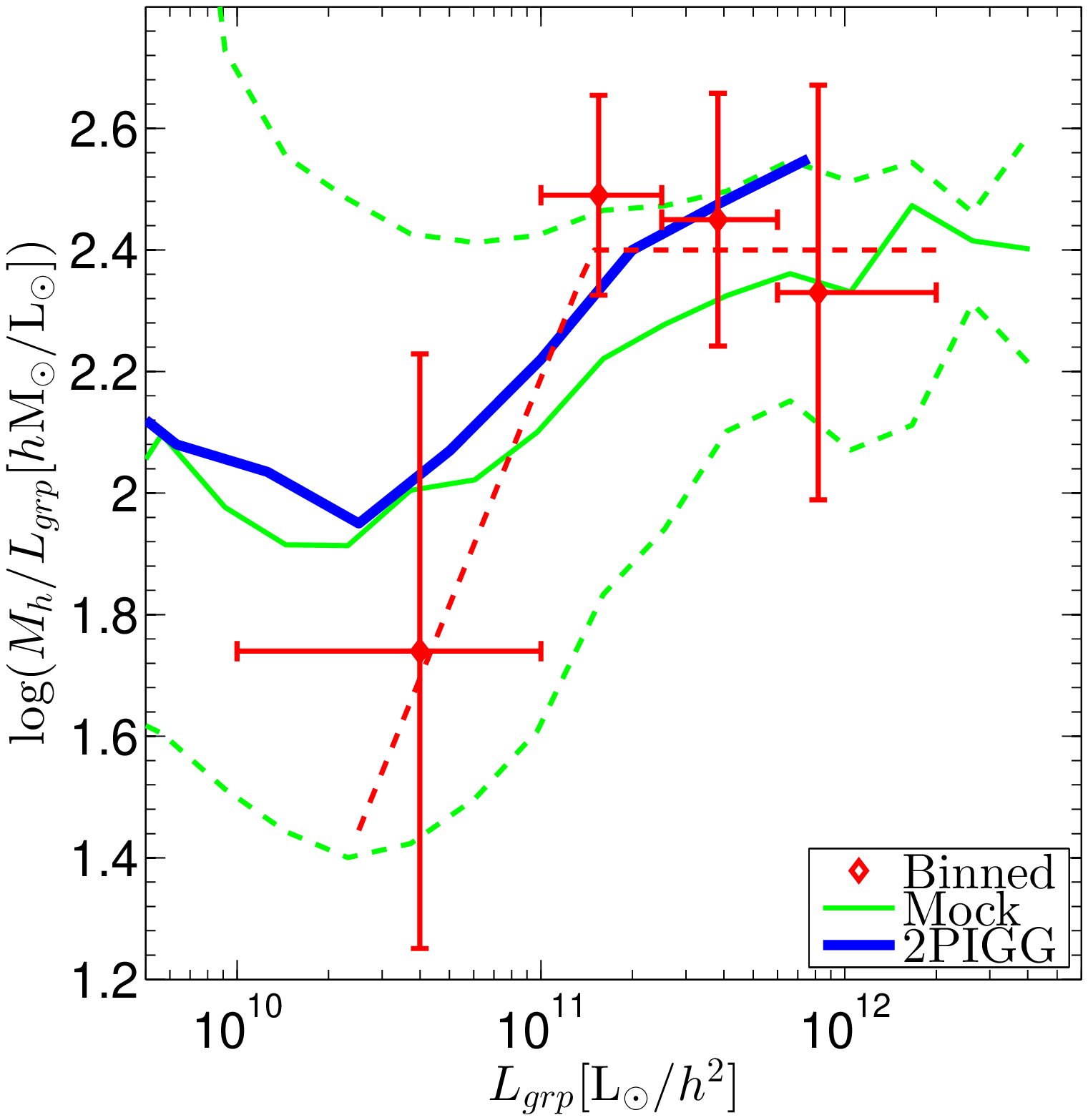}
\caption{The mass-to-light ratio of groups. In the left-hand panel, the
  points with error bars are the MLWL fitted $M_h/L$ within each luminosity
  bin. A green solid line shows the median $M_h/L$ at fixed luminosity in
  the mocks, while the green dashed lines are the 16th and 84th (i.e., $\pm 1\sigma$)
  percentiles. The blue thick line is the 2PIGG-inferred $M_h/L$ \citep{2PIGGML}.
  The right-hand panel is like the left, just with the groups in
  both real and mock samples further selected to mimic the 2PIGG
  selection function. The red dashed line is an
  unbinned broken power-law fit of the form $M_h/L=A
  \min(L_{\rm grp}/L_p,1)^b$.}\label{fig_MLrat} 
\end{figure*}

In the left-hand panel of Fig.~\ref{fig_MLrat}, we compare the measured
group mass-to-light ratios with those from the
2 degree Field Galaxy Redshift Survey \citep[2dFGRS, ][]{2dF}
Percolation-Inferred Galaxy Group catalogue \citep[][2PIGG
  hereafter]{2PIGGML}. The mass-to-light ratios in 2PIGG are derived from group dynamical masses and $r_F$ band total luminosities. We have calibrated their $r_F$ band luminosity to $r$ band in the comparison. This time our measurement from the G$^3$Cv5 is
done by fitting a 
constant $M_h/L$ value to all groups within each luminosity
bin. For the mocks, we measure the median and 16th and 84th (i.e., $\pm 1\sigma$) percentiles of
$M_h/L$ within each luminosity bin. \footnote{The results
  are quite similar if we convert the $M_h(L_{\rm grp})$ scaling relations
  obtained in the previous section to $M_h/L$.} Again our measured $M_h/L$ agrees
very well with the mock prediction. Below $L_{\rm grp}=2\times
10^{11}\lsunhh$, it appears that our measurement is also in good
agreement with that from the 2PIGG catalogue. However, for brighter groups, our
measured $M_h/L$ stays almost constant, while the 2PIGG $M_h/L$
continues to increase with $L_{\rm grp}$. This difference can be largely
explained by the different depths of the two surveys. 
Since groups are selected to have a minimum
number of $N_{\rm min}$ galaxies in both catalogues, the group selection
function can be described as
\begin{equation}\label{eq_VolMultLim}
V_N^{\rm lim}(z)=N_{\rm min}/n_{<M_{\rm lim}(z)},
\end{equation}
where $n_{<M_{\rm lim}(z)}$ is the number density of galaxies above the survey flux limit.
As the GAMA survey is $\sim 2$ magnitudes deeper
than the 2dFGRS, we expect 2PIGGs to behave like poorly sampled
GAMA groups. Note that the completeness of the
2dFGRS is not as uniform as in GAMA, so the estimated $n_{<M_{\rm lim}(z)}$
varies across the sky. We model this $n_{<M_{\rm lim}(z)}$
with a Gaussian distribution at given $z$, and generate a random
$n_{<M_{\rm lim}(z)}$ for each GAMA group at $z$ to account for the
variation of completeness in the 2dFGRS.
Repeating the $M_h/L$ calculation on a sub-sample of
our groups selected with the 2dFGRS depth, $V_N>V_N^{\rm lim,2dF}$, which
constitutes $\sim 1/3$ of our standard sample, gives the results in
the right-hand panel of Fig.~\ref{fig_MLrat}. Due to the reduced
signal-to-noise, we also show an unbinned broken power-law fit of the
form $M/L=A \min(L_{\rm grp}/L_p,1)^b$, where $L_p$ and $b$ are the
parameters to be fitted.
Using the 2dFGRS selection function decreases the measured $M_h/L$ at
low $L_{\rm grp}$ due to the inclusion of $N=2$ groups, while
that at high $L_{\rm grp}$ is increased due to the exclusion of low $V_N$
groups at given redshift. This time our measurement largely agrees
with the 2PIGG result for groups around $L_{\rm grp}=3\times
10^{11}\msunhh$, showing the importance of sample selection when
comparing observed group properties with other results. %Our measured $M_h-L$ relation is in broad agreement with some HOD models of galaxy clustering \citep[e.g.,][]{Brown08} which can give the average luminosity at a fixed halo mass. However, a proper comparison will involve predicting the median halo mass at fixed observable value from the HOD model, and account for the sample selection, which we demonstrate below when comparing the stellar mass-halo mass relation with HOD models.

\subsection{Comparison of the halo mass-stellar mass relation with other works}\label{sec_SMHM}
\begin{figure*}
\myplottwo{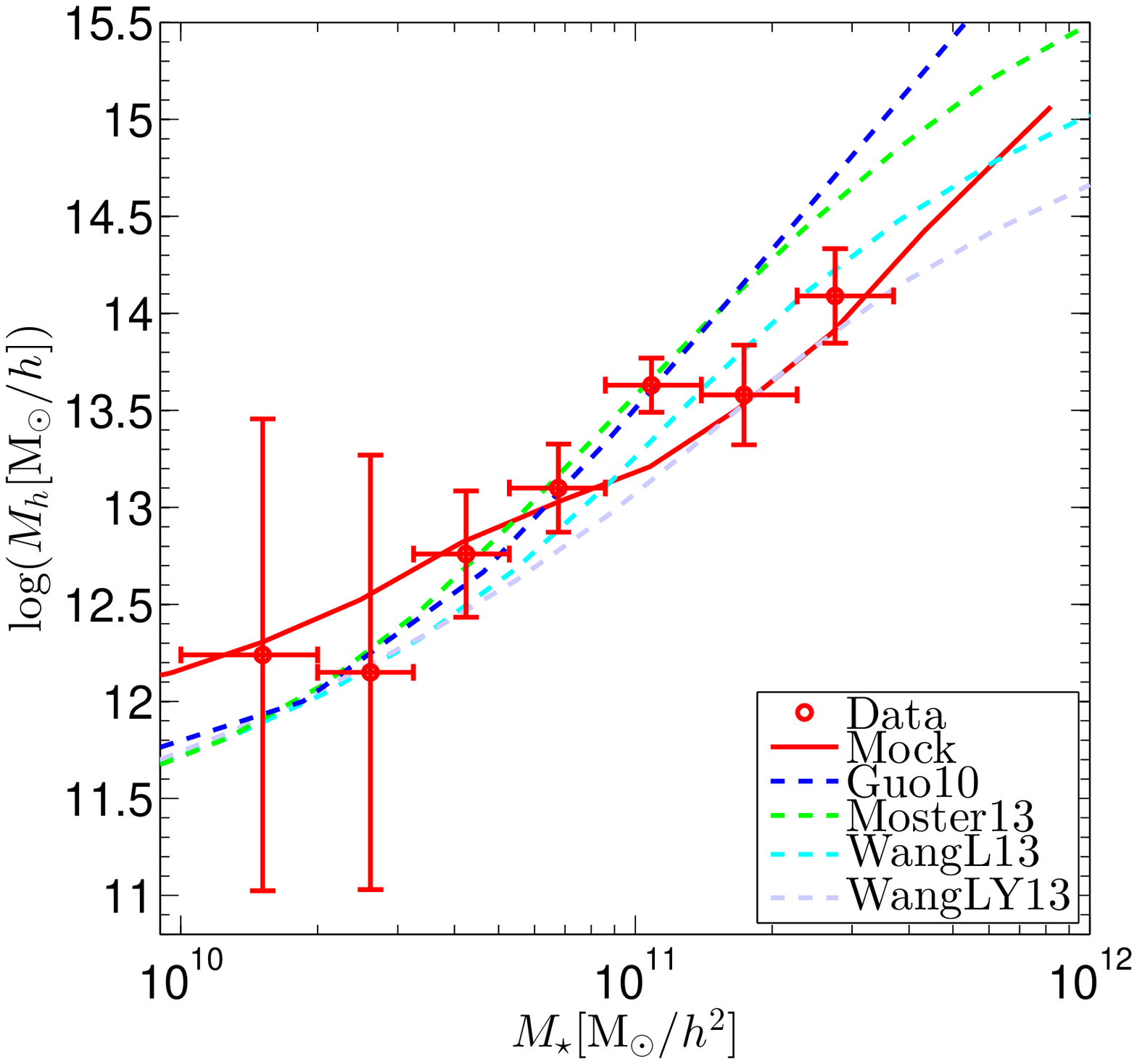}{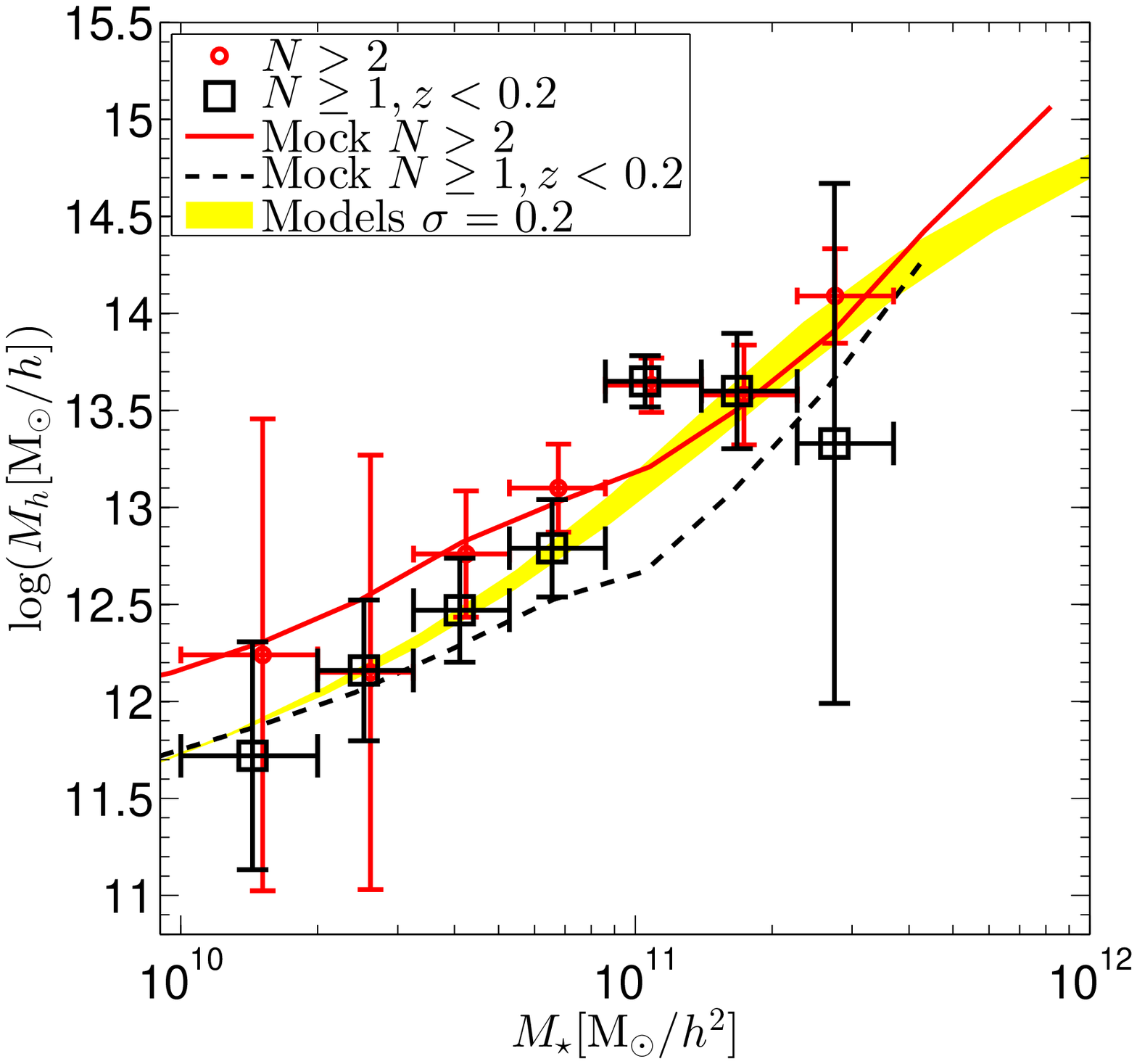}
\caption{The measured halo mass-stellar mass relation compared with
  several HOD prescriptions and mock predictions. In both panels, the red
  circles with error bars are the MLWL measured $M_h(M_\star)$ relation for
  multiplicity $N>2$ groups, and the red solid line is the median
  relation for $N>2$ groups in the mocks. In the left-hand panel, the
  dashed lines are the converted $M_h(M_\star)$ relations from the
  various HOD distributions. From top to bottom on the top-right corner, they are \citet{Guo10}(blue), \citet{Moster13}(green), \citet{WangL13}(cyan) and \citet{WangLY13}(grey). In the right-hand panel, the black squares with
  errorbars are the measured relation for a volume-limited central galaxy sample that is complete up to $z=0.2$ and covers central galaxies down to $N=1$. The black dashed line is the median relation for the
  central galaxies in the mock, similarly selected to be volume limited. A yellow band shows the range of
  converted $M_h(M_\star)$ relations from the four HOD models when a
  common dispersion of $\sigma_{\log(M_\star)}=0.2$ is adopted. %The cyan band is the same as the yellow but for $\sigma_{\log(M_\star)}=0.3$.
  }\label{fig_MMstar_comp} 
\end{figure*}

\begin{figure}
\myplot{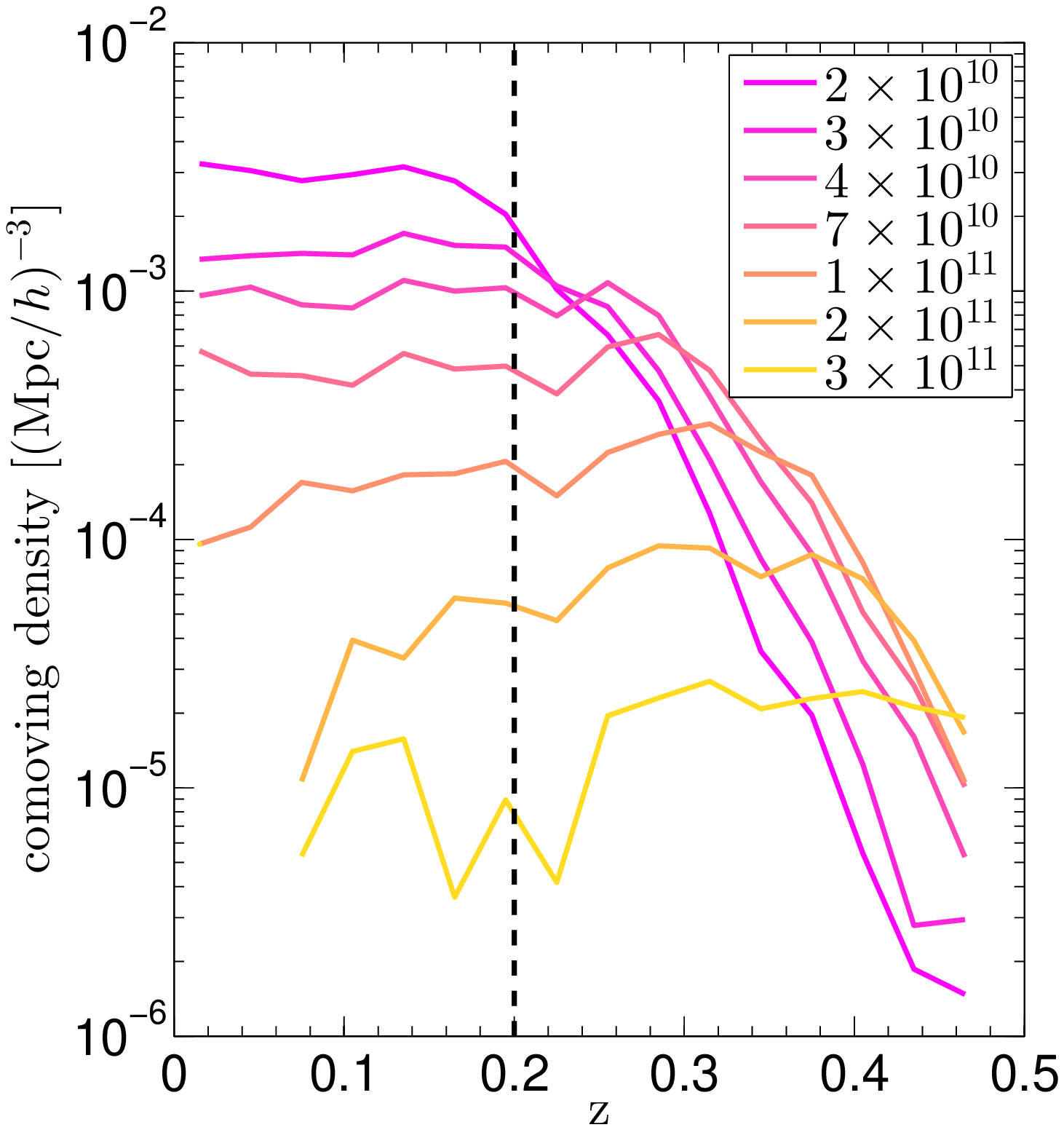}
\caption{The redshift distribution of central galaxies inside
  different stellar mass bins. The sample covers all the central
  galaxies in groups down to $N=2$, and all the ungrouped galaxies
  down to $r_{AB}<19.4$. Different coloured lines represent different
  stellar mass bins, with increasing stellar mass from top to bottom
  at $z<0.2$. The binning in stellar mass is the same as used in Fig.~\ref{fig_MMstar_comp}. Up to $z=0.2$, all the stellar mass bins are complete, except for a slight incompleteless in the smallest mass bin. }\label{fig_dndz} 
\end{figure}
The measured $M_h(M_\star)$ is compared
with several recent halo occupation distribution (HOD) descriptions for
group central galaxies in Fig.~\ref{fig_MMstar_comp}. We calibrate
the units and mass definitions in 
these relations to those used here, and list them in Table~\ref{table_hod}
along with the dispersion in stellar mass at fixed halo mass used
for each of these studies. Note that all the average $M_\star(M_h)$
relations in these HOD descriptions are measuring the median stellar
mass of the central galaxy for haloes of a given mass.
What we measure is the median halo mass for
groups of a given central stellar mass, assuming a log-normal distribution in halo mass at fixed central stellar mass. To make the HOD-based relations comparable with our measurements, we make use of
\begin{equation}\label{eq_Bayesian}
dP(M_h|M_\star)=\frac{dP(M_\star|M_h) \phi(M_h)dM_h}{\int dP(M_\star|M_h) \phi(M_h)dM_h},
\end{equation}
where $\phi(M_h)=dN(M_h)/dM_h$ is the halo mass function. When
$dP(M_\star|M_h)$ follows a log-normal distribution, as is assumed
when the HOD-based relations are inferred, we find that the
converted distribution, $dP(M_h|M_\star)$, is also very well described by a
log-normal distribution, consistent with what had been assumed in
Equation~\eqref{eq_Mprior}. The median halo mass, or mean
logarithmic mass, can then be found through
\begin{equation}\label{eq_MhMean}
\log M_h(M_\star)=\int \log(M_h) dP(M_h|M_\star).
\end{equation}
We adopt the \citet{Sheth2001} mass function in our
conversion\footnote{Calculated with hmf \citep{ HMF}:\\\url{http://hmf.icrar.org}}, and
have checked that adopting the \citet{Tinker2008} mass function or a different
cosmology (Millennium/WMAP9/Planck) produces no more than a $10\%$
difference in the converted relations at the high mass end, much smaller
than model to model variations. %primarily driven by different
				%$\sigma_8$  

At low masses ($M_\star<10^{11}\msunhh$), it appears that the \cite{Guo10}
and \cite{Moster13} results agree best with our measurement for
groups with more than two members, while at the high mass end the average
halo mass in our sample is lower than their predictions.  
However, we emphasize that our standard sample is limited to groups
with three or more members. This introduces a redshift dependent
selection in $V_N$, given by $V_N^{\rm lim}(z)=N_{\rm min}/n_{<M_{\rm lim}(z)}$. Since
$V_N$ is strongly correlated with halo mass, in principle all our measured
relations could be biased with respect to the relation in a volume
limited halo sample. We have tested a different multiplicity cut of
$N>5$, and it does give a systematically higher $M_h(M_\star)$ relation.
In Appendix~\ref{app_Nabs} we explicitly show how the halo mass depends on
$N_{\rm abs}$ at fixed $M_\star$ or $L_{\rm grp}$ in a volume limited mock
catalogue. With the knowledge that our observed halo mass-stellar mass
relation is strongly multiplicity dependent, including central
galaxies from $N<3$ groups will only lower our measurement. %This is explicitly shown in the right panel of Fig.~\ref{fig_MMstar_comp} with black data points, by extending our measurement down to $N=1$ groups. 
To derive a selection-free stellar mass (SM)-halo mass (HM) relation, we
extend our sample to include $N=2$ groups as well as ungrouped
galaxies representing $N=1$ groups. This gives us a flux-limited
central galaxy sample. As we show in Fig.~\ref{fig_dndz}, by further
restricting to $z<0.2$, we get a volume limited sample of central
galaxies with $M_\star>10^{10}\msunhh$. The measured halo masses are
shown with black squares in the right-hand panel of Fig.~\ref{fig_MMstar_comp}.
We also show the median halo mass for mock galaxies with the
same selection. The match between data and mocks improves slightly if
we shift the measured halo masses down by $0.07$ dex, the typical
size of systematic correction estimated in Table~\ref{table_par}. 
%Down to the single-member limit the sample becomes flux-limited.  and the measurement will be truly selection-free if the halo mass is not sensitive to galaxy luminosity at fixed stellar mass. In Appendix~\ref{app_Nabs} we show that in the mocks this is indeed the case. We also verified the weak luminosity dependence of the stellar mass (SM)-halo mass(HM) relation in the real data. To do this we first fit $M_h=AM_\star^{\alpha}$ to the central galaxy sample with MLWL, obtaining $\alpha_0=0.84$. We then fit again with $M_h=A M_\star^{\alpha} L_{\rm cen}^\beta$, with $\alpha$ fixed to $\alpha_0$. This gives us a weak residual dependence of the halo mass on central galaxy luminosity $L_{\rm cen}$, with $\beta=0.28\pm 0.20$. This dependence is marginally consistent with no dependence.
%The sharp transition at $M_\star\sim 10^{11}\msunhh$ in the mock measurements is due to the lack of star-forming galaxies at $M_\star\gtrsim 10^{11}\msunhh$ because of AGN feedback in massive haloes.

The quoted HOD models, though differing substantially in their
predicted stellar mass for a given halo mass, all give a satisfactory
fit to the stellar mass function with their adopted
dispersions. Hence, deriving the SM-HM relation from pure abundance
matching inevitably faces a degeneracy between the average SM-HM
relation and the SM dispersion at fixed HM~\citep[see, e.g.,][]{WangL06}.

We find the converted relation is more sensitive to the
model dispersion than to the mean relation. We show in the right panel
of Fig.~\ref{fig_MMstar_comp} that when a common dispersion value
is adopted, all of the converted relations are very similar.
With $\sigma_{\log(M_\star)}=0.2$, the volume-limited measurement
can be well reproduced by any of the HOD models. This is consistent with the values of $0.16-0.2$ dex found by many previous works~\citep[e.g.,][]{Yang09,Li12,COSMOS12,Reddick13,Behroozi13, Kravtsov14}. Note this dispersion includes both the
intrinsic stellar mass variation at a given halo mass and the stellar
mass measurement error. Subtracting the typical stellar mass
measurement error of $0.13$ dex for our sample, the intrinsic
dispersion is found to be $\sigma_{\log(M_\star),{\rm intrinsic}}\sim
0.15$.  
%Considering a systematic uncertainty in our measurement which can
%lower our measurement by typically $\sim 0.07$~dex, this e stimate could be slightly increased.

\begin{figure}
 \myplot{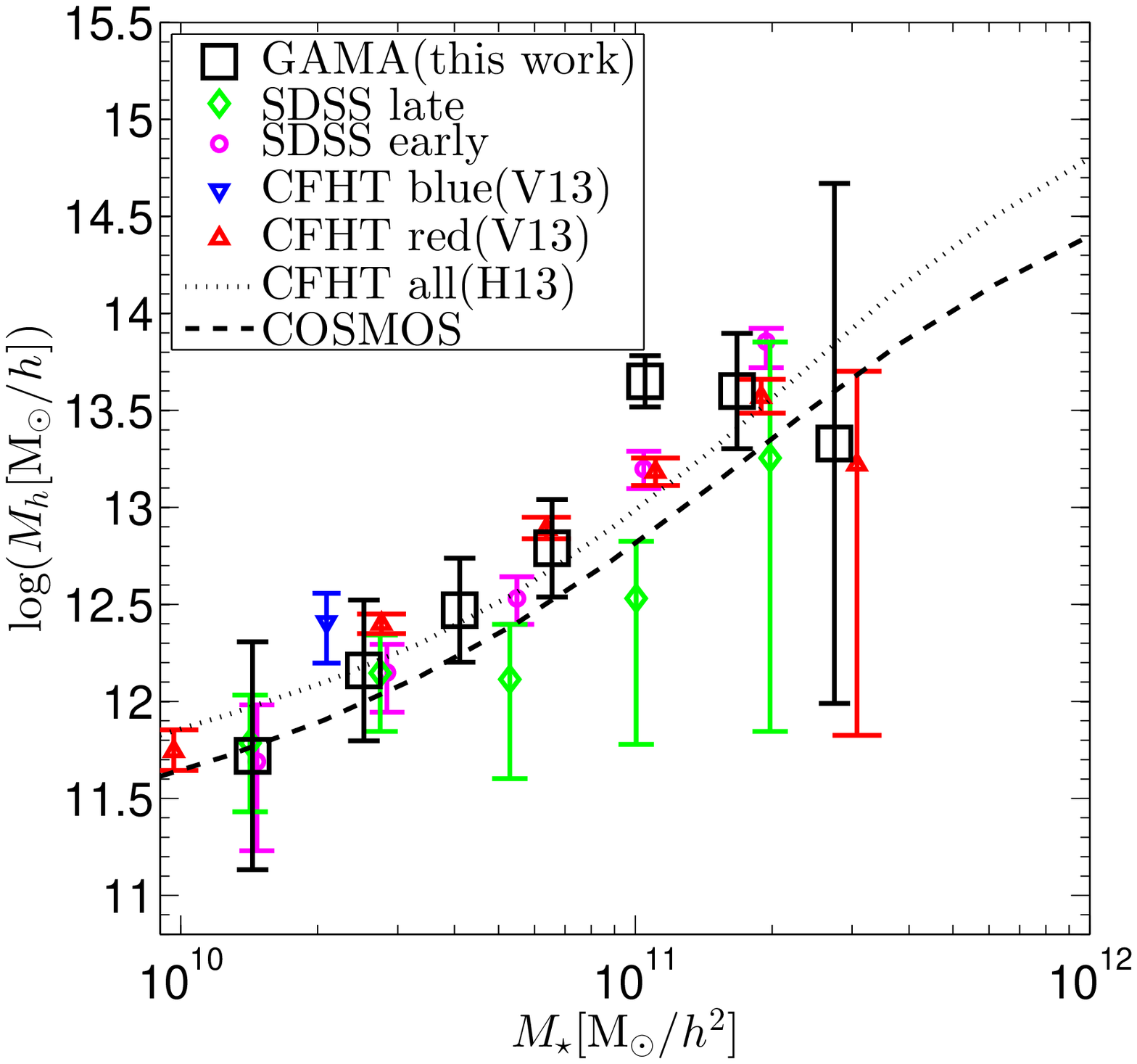}
 \caption{The central halo mass-stellar mass relation measured from weak lensing. We compare our measurements derived from the volume limited stellar mass sample to several other galaxy-galaxy lensing measurements, including from SDSS~\citep[split according to early and late types;][]{Mandelbaum06a}, CFHTLenS (split according to colour by~\citealp{CFHTVelander}, V13, or a fit to the full sample~\citealp[H13]{CFHTHudson}), and COSMOS~\citep[][$z\sim 0.37$]{COSMOS12}.  As in Figure~\ref{fig_MMstar_comp}, we plot the average logarithmic halo mass at fixed stellar mass. The COSMOS measurement and is provided as a fitted $<\log(M_\star)|M_h>$ relation, which we have converted to our convention according to Equations~\eqref{eq_Bayesian} and \eqref{eq_MhMean}. The same conversion is done for the results from \citet{CFHTHudson}. }\label{fig_LensComp}
\end{figure}

In Figure~\ref{fig_LensComp}, we compare our SM-HM relation derived from the volume limited sample to that measured in several other galaxy-galaxy lensing studies, including the measurements in SDSS~\citep{Mandelbaum06a}, CFHTLenS~\citep{CFHTVelander, CFHTHudson} and COSMOS~\citep{COSMOS12}. As in Figure~\ref{fig_MMstar_comp}, we plot the average logarithmic halo mass at fixed stellar mass. The halo mass definitions are either in or converted to our $M_{200b}$ convention according to the scaling relation in \citet{Giocoli10}, except for the SDSS measurements which adopt $M_{180b}$. The difference between $M_{200b}$ and $M_{180b}$ is only around $3$ percent, so we do not correct for it here. The COSMOS result is provided in the form of a HOD model, which we have converted to our convention according to Equations~\eqref{eq_Bayesian} and \eqref{eq_MhMean}. CFHTLens results are given by \citet{CFHTVelander} and \citet{CFHTHudson} independently. To avoid overcomplicating the figure, in the case of \citet{CFHTHudson} we only plot their best-fit relation for the full sample evaluated at $z=0.3$, the redshift of their lowest redshift bin. 

It is worth noting that our measurements are based on a sample of central galaxies, while others use samples mixing both central and satellite galaxies. They all rely on HOD modelling to extract a stellar mass-halo mass relation, typically mixing both populations. They can still differ in how satellite galaxies are modelled. In the COSMOS model, a satellite galaxy has no subhaloes, and contributes to the lensing signal only through the displaced density profile of its host halo. In the other three studies shown in Fig.~\ref{fig_LensComp}, satellites are associated with subhaloes with truncated density profiles, parametrized by a halo mass parameter that follows the same stellar mass-halo mass relation as that of the central galaxies. One should also note that while both the SDSS and our lens samples have spectroscopic redshifts, the CFHTLens and COSMOS lens samples rely on photometric redshifts. Lastly, we point out that both our stellar mass sample and the COSMOS sample are volume-limited, unlike the other samples being considered here, which are flux-limited and hence subject to stellar mass incompleteness. However, we find little difference in our results between flux-limited and volume-limited samples. Overall, despite the different methods and datasets, good agreement is found among the various results considered here. 

Note that our measured masses may be systematically ($0.1-0.2$ dex) above the median halo mass of the underlying distribution at a fixed stellar mass, primarily as a consequence of the scatter in the $M_h(M_*)$ relation. A correction for this would bring our measurements closer to the COSMOS line. Similar corrections are already included in the quoted relations from the other studies, except for \citet{CFHTHudson}. It is worth emphasizing that this correction refers to the difference between the best-fitting mass and the median mass of the stacked haloes in each stellar mass bin, but not the Bayesian conversion between $<M_*|M_h>$ and $<M_h|M_*>$ (i.e, Equation~\eqref{eq_Bayesian}). In \citet{CFHTHudson}, even though the Bayesian conversion is performed when fitting a parametric SM-HM relation, the correction from the best-fit lensing mass to the median mass is not given. In some other works, for example when the COSMOS measurement is compared with that from the CFHTLens in \citet{CFHTVelander} and with SDSS measurements in \citet{COSMOS12}, the Bayesian conversion is not carried out, resulting in an apparent discrepancy between the COSMOS result and other measurements at the high mass end.

There could also be systematic uncertainties in the stellar mass estimates across different studies. \citet{Mandelbaum06b} adopt a Kroupa IMF in the stellar mass estimate, while all the others assume a Chabrier IMF. The difference in stellar mass caused by these two IMFs is typically $0.05$ dex. However, one should keep in mind that the systematic uncertainties in stellar mass estimates can be as large as $0.25$ dex, depending on the detailed implementation of the stellar population synthesis models~\citep{Behroozi10}.

Finally, we note that the general trend observed between halo and stellar mass in this work is similar to those obtained in, e.g., \citet{Lin04,Lin04b,Zheng07,Brown08,Guo14,Oliva14}, using a variety of methods and galaxy samples.

% HOD, satellite truncated profile, and a occupuation term; no volume limited sample has been tried; no central sample. Only COSMOS addressed stellar mass completeness. COSMOS has no subhalo in the HOD, satellite only contribute through offset-host.

\begin{table*}
\caption{MLWL Calibrated Mass-Observable Relations. The parameter
  values are listed as $x=\hat{x}+\Delta_x \pm \sigma_x$, where
  $\hat{x}$, $\Delta_x$ and $\sigma_x$ are the best-fitting value, bias
  and error for parameter $x$. $TS$ is the test statistic defined in
  Equation~\eqref{eq_TS}, where the null model is the one with no
  gravitational shear. In general a larger $TS$ means a better fit. Additional systematic uncertainties can lead to reduced $TS$ values, and a dilution factor $b$ is estimated to account for these uncertainties. We list the test statistic together with its
  estimated dilution factor $b$ as $TS/b$. The significance value is derived from the $TS$ value, taking into account the degrees of freedom in the fitting, and describes how significantly the best-fit differs from statistical fluctuations of the null model. We derive the significances  for both the raw $TS$ and the diluted $TS/b$, and list the diluted
  significance in parenthesis beside the raw significance. In estimating the parameter bias, we adopt a mass dispersion of 0.5 dex by default, but use 0.7 dex for the dynamical mass estimators (marked with \dag), leading to larger estimated biases in their parameters. The
  power-law pivot scales are: $M_0\equiv 10^{14}\msunh$, $M_{*0}\equiv
  10^{11}\msunhh$, $L_0\equiv 2\times 10^{11}\lsunhh$, $v_0 \equiv 500 \,{\rm km\,s}^{-1}$, $V_0 \equiv 1000 (h^{-1}\mathrm{Mpc})^3$, $R_0 \equiv
  0.3 h^{-1}\mathrm{Mpc}$. $M_p$ is in units of $\msunh$, while $A$ is dimensionless. $M_{\rm dyn}$ and $M_{\rm lum}$ refer to the G$^3$Cv5 mass estimator defined in Equation~\eqref{eq_dynmass} and \eqref{eq_lummass}. The $\log()$ function
  is base 10 throughout this paper. $C$
  represents the correlation coefficient, inferred from the
  Hessian matrix of the log-likelihood, for the two parameters listed as subscripts.
  Note these results are derived from a flux-limited group
  catalogue (G$^3$Cv5) and are subject to the group selection
  function.}\label{table_par}
\begin{center}
\begin{tabular}{cp{4.0cm}p{2.8cm}ccc}
\hline
\hline  Halo Mass Estimator & Fitted Parameters & Parameter correlation& $TS/b$ & Significance & Reference\\ 
\hline $M_p \left(\frac{V_N}{V_0}\right)^\alpha$ & $\log(M_p)=13.58 - 0.07 \pm 0.13 $ \par $\alpha=1.03 +0.01\pm0.23$ & $C_{\log(M_p) \alpha}=0.18$ & 46.1/1.5 & 6.5(5.1) & Fig.~\ref{fig_scaling}\\ 
\hline $M_p \left(\frac{L_{\rm grp}}{L_0}\right)^\alpha$ & $\log(M_p)=13.48 -0.08\pm 0.12$ \par $\alpha=1.08+0.01\pm 0.22$ & $C_{\log(M_p) \alpha}=-0.16$ & 53.9/2.2 & 7.0(4.6) & Fig.~\ref{fig_scaling}\\
\hline $M_{p}\left(\frac{M_*}{M_{*0}}\right)^\alpha$ & $\log(M_p)=13.34 - 0.07 \pm 0.12$\par $\alpha=1.08 +0.02 \pm 0.28$ & $C_{\log(M_p)\alpha}=0.07$ & 42.7/1.4 & 6.2(5.2) & Fig.~\ref{fig_scaling}\\
\hline \dag \hspace{0.3cm} $M_p\left(\frac{\sigma_v}{v_0}\right)^\alpha$ & $\log(M_p)=13.67 - 0.21 \pm 0.08 $ \par $\alpha=2.09 +0.08\pm 0.34$ & $C_{\log(M_p) \alpha}=0.32$ & 46.9/5 & 6.5(2.6) & Fig.~\ref{fig_scaling2}\\
\hline $M_p \left(\frac{R_{50}}{R_0}\right)^\alpha$ & $\log(M_p)=13.34 - 0.06 \pm 0.13 $ \par $\alpha=0.98 +0.05 \pm 0.38$ & $C_{\log(M_p) \alpha}=0.49$  & 32.6/1.4 & 5.4(4.4) & Fig.~\ref{fig_scaling2}\\ 
\hline $M_p\left(\frac{\sigma_v}{v_0}\right)^{\alpha_\sigma} \left(\frac{V_N}{V_0}\right)^{\alpha_V}$ & $\log(M_p)=13.78 - 0.07 \pm 0.17 $ \par $\alpha_\sigma=1.28 +0.00 \pm0.45$\par $\alpha_V=0.61+0.02\pm0.24$ & $C_{\log(M_p)\alpha_\sigma}=0.24$\par $C_{\log(M_p)\alpha_V}=-0.05$ \par $C_{\alpha_\sigma\alpha_V}=-0.65$ & 54.6/1.7 & 6.8(5.0) & Fig.~\ref{fig_estimator}\\ 
\hline $M_p \left(\frac{L_{\rm grp}}{L_0}\right)^{\alpha_L} \left(\frac{V_N}{V_0}\right)^{\alpha_V}$ & $\log(M_p)=13.31 - 0.03 \pm 0.28 $ \par $\alpha_L=1.99-0.10\pm 0.98$\par $\alpha_V=-0.92+0.10\pm 0.90$ & $C_{\log(M_p)\alpha_L}=-0.73$\par $C_{\log(M_p)\alpha_V}=0.74$ \par $C_{\alpha_L\alpha_V}=-0.95$ & 56.2/1.6 & 6.9(5.3) & Fig.~\ref{fig_estimator}\\ 
\hline $M_p \left(\frac{L_{\rm grp}}{L_0}\right)^{\alpha_L} \left(\frac{\sigma_v}{v_0}\right)^{\alpha_\sigma} (1+z)^{\alpha_z}$ & $\log(M_p)=14.15 -0.07 \pm 0.30$ \par $\alpha_L=0.78+0.02\pm 0.29$ \par $\alpha_\sigma=1.31+0.03\pm 0.52$ \par $\alpha_z=-5.79+0.18\pm3.64$ & $C_{\log(M_p) \alpha_L}=-0.20$\par $C_{\log(M_p) \alpha_\sigma}=0.53$ \par $C_{\log(M_p) \alpha_z}=-0.91$ \par $C_{\alpha_L \alpha_\sigma}=-0.67$ \par $C_{\alpha_L \alpha_z}=0.01$ \par $C_{\alpha_\sigma \alpha_z}=-0.37$ & 63.7/2.4& 7.2(4.2) & Fig.~\ref{fig_estimator}\\
%\hline \multicolumn{5}{|c|}{Bias Parametrization}\\
\hline \dag \hspace{0.3cm}  $A\left(\frac{M_{\rm dyn}}{M_0}\right)^\alpha M_{\rm dyn}$ &  $\log(A)=-0.54 - 0.22\pm 0.10$\par $\alpha=-0.31+0.04\pm 0.15$ & $C_{\log(A)\alpha}=0.12$ & 43.9/5.3 & 6.3(2.4) &Fig.~\ref{fig_G3Cbias}\\ 
\hline $A\left(\frac{M_{\rm lum}}{M_0}\right)^\alpha M_{\rm lum}$ & $\log(A)=-0.28 -0.07\pm 0.12$\par $\alpha=-0.01 + 0.01\pm 0.19$ & $C_{\log(A)\alpha}=0.28$ & 52.2/1.9 & 6.9(4.9)& Sec.~\ref{sec_diagnose}\\
\hline $A M_{\rm lum}$ & $\log(A)=-0.28 -0.09 \pm 0.09$ & - & 52.2/2.2 & 7.2(4.9)& Sec.~\ref{sec_result_sys}; Fig.~\ref{fig_G3Cbias}\\  
%\hline $[A_0+\frac{A_N}{\sqrt{N}}+\frac{A_z}{\sqrt{z}}]\sigma_v^2 R_{50}$ &{$A_0=8.4+0.2\pm 5.4$\par $A_N=-16.1+0.5 \pm 7.6$\par $A_z=0.9 -0.4 \pm 1.4$} & $C_{0N}=-0.8$\par $C_{0z}=-0.8$\par $C_{Nz}=0.3$ &61.2 & 7.3\\
\hline
\end{tabular} 
\end{center}
\end{table*}

\section{Calibration and Construction of Mass Estimators}\label{sec_estimators}
In this section we apply our MLWL method to calibrate the existing halo mass estimators from the G$^3$Cv5 catalogue. We also try to construct some new estimators combining the various mass observables we have studied above, and select the best combinations according to their performance in the MLWL fitting. These results are also summarized in Table~\ref{table_par}.

\subsection{Diagnosing G$^3$Cv5 Mass Estimators}\label{sec_diagnose}
\begin{figure*}
\includegraphics[width=0.33\textwidth, height=0.33\textwidth]{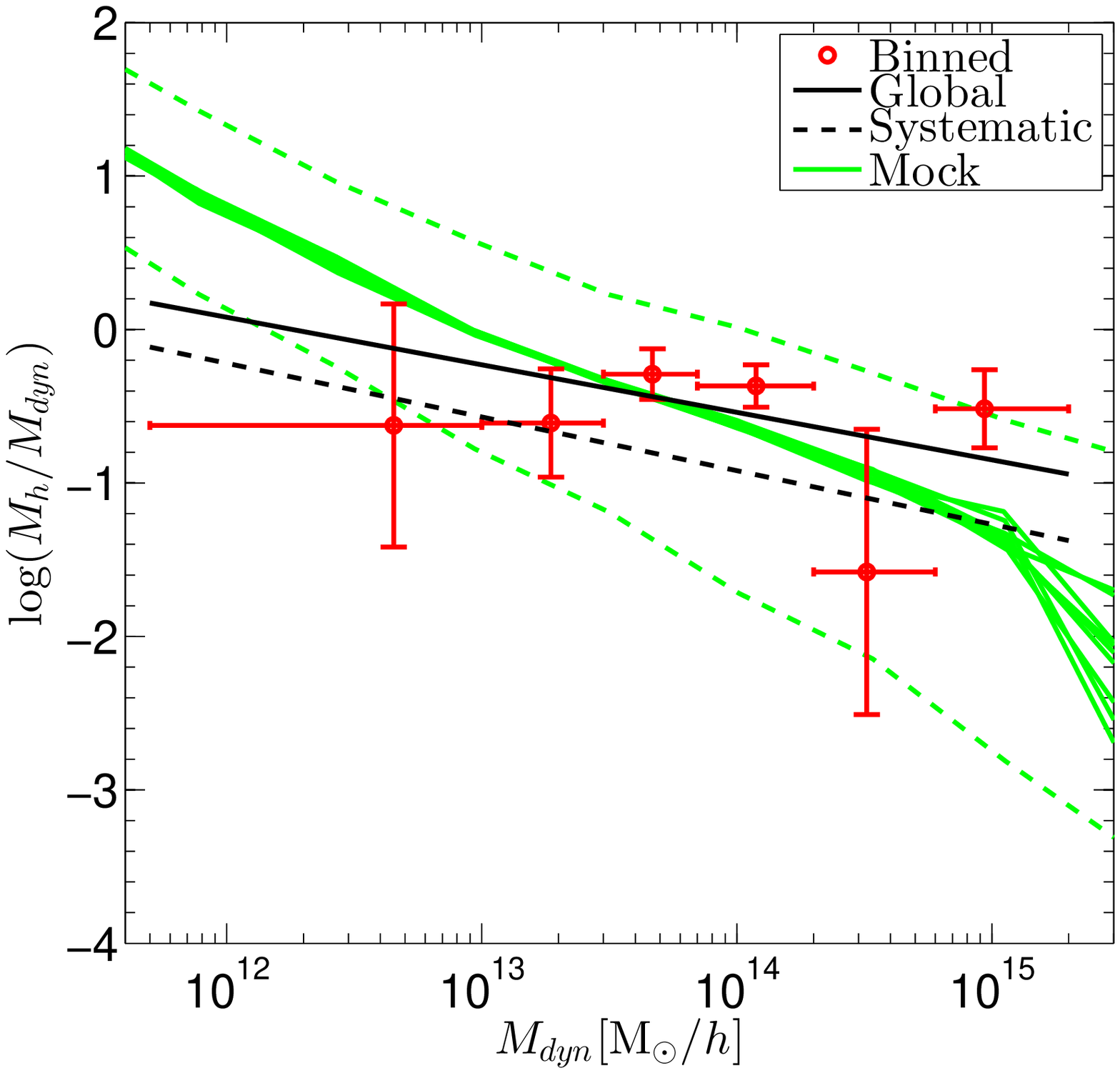}
\includegraphics[width=0.33\textwidth, height=0.33\textwidth]{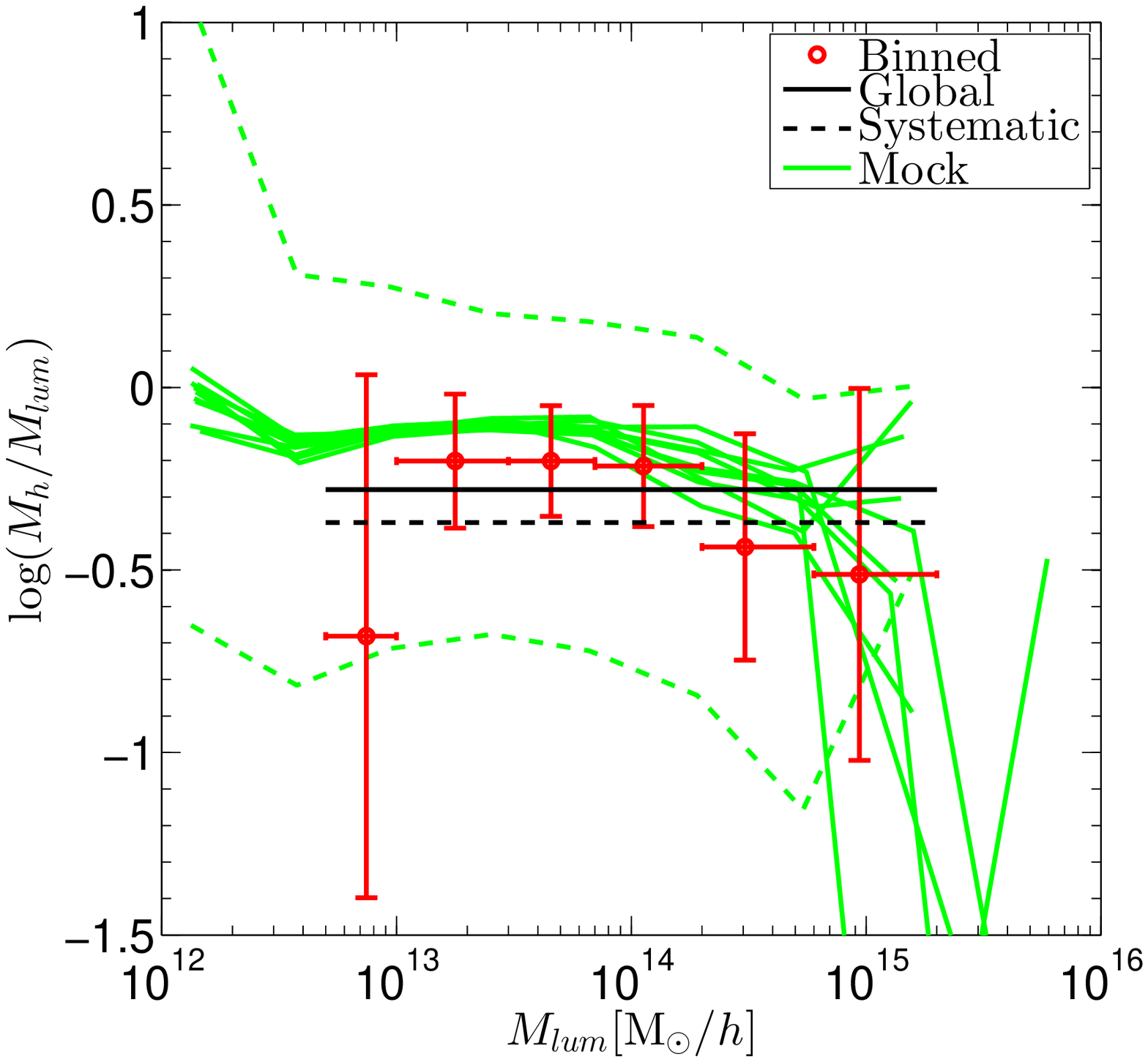}
\includegraphics[width=0.33\textwidth, height=0.33\textwidth]{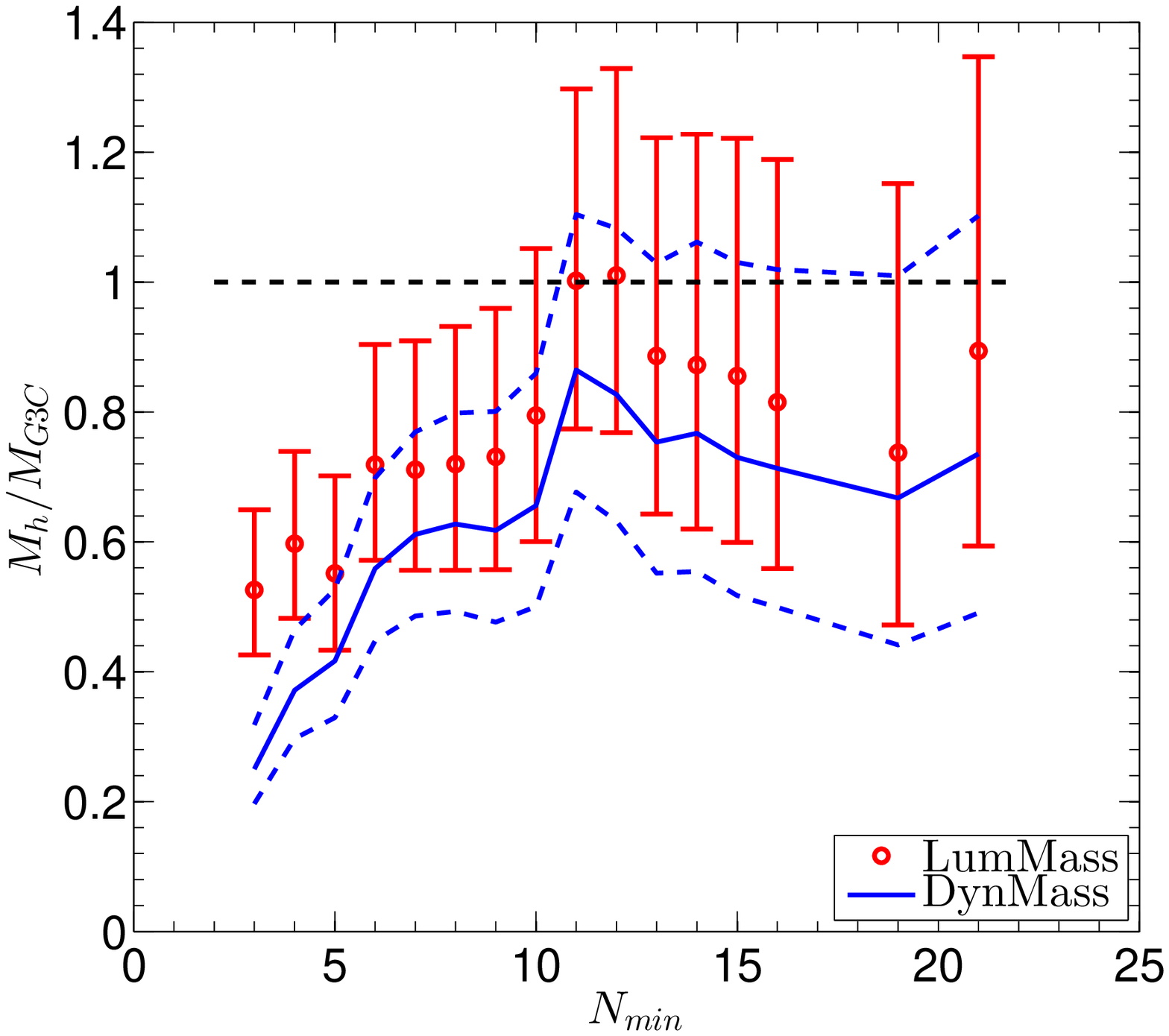}
\caption{Left and middle panels: the biases of halo mass with respect to the G$^3$Cv5 mass
  estimates, where G$^3$Cv5 mass refers to dynamical (left) and
  luminosity (middle) mass in the two panels. In both cases, the red
  data points are the MLWL measured mass ratio within each G$^3$Cv5 mass
  bin; the black solid line is a power-law MLWL fit to the mass
  ratio-G$^3$Cv5 mass relation in the whole sample; the black dashed line
  is the fit with systematic corrections; the green solid lines are
  the median relations in the nine mock catalogues, while the green
  dashed lines mark the typical 16th and 84th (i.e., $\pm 1\sigma$) percentiles in one
  mock. Note that for luminosity mass the power-law fit gives a slope
  so close to zero that we have fixed it to be
  zero. Right panel: The measured bias of the G$^3$Cv5 mass estimates adopting different
  multiplicity cuts $N_{\rm min}$. The red points with error bars show the
  bias of the G$^3$Cv5 luminosity mass. The blue solid line is the measured
  bias of the G$^3$Cv5 dynamical mass, and the blue dashed lines are the
  error bounds. }\label{fig_G3Cbias}
\end{figure*}

The G$^3$Cv5 catalogue comes with two mass estimates. A typical usage
of these estimates involves investigating other group properties at fixed
group mass \citep[e.g.,][]{Guo14,Oliva14}. In this section we compare the weak
lensing measured masses within these bins with the G$^3$Cv5 estimates. To
this end, we measure the ratio of halo mass to G$^3$Cv5 mass within each
bin, and also try a power-law fit to the relation between
the ratio and the G$^3$Cv5 mass, i.e.
$M_h/M_{G3C}=A(M_{G3C}/10^{14}\msunh)^\alpha$. The results are shown in the left and middle panels of 
Fig.~\ref{fig_G3Cbias}. In general, the WL-measured masses
are smaller than the G$^3$Cv5 masses. At their closest, for haloes around 
$10^{13}-10^{14}\msunh$, the G$^3$Cv5 mass estimates are still larger
than the WL ones by $0.1-0.2$ dex. The global
power-law fit to the dynamical mass bias yields $\log(A)=-0.54-0.22\pm 0.10$ and
$\alpha=-0.31+0.04\pm 0.15$. The fit to the luminosity mass bias
gives a slope that is consistent with zero, so we fix it to be zero and
find $\log(A)=-0.28-0.09\pm 0.09$. This means that the weak lensing mass measurement is
$3.5$ times smaller than the dynamical mass estimates near
$10^{14}\msunh$, or $\sim 2$ times smaller than the luminosity mass
globally. Similar biases are observed in the mock catalogues when
comparing the input halo masses with those from the G$^3$Cv5 mass
estimators. The slope for the dynamical mass bias in the mocks is
somewhat steeper than that shown by the groups in the G$^3$Cv5 itself,
reflecting the different mass-velocity dispersion
relation that we observed in Section~\ref{sec_scaling}. The agreement
between luminosity mass and lensing mass is slightly better, although the
discrepancy is amplified when systematic corrections are taken into
account.

The presence of bias at fixed estimated mass does not conflict with
the G$^3$Cv5 claim of a global median unbiased mass calibration.
The G$^3$Cv5 calibration is done ensuring that the estimated masses are
unbiased with respect to the real group masses in their global median
value. Also, only the dynamical mass is calibrated with halo masses in the
mock. The luminosity mass is subsequently calibrated against the
dynamical mass. While \cite{G3C} split the G$^3$Cv5 groups into multiplicity
and redshift bins, they did not find unbiased mass
estimates for each dynamical or luminosity mass bin. This 
calibration thus leaves room for a mass-dependent bias both below
and above the median mass value. The problem can become more severe if
the mass-velocity dispersion-radius relation, which is used as a primary mass
estimator, differs between the data and the mock. Unfortunately, such a
difference is just what we have observed using our lensing
measurement--a conclusion that can only be reached using
an independent mass measurement such as lensing. Finally, the G$^3$Cv5
calibration is only done using FoF groups that are 
bijectively matched with particular haloes. As a result, an overall
bias could also show up when one examines the masses of the entire
group sample. These three 
effects combined result in both an overall and a mass-dependent bias
of our mass measurement with respect to the G$^3$Cv5 estimates. This
bias also propagates to the G$^3$Cv5 luminosity mass, which is a secondary
estimator. 

The G$^3$Cv5 masses become less biased with higher multiplicity cuts. In the right panel of Figure~\ref{fig_G3Cbias}, we show the measured bias of the G$^3$Cv5 dynamical and luminosity masses when adopting different multiplicity cuts $N\geq N_{\rm min}$. With higher
multiplicity cuts, the biases become weaker, and are consistent with unity for $N>10$ groups. 

\subsection{Constructing Mass Estimators}\label{sec_estimator}
\begin{figure}
\includegraphics[scale=0.5]{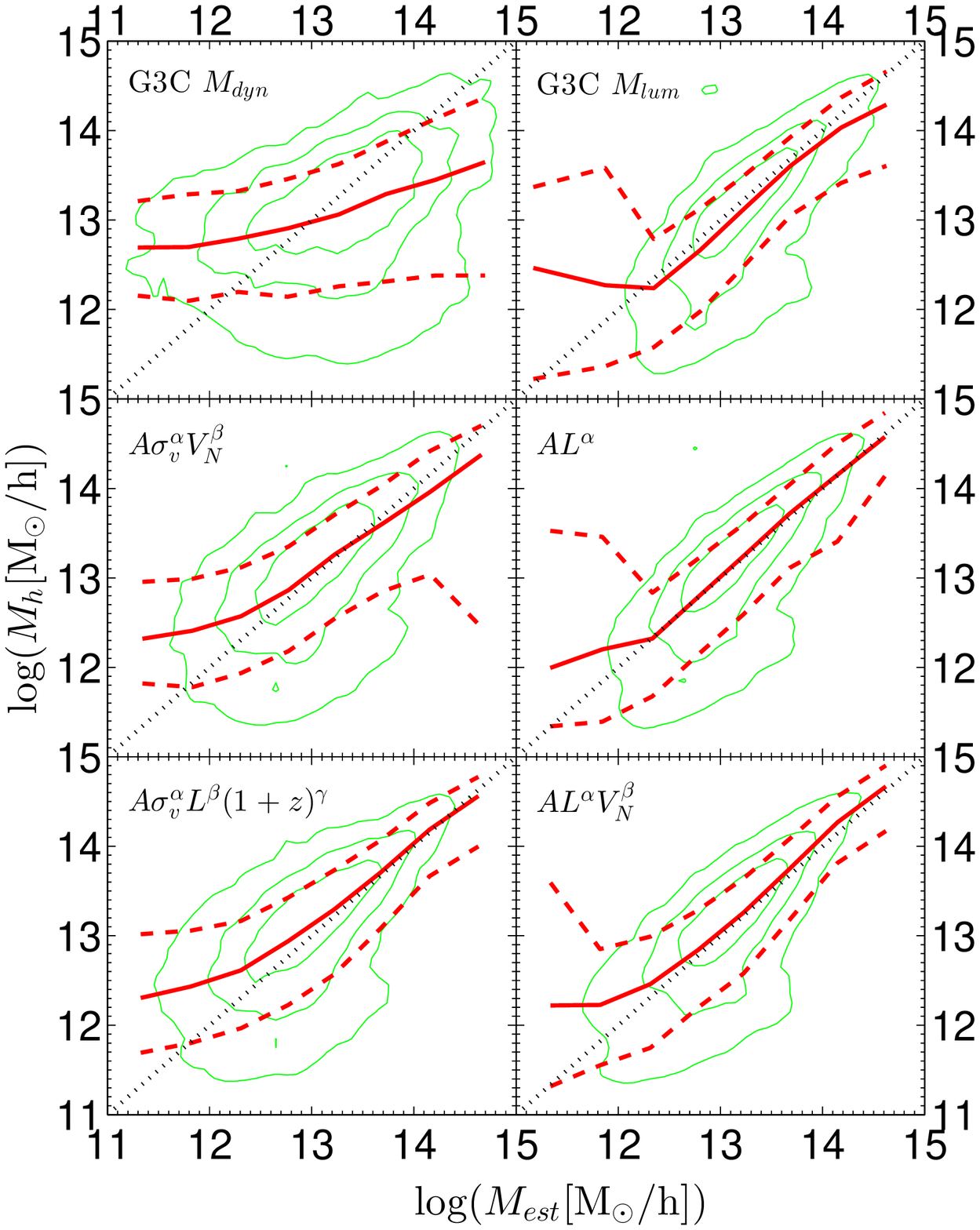}
\caption{The performance of different mass estimators applied to the
  mock catalogue for groups with $N\geq 3$. In each panel, the contour
  lines mark the 30th, 60th and 90th percentiles of the 2D density
  distribution. The red solid line is the median distribution of
  actual halo mass, conditioned on observed mass, with red dashed
  lines showing the 16th and 84th (i.e., $\pm 1\sigma$) percentiles.
  The top panels are for the G$^3$Cv5 estimators as calibrated in
  \citet{G3C}, while the others show our new estimators from
  Table~\ref{table_par}. Note that systematic corrections have been
  applied to the parameters. }\label{fig_estimator}
\end{figure}
\begin{figure}
\includegraphics[scale=0.5]{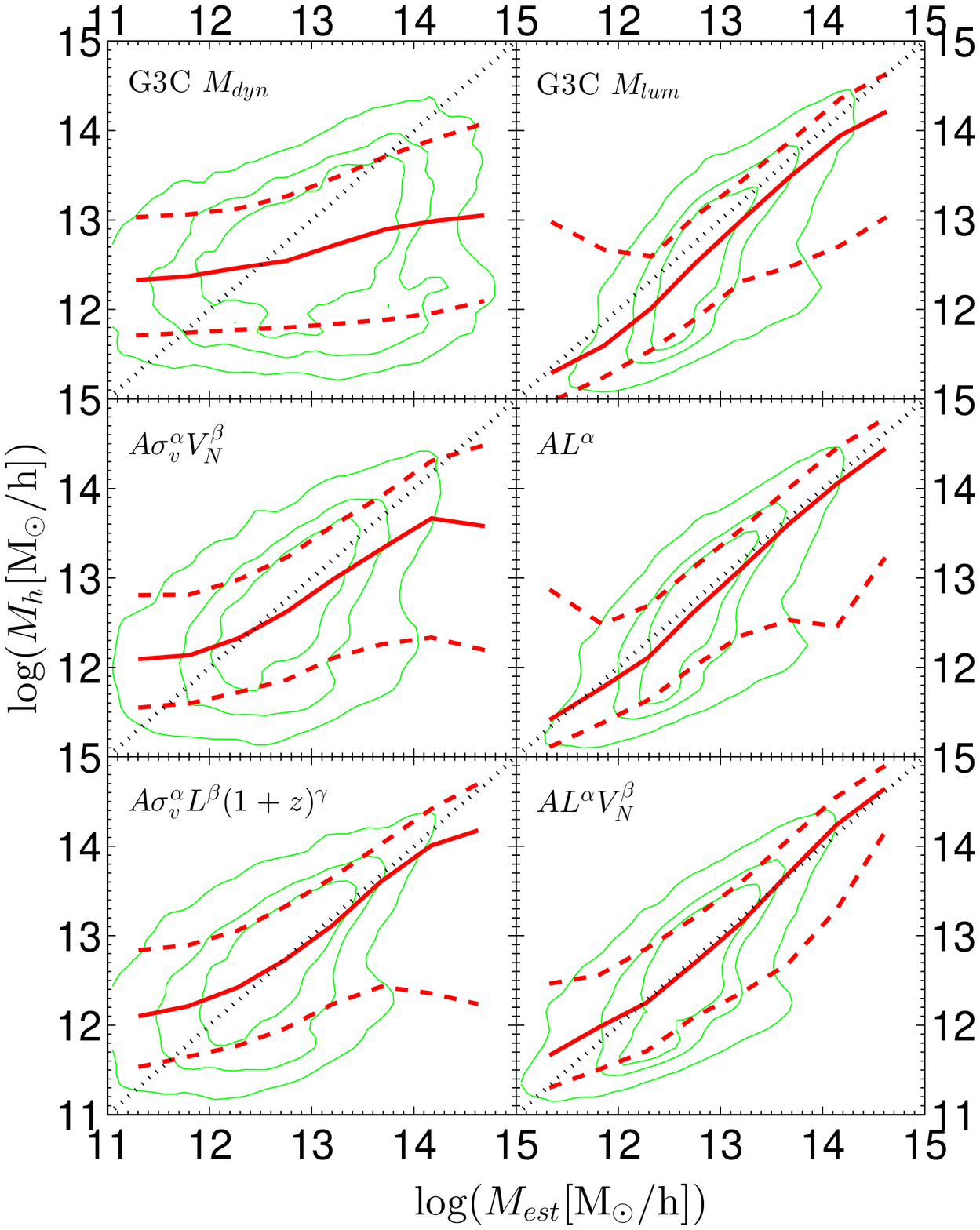}
\caption{Same as Fig.~\ref{fig_estimator} but for $N\geq 2$
  groups.}\label{fig_estimator2} 
\end{figure}
To allow a more general parametrization of the dynamical mass, we
consider power-law combinations of six physical observables: $V_N$,
$(1+z)$, $\sigma_v$, $R_{50}$, $L_{\rm grp}$ of the groups and $M_\star$
of group central galaxies. While we have considered simultaneous independent variations of all of
the six power-law exponents, appropriate subsets of these variables,
with all other exponents fixed to zero, are able to provide a good mass estimator. 

According to Equation~\eqref{eq_sigtonoi}, the model with the highest
$TS$ would also be the one with the \emph{least scatter},
$\sum_i(\frac{\Delta S_i}{N_i})^2$. In other words, for a given sample, $TS$ is a measurement of the intrinsic tightness of each mass-observable relation. To account for both the
improvement in fit and the number of degrees of freedom in the model,
the lowest $p$-value (or highest significance) combination of these
parameters is found. With a significance of $7.23 \sigma$ derived from
$TS$ relative to a null model with no lensing mass present, the best
choice is of the form
\begin{equation}\label{eq_EstCombo}
M\propto L_{\rm grp}^{\alpha_L} \sigma_v^{\alpha_\sigma} (1+z)^{\alpha_z}.
\end{equation} The best-fit parameters are listed in Table~\ref{table_par}.

Since the $TS$ value can be diluted in the presence of systematics, we also
tried the above selection process with $TS'=TS/b$. As listed in
Table~\ref{table_par}, the estimated $b$ is $\sim 2-5$, and is
primarily set by the systematic uncertainty in the mass dispersion,
$\sigma_{\log M}$. Given the uncertainty in how well the modelled mass
dispersion mimics that in the real data, it is unclear which of $TS$
or $TS'$ provides a better measure of significance. Hence we consider
both in an effort to provide insight into the robustness of the results.
If we assume a
common value of $b$ for all estimators, then for $b$ in the range
$2-20$, the best construction is simply 
\begin{equation}
M\propto L_{\rm grp}^{\alpha_L}.
\end{equation}
The significance according to the raw $TS$ is $7.0 \sigma$. 

Simple combinations of $V_N$ with $\sigma_v$ or $L_{\rm grp}$
can achieve comparable significance to the best combinations given
above, in the forms
\begin{equation}
M\propto \sigma_v^{\alpha_\sigma} V_N^{\alpha_V}
\end{equation}
and 
\begin{equation}
M\propto L_{\rm grp}^{\alpha_L} V_N^{\alpha_V}.
\end{equation}
Note that $V_N\propto L_{\rm grp}$ if we assume a universal luminosity
function for both group and field galaxies. Consequently, the
estimator $M(V_N,\sigma,z)$ has a comparable 
significance to $M(L_{\rm grp},\sigma,z)$. In addition, estimators that 
explicitly depend on $L_{\rm grp}$ and $z$ are expected to be robust to
changes in the survey selection function described by $V_N^{\rm lim}(z)$. Again the best-fit parameters for all the above estimators can be found in Table~\ref{table_par}.

In Fig.~\ref{fig_estimator} we show the joint distributions of
true mass in the mocks, for $N\geq 3$ groups, and estimated mass
derived from our mass estimators, calibrated
with real lensing measurements, as well as the G$^3$Cv5 mass estimators with
their official calibration. 
The G$^3$Cv5 dynamical mass estimator has both the
largest bias and scatter. As shown
in Section~\ref{sec_diagnose}, the G$^3$Cv5 luminosity mass estimator is
also biased, despite having been carefully tuned at different
multiplicities and redshifts. Combining velocity dispersion with $V_N$
instead of $R_{50}$ results in a much improved dynamical
estimator. The performance can be further improved when $\sigma_v$ is
combined with $L_{\rm grp}$ and $z$, but the estimation of
$L_{\rm grp}$ from observed group properties does involve many more steps than
is the case for $V_N$. 
%Actually when $L_{\rm grp}$ is available, a simple globally
%calibrated luminosity estimator shows remarkably good performance in
%the mocks. When $L_{\rm grp}$ is further combined with $V_N$, the
%dispersion at high mass is slightly decreased. We also note that when
Bravely applying our $N\geq 3$ calibrated estimators to $N\ge2$ groups in
the mock still produces good results for
all the new estimators except $M(\sigma, V_N)$, as seen in
Fig.~\ref{fig_estimator2}. 

We caution that the performance of different estimators in the mocks
should not be taken as conclusive, since the mocks may not be an
appropriate realization of the real universe. In particular, 
haloes below $10^{12}\msunh$ are resolved by at most $\sim 1000$
particles in the Millennium simulation. In this
mass range, fewer than $100$ particles will typically lie within
subhaloes in any given halo, and any galaxies associated with these
subhaloes will be less numerically reliable than one might wish. Hence
one expects that the mock catalogue will provide a deficient representation
for haloes below $10^{12}\msunh$. 

%\FloatBarrier

\section{Conclusions}\label{sec_conclusion}
We have carried out a maximum-likelihood weak lensing analysis on a
set of SDSS source galaxies located in the GAMA survey regions, in order to derive halo masses for the GAMA galaxy groups.
The group mass distribution is modelled with an NFW profile, with a
mass-concentration relation fixed by previous simulation results. This
enables us to predict the gravitational shear produced by each halo
with a single parameter, namely halo mass. Comparing the predicted
shear with the observed shapes of background galaxies allows us to fit
the halo mass of our foreground lenses. By splitting the G$^3$Cv5 group
sample according to various observed properties, we have explored the
scaling relations between halo mass and these observables. With
power-law parametrization of these relations, global fits over the
entire sample are also performed. The resulting likelihood ratios quantify the intrinsic tightness of each mass-observable relation. All the fitted results are summarized in Table~\ref{table_par}.  The dominant systematic
uncertainty in our measured mass-observable relations comes from the
assumed halo mass dispersion around the median, modelled as a lognormal distribution in mass.

We emphasize that the majority of our measurements are based on the multiplicity-limited G$^3$Cv5 group sample, and are subject to the group selection function described by $V_N\leq V_N^{\rm lim}(z)$. The only exception is the measured halo mass-stellar mass relation where a volume-limited stellar mass sample is specially constructed. Proper comparison of our results with theory or other measurements have to take the selection effect into account. To help interpret our results and to compare with theoretical
predictions we have constructed mock catalogues based on the
application of the GALFORM semianalytic model of galaxy formation
\citep{Bower06} to halo merger trees in the $\Lambda$CDM Millennium
N-body simulation \cite{Millennium}.  The mock catalogues are generated
using the selection function of the real GAMA survey. For the first time, identifical group finding algorithms and selection functions have been applied to both observational data and lightcone galaxy mocks to enable side-by-side comparisons between lensing measurements and a semi-analytic model. Overall there is very good agreement between our measured mass-observable relations
and those predicted by the galaxy formation model. In particular, we
find that:

\begin{itemize}
\item The halo mass scales roughly in proportion to group luminosity, multiplicity 
  volume and central galaxy stellar mass in the multiplicity limited G$^3$Cv5 sample. These relations are in 
  excellent agreement with predictions from the mocks. 

\item For given stellar mass of the central galaxy, the halo mass
  strongly depends on the number of galaxies in the group. To compare
  our measurement with existing HOD models, we have constructed a
  volume limited central galaxy catalogue, and measured the stellar
  mass-halo mass relation free from selection effects. We find the
  measurement of the $M_h(M_\star)$ relation provides a very powerful
  constraint on the HOD scatter of the $M_\star(M_h)$ relations. A
  dispersion $\sigma_{\log(M_\star)}=0.2$, or $0.15$ after subtracting
  the stellar mass measurement noise, is able to yield a good
  agreement between our measurement and all the HOD predictions that
  we considered.

\item The measured $M_h(\sigma_v)$ relation shows a slightly different
  slope from that in the mocks, which could originate from the
  velocity bias of galaxies with respect to dark matter. The measured
  $M_h(R_{50})$ relation is also in slight tension with those in the
  mock catalogues at small $R_{50}$. Such a small scale discrepancy is
  also obvious in the $L_{\rm grp}(R_{50})$ scaling of groups.  It can
  be partly explained by the limited spatial resolution of the
  Millennium simulation, and may also reflect the treatment of orphan
  galaxies in the model.

\item The G$^3$Cv5 mass estimators are biased when used for
  stacking. Luminosity mass has a small but constant bias, while
  dynamical mass can have a large and mass-dependent bias. A globally 
  calibrated mass-to-light relation can serve as a very
  good mass estimator for groups, and is the tightest halo mass to single observable relation in our sample.  
  The estimation can be slightly
  improved when combined with $V_N$. The mass estimates from dynamical
  measurements can be much improved when combining $\sigma_v$ with
  $V_N$ instead of $R_{50}$, or when combined with group luminosity
  and redshift. 

%\item We find that residuals in the ellipticity measurement that persists over the survey region could result in a systematic bias near survey boundaries. This bias can be described by a model-noise correlation term in our MLWL, and shows up in stacked lensing as the so-called systematic shear. Adopting close-circle cuts completely removes such a boundary effect.
  
\item The dominant source of systematic uncertainty in our mass
  estimators comes from the assumed dispersion in halo mass about the
  median value, modelled with a lognormal distribution in mass. For a mass dispersion of $0.5-0.7$ dex, the resulting
  overestimation in median lens mass is typically $0.2-0.3$ dex. This
  is slightly counteracted by smaller underestimations caused by
  uncertainties in the redshifts of background photometric galaxies
  and the positions of gravitational centres of foreground
  lenses. Selection cuts 
  in the data do not cause significant biases in the results. The
  systematic uncertainties considered here change the slopes of the
  mass-observable relations by only $0.01$, but do have a greater
  impact on the significance of the results, reducing $TS$ for the
  fits by a factor of $2-5$.

\end{itemize}

In this work we have taken a galaxy-by-galaxy maximum-likelihood approach
to extract the lensing signal of galaxy groups.
%, primarily driven by
%the fact that our lens and source samples are both small compared to
%some other works where stacked weak lensing was adopted. 
Compared with
stacked weak lensing, our approach makes much more efficient use of
the information contained in individual galaxy shapes. In addition, our
utilization of the information carried by individual lenses is also
more efficient, since our fitting can be done free from binning. In
contrast, stacked weak lensing usually measures a weighted average
density profile of the underlying, to be modelled, matter
distribution. This involves averaging over the distribution of halo masses
and redshift. A direct fitting without knowing the underlying sample
distribution and the stacking weights leaves the result difficult to
interpret, or gives biased results if bravely interpreted as the
average mass of the sample. A further complication in stacked lensing
comes from the redshift evolution of halo profiles. Haloes
evolve with redshift, as do the definitions of the halo mass and edge,
so the same halo mass does not correspond to the same
profile in either physical or comoving coordinates. It is not clear
what is the best coordinate system for stacking. In contrast, our
likelihood fitting deals with each halo separately, and can properly
incorporate any distribution and evolution in halo density
profiles. We note that stacked lensing could complement MLWL by providing a non-parametric measurement of the average density profile. In this work we only do stacked weak lensing for visualization of the measured and fitted profiles. 

We plan to explore the mass-concentration relation and the halo mass
function probed by GAMA in subsequent papers. This methodology would
also be well suited for higher redshift, using the combination of the
VIPERS survey \citep{VIPERS}, which has 100,000 spectroscopic galaxies
with $0.5<z<1.2$, and the CFHTLens source catalogue, which has a
median redshift $\sim0.75$ and a source density of
$17~\rm{arcmin}^{-2}$. The KiDS survey \citep{KIDS} has just come to its
first data release of $50$ square degree data overlapping with
GAMA. Adopting the KiDS shear catalogue, we expect to have more than a
factor of 3 improvement in signal to noise ratio.

\section*{Acknowledgments}
We thank Richard Massey, Lingyu Wang, Qi Guo, Lan Wang, Shaun Cole and Yanchuan
Cai for helpful comments and discussions. 

CSF acknowledges an ERC Advanced Investigator grant (COSMIWAY). 
PN acknowledges the support of the Royal Society through the award of
a University Research Fellowship and the European Research Council,
through receipt of a Starting Grant (DEGAS-259586). MJB acknowledges funding from an Australian Research Council Future Fellowship FT100100280.

The likelihood optimization
is done with the software package \texttt{IMINUIT}\footnote{\url{http://iminuit.github.io/iminuit}}, an interactive python
interface to the \texttt{MINUIT}\citep{MINUIT} minimizer developed at
CERN. This work used the DiRAC Data Centric system at Durham University,
operated by the Institute for Computational Cosmology on behalf of the
STFC DiRAC HPC Facility (www.dirac.ac.uk). This equipment was funded
by BIS National E-infrastructure capital grant ST/K00042X/1, STFC
capital grant ST/H008519/1, and STFC DiRAC Operations grant
ST/K003267/1 and Durham University. DiRAC is part of the National
E-Infrastructure. This work was supported by the Science and
Technology Facilities Council [grant number ST/F001166/1]. 

GAMA is a joint European-Australasian project based around a
spectroscopic campaign using the Anglo-Australian Telescope. The GAMA
input catalogue is based on data taken from the Sloan Digital Sky
Survey and the UKIRT Infrared Deep Sky Survey. Complementary imaging
of the GAMA regions is being obtained by a number of independent
survey programs including GALEX MIS, VST KiDS, VISTA VIKING, WISE,
Herschel-ATLAS, GMRT and ASKAP providing UV to radio coverage. GAMA is
funded by the STFC (UK), the ARC (Australia), the AAO, and the
participating institutions. The GAMA website is
\url{http://www.gama-survey.org/} .

\bibliographystyle{\mybibstyle}
\setlength{\bibhang}{2.0em}
\setlength\labelwidth{0.0em}
\bibliography{lens_ref}

\begin{thebibliography}{96}
\expandafter\ifx\csname natexlab\endcsname\relax\def\natexlab#1{#1}\fi

\bibitem[{{Adachi} \& {Kasai}(2012)}]{DLfit}
{Adachi} M., {Kasai} M., 2012, Progress of Theoretical Physics, 127, 145,
  \eprint{arXiv:1111.6396}

\bibitem[{{Adelman-McCarthy} {et~al}\mbox{.}(2008){Adelman-McCarthy},
  {Ag{\"u}eros}, {Allam}, {Allende Prieto}, {Anderson}, {Anderson}, {Annis},
  {Bahcall}, {Bailer-Jones}, {Baldry}, {Barentine}, {Bassett}, {Becker},
  {Beers}, {Bell}, {Berlind}, {Bernardi}, {Blanton}, {Bochanski}, {Boroski},
  {Brinchmann}, {Brinkmann}, {Brunner}, {Budav{\'a}ri}, {Carliles}, {Carr},
  {Castander}, {Cinabro}, {Cool}, {Covey}, {Csabai}, {Cunha}, {Davenport},
  {Dilday}, {Doi}, {Eisenstein}, {Evans}, {Fan}, {Finkbeiner}, {Friedman},
  {Frieman}, {Fukugita}, {G{\"a}nsicke}, {Gates}, {Gillespie}, {Glazebrook},
  {Gray}, {Grebel}, {Gunn}, {Gurbani}, {Hall}, {Harding}, {Harvanek}, {Hawley},
  {Hayes}, {Heckman}, {Hendry}, {Hindsley}, {Hirata}, {Hogan}, {Hogg}, {Hyde},
  {Ichikawa}, {Ivezi{\'c}}, {Jester}, {Johnson}, {Jorgensen}, {Juri{\'c}},
  {Kent}, {Kessler}, {Kleinman}, {Knapp}, {Kron}, {Krzesinski}, {Kuropatkin},
  {Lamb}, {Lampeitl}, {Lebedeva}, {Lee}, {Leger}, {L{\'e}pine}, {Lima}, {Lin},
  {Long}, {Loomis}, {Loveday}, {Lupton}, {Malanushenko}, {Malanushenko},
  {Mandelbaum}, {Margon}, {Marriner}, {Mart{\'{\i}}nez-Delgado}, {Matsubara},
  {McGehee}, {McKay}, {Meiksin}, {Morrison}, {Munn}, {Nakajima}, {Neilsen},
  {Newberg}, {Nichol}, {Nicinski}, {Nieto-Santisteban}, {Nitta}, {Okamura},
  {Owen}, {Oyaizu}, {Padmanabhan}, {Pan}, {Park}, {Peoples}, {Pier}, {Pope},
  {Purger}, {Raddick}, {Re Fiorentin}, {Richards}, {Richmond}, {Riess}, {Rix},
  {Rockosi}, {Sako}, {Schlegel}, {Schneider}, {Schreiber}, {Schwope}, {Seljak},
  {Sesar}, {Sheldon}, {Shimasaku}, {Sivarani}, {Smith}, {Snedden}, {Steinmetz},
  {Strauss}, {SubbaRao}, {Suto}, {Szalay}, {Szapudi}, {Szkody}, {Tegmark},
  {Thakar}, {Tremonti}, {Tucker}, {Uomoto}, {Vanden Berk}, {Vandenberg},
  {Vidrih}, {Vogeley}, {Voges}, {Vogt}, {Wadadekar}, {Weinberg}, {West},
  {White}, {Wilhite}, {Yanny}, {Yocum}, {York}, {Zehavi}, \&
  {Zucker}}]{SDSSDR6}
{Adelman-McCarthy} J.~K. {et~al.}, 2008, \apjs, 175, 297,
  \eprint{arXiv:0707.3413}

\bibitem[{{Bartelmann} \& {Schneider}(2001)}]{Bartelmann01}
{Bartelmann} M., {Schneider} P., 2001, \physrep, 340, 291,
  \eprint{astro-ph/9912508}

\bibitem[{{Beers} {et~al}\mbox{.}(1990){Beers}, {Flynn}, \&
  {Gebhardt}}]{Gapper}
{Beers} T.~C., {Flynn} K., {Gebhardt} K., 1990, \aj, 100, 32

\bibitem[{{Behroozi} {et~al}\mbox{.}(2010){Behroozi}, {Conroy}, \&
  {Wechsler}}]{Behroozi10}
{Behroozi} P.~S., {Conroy} C., {Wechsler} R.~H., 2010, \apj, 717, 379,
  \eprint{arXiv:1001.0015}

\bibitem[{{Behroozi} {et~al}\mbox{.}(2013){Behroozi}, {Wechsler}, \&
  {Conroy}}]{Behroozi13}
{Behroozi} P.~S., {Wechsler} R.~H., {Conroy} C., 2013, \apj, 770, 57,
  \eprint{arXiv:1207.6105}

\bibitem[{{Bernstein} \& {Jarvis}(2002)}]{BJ02}
{Bernstein} G.~M., {Jarvis} M., 2002, \aj, 123, 583, \eprint{astro-ph/0107431}

\bibitem[{{Blanton} {et~al}\mbox{.}(2003){Blanton}, {Hogg}, {Bahcall},
  {Brinkmann}, {Britton}, {Connolly}, {Csabai}, {Fukugita}, {Loveday},
  {Meiksin}, {Munn}, {Nichol}, {Okamura}, {Quinn}, {Schneider}, {Shimasaku},
  {Strauss}, {Tegmark}, {Vogeley}, \& {Weinberg}}]{Blanton03}
{Blanton} M.~R. {et~al.}, 2003, \apj, 592, 819, \eprint{astro-ph/0210215}

\bibitem[{{Bower} {et~al}\mbox{.}(2006){Bower}, {Benson}, {Malbon}, {Helly},
  {Frenk}, {Baugh}, {Cole}, \& {Lacey}}]{Bower06}
{Bower} R.~G., {Benson} A.~J., {Malbon} R., {Helly} J.~C., {Frenk} C.~S.,
  {Baugh} C.~M., {Cole} S., {Lacey} C.~G., 2006, \mnras, 370, 645,
  \eprint{astro-ph/0511338}

\bibitem[{{Brown} {et~al}\mbox{.}(2008){Brown}, {Zheng}, {White}, {Dey},
  {Jannuzi}, {Benson}, {Brand}, {Brodwin}, \& {Croton}}]{Brown08}
{Brown} M.~J.~I. {et~al.}, 2008, \apj, 682, 937, \eprint{arXiv:0804.2293}

\bibitem[{{Budzynski} {et~al}\mbox{.}(2012){Budzynski}, {Koposov}, {McCarthy},
  {McGee}, \& {Belokurov}}]{Budzynski12}
{Budzynski} J.~M., {Koposov} S.~E., {McCarthy} I.~G., {McGee} S.~L.,
  {Belokurov} V., 2012, \mnras, 423, 104, \eprint{arXiv:1201.5491}

\bibitem[{{Choi} {et~al}\mbox{.}(2012){Choi}, {Tyson}, {Morrison}, {Jee},
  {Schmidt}, {Margoniner}, \& {Wittman}}]{DLS}
{Choi} A., {Tyson} J.~A., {Morrison} C.~B., {Jee} M.~J., {Schmidt} S.~J.,
  {Margoniner} V.~E., {Wittman} D.~M., 2012, \apj, 759, 101,
  \eprint{arXiv:1208.3904}

\bibitem[{{Colless} {et~al}\mbox{.}(2001){Colless}, {Dalton}, {Maddox},
  {Sutherland}, {Norberg}, {Cole}, {Bland-Hawthorn}, {Bridges}, {Cannon},
  {Collins}, {Couch}, {Cross}, {Deeley}, {De Propris}, {Driver}, {Efstathiou},
  {Ellis}, {Frenk}, {Glazebrook}, {Jackson}, {Lahav}, {Lewis}, {Lumsden},
  {Madgwick}, {Peacock}, {Peterson}, {Price}, {Seaborne}, \& {Taylor}}]{2dF}
{Colless} M. {et~al.}, 2001, \mnras, 328, 1039, \eprint{astro-ph/0106498}

\bibitem[{{Cooray} \& {Sheth}(2002)}]{HODrev}
{Cooray} A., {Sheth} R., 2002, \physrep, 372, 1, \eprint{astro-ph/0206508}

\bibitem[{{de Jong} {et~al}\mbox{.}(2013){de Jong}, {Verdoes Kleijn},
  {Kuijken}, \& {Valentijn}}]{KIDS}
{de Jong} J.~T.~A., {Verdoes Kleijn} G.~A., {Kuijken} K.~H., {Valentijn} E.~A.,
  2013, Experimental Astronomy, 35, 25, \eprint{arXiv:1206.1254}

\bibitem[{{Driver} {et~al}\mbox{.}(2011){Driver}, {Hill}, {Kelvin}, {Robotham},
  {Liske}, {Norberg}, {Baldry}, {Bamford}, {Hopkins}, {Loveday}, {Peacock},
  {Andrae}, {Bland-Hawthorn}, {Brough}, {Brown}, {Cameron}, {Ching}, {Colless},
  {Conselice}, {Croom}, {Cross}, {de Propris}, {Dye}, {Drinkwater}, {Ellis},
  {Graham}, {Grootes}, {Gunawardhana}, {Jones}, {van Kampen}, {Maraston},
  {Nichol}, {Parkinson}, {Phillipps}, {Pimbblet}, {Popescu}, {Prescott},
  {Roseboom}, {Sadler}, {Sansom}, {Sharp}, {Smith}, {Taylor}, {Thomas},
  {Tuffs}, {Wijesinghe}, {Dunne}, {Frenk}, {Jarvis}, {Madore}, {Meyer}, {
  Seibert}, {Staveley-Smith}, {Sutherland}, \& {Warren}}]{GAMA}
{Driver} S.~P. {et~al.}, 2011, \mnras, 413, 971, \eprint{arXiv:1009.0614}

\bibitem[{{Duffy} {et~al}\mbox{.}(2008){Duffy}, {Schaye}, {Kay}, \& {Dalla
  Vecchia}}]{Duffy08}
{Duffy} A.~R., {Schaye} J., {Kay} S.~T., {Dalla Vecchia} C., 2008, \mnras, 390,
  L64, \eprint{arXiv:0804.2486}

\bibitem[{{Eke} {et~al}\mbox{.}(2004{\natexlab{a}}){Eke}, {Baugh}, {Cole},
  {Frenk}, {Norberg}, {Peacock}, {Baldry}, {Bland-Hawthorn}, {Bridges},
  {Cannon}, {Colless}, {Collins}, {Couch}, {Dalton}, {de Propris}, {Driver},
  {Efstathiou}, {Ellis}, {Glazebrook}, {Jackson}, {Lahav}, {Lewis}, {Lumsden},
  {Maddox}, {Madgwick}, {Peterson}, {Sutherland}, \& {Taylor}}]{2PIGG}
{Eke} V.~R. {et~al.}, 2004{\natexlab{a}}, \mnras, 348, 866,
  \eprint{astro-ph/0402567}

\bibitem[{{Eke} {et~al}\mbox{.}(2004{\natexlab{b}}){Eke}, {Frenk}, {Baugh},
  {Cole}, {Norberg}, {Peacock}, {Baldry}, {Bland-Hawthorn}, {Bridges},
  {Cannon}, {Colless}, {Collins}, {Couch}, {Dalton}, {de Propris}, {Driver},
  {Efstathiou}, {Ellis}, {Glazebrook}, {Jackson}, {Lahav}, {Lewis}, {Lumsden},
  {Maddox}, {Madgwick}, {Peterson}, {Sutherland}, \& {Taylor}}]{2PIGGML}
{Eke} V.~R. {et~al.}, 2004{\natexlab{b}}, \mnras, 355, 769,
  \eprint{astro-ph/0402566}

\bibitem[{{Farrow}(2013)}]{Daniel}
{Farrow} D., 2013, PhD thesis, Durham University

\bibitem[{{Feldmann} {et~al}\mbox{.}(2006){Feldmann}, {Carollo}, {Porciani},
  {Lilly}, {Capak}, {Taniguchi}, {Le F{\`e}vre}, {Renzini}, {Scoville},
  {Ajiki}, {Aussel}, {Contini}, {McCracken}, {Mobasher}, {Murayama}, {Sanders},
  {Sasaki}, {Scarlata}, {Scodeggio}, {Shioya}, {Silverman}, {Takahashi},
  {Thompson}, \& {Zamorani}}]{ZEBRA}
{Feldmann} R. {et~al.}, 2006, \mnras, 372, 565, \eprint{astro-ph/0609044}

\bibitem[{{George} {et~al}\mbox{.}(2012){George}, {Leauthaud}, {Bundy},
  {Finoguenov}, {Ma}, {Rykoff}, {Tinker}, {Wechsler}, {Massey}, \&
  {Mei}}]{George12}
{George} M.~R. {et~al.}, 2012, \apj, 757, 2, \eprint{arXiv:1205.4262}

\bibitem[{{Giocoli} {et~al}\mbox{.}(2010){Giocoli}, {Tormen}, {Sheth}, \& {van
  den Bosch}}]{Giocoli10}
{Giocoli} C., {Tormen} G., {Sheth} R.~K., {van den Bosch} F.~C., 2010, \mnras,
  404, 502, \eprint{arXiv:0911.0436}

\bibitem[{{Guo} {et~al}\mbox{.}(2013){Guo}, {Cole}, {Eke}, {Frenk}, \&
  {Helly}}]{Guo13}
{Guo} Q., {Cole} S., {Eke} V., {Frenk} C., {Helly} J., 2013, \mnras, 434, 1838,
  \eprint{arXiv:1301.3134}

\bibitem[{{Guo} {et~al}\mbox{.}(2014){Guo}, {Lacey}, {Norberg}, {Cole},
  {Baugh}, {Frenk}, {Cooray}, {Dye}, {Bourne}, {Dunne}, {Eales}, {Ivison},
  {Maddox}, {Alpasan}, {Baldry}, {Driver}, \& {Robotham}}]{Guo14}
{Guo} Q. {et~al.}, 2014, arXiv:1401.0986, \eprint{arXiv:1401.0986}

\bibitem[{{Guo} {et~al}\mbox{.}(2010){Guo}, {White}, {Li}, \&
  {Boylan-Kolchin}}]{Guo10}
{Guo} Q., {White} S., {Li} C., {Boylan-Kolchin} M., 2010, \mnras, 404, 1111,
  \eprint{arXiv:0909.4305}

\bibitem[{{Guzzo} {et~al}\mbox{.}(2013){Guzzo}, {Scodeggio}, {Garilli},
  {Granett}, {Abbas}, {Adami}, {Arnouts}, {Bel}, {Bolzonella}, {Bottini},
  {Branchini}, {Cappi}, {Coupon}, {Cucciati}, {Davidzon}, {De Lucia}, {de la
  Torre}, {Fritz}, {Franzetti}, {Fumana}, {Hudelot}, {Ilbert}, {Iovino},
  {Krywult}, {Le Brun}, {Le F{\`e}vre}, {Maccagni}, {Ma{\l}ek}, {Marulli},
  {McCracken}, {Paioro}, {Peacock}, {Polletta}, {Pollo}, {Schlagenhaufer},
  {Tasca}, {Tojeiro}, {Vergani}, {Zamorani}, {Zanichelli}, {Burden}, {Di
  Porto}, {Marchetti}, {Marinoni}, {Mellier}, {Moscardini}, {Nichol},
  {Percival}, {Phleps}, \& {Wolk}}]{VIPERS}
{Guzzo} L. {et~al.}, 2013, arXiv:1303.2623, \eprint{arXiv:1303.2623}

\bibitem[{{Han} {et~al}\mbox{.}(2012){Han}, {Frenk}, {Eke}, {Gao}, {White},
  {Boyarsky}, {Malyshev}, \& {Ruchayskiy}}]{Han2012}
{Han} J., {Frenk} C.~S., {Eke} V.~R., {Gao} L., {White} S.~D.~M., {Boyarsky}
  A., {Malyshev} D., {Ruchayskiy} O., 2012, \mnras, 427, 1651,
  \eprint{arXiv:1207.6749}

\bibitem[{{Hayashi} \& {White}(2008)}]{Hayashi08}
{Hayashi} E., {White} S.~D.~M., 2008, \mnras, 388, 2, \eprint{arXiv:0709.3933}

\bibitem[{{Hilbert} \& {White}(2010)}]{Hilbert10}
{Hilbert} S., {White} S.~D.~M., 2010, \mnras, 404, 486,
  \eprint{arXiv:0907.4371}

\bibitem[{{Hill} {et~al}\mbox{.}(2011){Hill}, {Kelvin}, {Driver}, {Robotham},
  {Cameron}, {Cross}, {Andrae}, {Baldry}, {Bamford}, {Bland-Hawthorn},
  {Brough}, {Conselice}, {Dye}, {Hopkins}, {Liske}, {Loveday}, {Norberg},
  {Peacock}, {Croom}, {Frenk}, {Graham}, {Jones}, {Kuijken}, {Madore},
  {Nichol}, {Parkinson}, {Phillipps}, {Pimbblet}, {Popescu}, {Prescott},
  {Seibert}, {Sharp}, {Sutherland}, {Thomas}, {Tuffs}, \& {van
  Kampen}}]{Hill11}
{Hill} D.~T. {et~al.}, 2011, \mnras, 412, 765, \eprint{arXiv:1009.0615}

\bibitem[{{Hirata} \& {Seljak}(2003)}]{Hirata03}
{Hirata} C., {Seljak} U., 2003, \mnras, 343, 459, \eprint{astro-ph/0301054}

\bibitem[{{Hirata} {et~al}\mbox{.}(2004){Hirata}, {Mandelbaum}, {Seljak},
  {Guzik}, {Padmanabhan}, {Blake}, {Brinkmann}, {Bud{\'a}vari}, {Connolly},
  {Csabai}, {Scranton}, \& {Szalay}}]{Hirata04}
{Hirata} C.~M. {et~al.}, 2004, \mnras, 353, 529, \eprint{astro-ph/0403255}

\bibitem[{{Hoekstra} {et~al}\mbox{.}(2003){Hoekstra}, {Franx}, {Kuijken},
  {Carlberg}, \& {Yee}}]{Hoekstra03}
{Hoekstra} H., {Franx} M., {Kuijken} K., {Carlberg} R.~G., {Yee} H.~K.~C.,
  2003, \mnras, 340, 609, \eprint{astro-ph/0211633}

\bibitem[{{Hoekstra} {et~al}\mbox{.}(2001){Hoekstra}, {Franx}, {Kuijken},
  {Carlberg}, {Yee}, {Lin}, {Morris}, {Hall}, {Patton}, {Sawicki}, \&
  {Wirth}}]{Hoekstra01}
{Hoekstra} H. {et~al.}, 2001, \apjl, 548, L5, \eprint{astro-ph/0012169}

\bibitem[{{Hoekstra} {et~al}\mbox{.}(2004){Hoekstra}, {Yee}, \&
  {Gladders}}]{Hoekstra04}
{Hoekstra} H., {Yee} H.~K.~C., {Gladders} M.~D., 2004, \apj, 606, 67,
  \eprint{astro-ph/0306515}

\bibitem[{{Hudson} {et~al}\mbox{.}(2013){Hudson}, {Gillis}, {Coupon},
  {Hildebrandt}, {Erben}, {Heymans}, {Hoekstra}, {Kitching}, {Mellier},
  {Miller}, {Van Waerbeke}, {Bonnett}, {Fu}, {Kuijken}, {Rowe}, {Schrabback},
  {Semboloni}, {van Uitert}, \& {Velander}}]{CFHTHudson}
{Hudson} M.~J. {et~al.}, 2013, arXiv:1310.6784, \eprint{arXiv:1310.6784}

\bibitem[{{Hudson} {et~al}\mbox{.}(1998){Hudson}, {Gwyn}, {Dahle}, \&
  {Kaiser}}]{Hudson98}
{Hudson} M.~J., {Gwyn} S.~D.~J., {Dahle} H., {Kaiser} N., 1998, \apj, 503, 531,
  \eprint{astro-ph/9711341}

\bibitem[{James \& Roos(1975)}]{MINUIT}
James F., Roos M., 1975, Comput.Phys.Commun., 10, 343

\bibitem[{{Jeong} \& {Komatsu}(2009)}]{Jeong09}
{Jeong} D., {Komatsu} E., 2009, \apj, 703, 1230, \eprint{arXiv:0904.0497}

\bibitem[{{Jiang} {et~al}\mbox{.}(2008){Jiang}, {Jing}, {Faltenbacher}, {Lin},
  \& {Li}}]{Jiang08}
{Jiang} C.~Y., {Jing} Y.~P., {Faltenbacher} A., {Lin} W.~P., {Li} C., 2008,
  \apj, 675, 1095, \eprint{arXiv:0707.2628}

\bibitem[{{Jiang} {et~al}\mbox{.}(2013){Jiang}, {Helly}, {Cole}, \&
  {Frenk}}]{Lilian}
{Jiang} L., {Helly} J.~C., {Cole} S., {Frenk} C.~S., 2013, arXiv:1311.6649,
  \eprint{arXiv:1311.6649}

\bibitem[{{Johnston} {et~al}\mbox{.}(2007){Johnston}, {Sheldon}, {Wechsler},
  {Rozo}, {Koester}, {Frieman}, {McKay}, {Evrard}, {Becker}, \&
  {Annis}}]{Johnston07}
{Johnston} D.~E. {et~al.}, 2007, arXiv:0709.1159, \eprint{arXiv:0709.1159}

\bibitem[{{Kaiser} {et~al}\mbox{.}(1995){Kaiser}, {Squires}, \&
  {Broadhurst}}]{KSB95}
{Kaiser} N., {Squires} G., {Broadhurst} T., 1995, \apj, 449, 460,
  \eprint{astro-ph/9411005}

\bibitem[{{Kelvin} {et~al}\mbox{.}(2012){Kelvin}, {Driver}, {Robotham}, {Hill},
  {Alpaslan}, {Baldry}, {Bamford}, {Bland-Hawthorn}, {Brough}, {Graham},
  {H{\"a}ussler}, {Hopkins}, {Liske}, {Loveday}, {Norberg}, {Phillipps},
  {Popescu}, {Prescott}, {Taylor}, \& {Tuffs}}]{Kelvin12}
{Kelvin} L.~S. {et~al.}, 2012, \mnras, 421, 1007, \eprint{arXiv:1112.1956}

\bibitem[{{Kilbinger} {et~al}\mbox{.}(2013){Kilbinger}, {Fu}, {Heymans},
  {Simpson}, {Benjamin}, {Erben}, {Harnois-D{\'e}raps}, {Hoekstra},
  {Hildebrandt}, {Kitching}, {Mellier}, {Miller}, {Van Waerbeke}, {Benabed},
  {Bonnett}, {Coupon}, {Hudson}, {Kuijken}, {Rowe}, {Schrabback}, {Semboloni},
  {Vafaei}, \& {Velander}}]{CFHTCosmicShear}
{Kilbinger} M. {et~al.}, 2013, \mnras, 430, 2200, \eprint{arXiv:1212.3338}

\bibitem[{{Kravtsov} {et~al}\mbox{.}(2014){Kravtsov}, {Vikhlinin}, \&
  {Meshscheryakov}}]{Kravtsov14}
{Kravtsov} A., {Vikhlinin} A., {Meshscheryakov} A., 2014, arXiv:1401.7329,
  \eprint{arXiv:1401.7329}

\bibitem[{{Leauthaud} {et~al}\mbox{.}(2012){Leauthaud}, {Tinker}, {Bundy},
  {Behroozi}, {Massey}, {Rhodes}, {George}, {Kneib}, {Benson}, {Wechsler},
  {Busha}, {Capak}, {Cort{\^e}s}, {Ilbert}, {Koekemoer}, {Le F{\`e}vre},
  {Lilly}, {McCracken}, {Salvato}, {Schrabback}, {Scoville}, {Smith}, \&
  {Taylor}}]{COSMOS12}
{Leauthaud} A. {et~al.}, 2012, \apj, 744, 159, \eprint{arXiv:1104.0928}

\bibitem[{{Li} {et~al}\mbox{.}(2012){Li}, {Jing}, {Mao}, {Han}, {Peng}, {Yang},
  {Mo}, \& {van den Bosch}}]{Li12}
{Li} C., {Jing} Y.~P., {Mao} S., {Han} J., {Peng} Q., {Yang} X., {Mo} H.~J.,
  {van den Bosch} F., 2012, \apj, 758, 50, \eprint{arXiv:1206.3566}

\bibitem[{{Lin} \& {Mohr}(2004)}]{Lin04}
{Lin} Y.-T., {Mohr} J.~J., 2004, \apj, 617, 879, \eprint{astro-ph/0408557}

\bibitem[{{Lin} {et~al}\mbox{.}(2004){Lin}, {Mohr}, \& {Stanford}}]{Lin04b}
{Lin} Y.-T., {Mohr} J.~J., {Stanford} S.~A., 2004, \apj, 610, 745,
  \eprint{astro-ph/0402308}

\bibitem[{{Loveday} {et~al}\mbox{.}(2012){Loveday}, {Norberg}, {Baldry},
  {Driver}, {Hopkins}, {Peacock}, {Bamford}, {Liske}, {Bland-Hawthorn},
  {Brough}, {Brown}, {Cameron}, {Conselice}, {Croom}, {Frenk}, {Gunawardhana},
  {Hill}, {Jones}, {Kelvin}, {Kuijken}, {Nichol}, {Parkinson}, {Phillipps},
  {Pimbblet}, {Popescu}, {Prescott}, {Robotham}, {Sharp}, {Sutherland},
  {Taylor}, {Thomas}, {Tuffs}, {van Kampen}, \& {Wijesinghe}}]{Loveday12}
{Loveday} J. {et~al.}, 2012, \mnras, 420, 1239, \eprint{arXiv:1111.0166}

\bibitem[{{Mandelbaum} {et~al}\mbox{.}(2012){Mandelbaum}, {Hirata},
  {Leauthaud}, {Massey}, \& {Rhodes}}]{Mandelbaum12}
{Mandelbaum} R., {Hirata} C.~M., {Leauthaud} A., {Massey} R.~J., {Rhodes} J.,
  2012, \mnras, 420, 1518, \eprint{arXiv:1107.4629}

\bibitem[{{Mandelbaum} {et~al}\mbox{.}(2005{\natexlab{a}}){Mandelbaum},
  {Hirata}, {Seljak}, {Guzik}, {Padmanabhan}, {Blake}, {Blanton}, {Lupton}, \&
  {Brinkmann}}]{Mandelbaum05}
{Mandelbaum} R. {et~al.}, 2005{\natexlab{a}}, \mnras, 361, 1287,
  \eprint{astro-ph/0501201}

\bibitem[{{Mandelbaum} {et~al}\mbox{.}(2010){Mandelbaum}, {Seljak}, {Baldauf},
  \& {Smith}}]{Mandelbaum10}
{Mandelbaum} R., {Seljak} U., {Baldauf} T., {Smith} R.~E., 2010, \mnras, 405,
  2078, \eprint{arXiv:0911.4972}

\bibitem[{{Mandelbaum} {et~al}\mbox{.}(2006{\natexlab{a}}){Mandelbaum},
  {Seljak}, {Cool}, {Blanton}, {Hirata}, \& {Brinkmann}}]{Mandelbaum06b}
{Mandelbaum} R., {Seljak} U., {Cool} R.~J., {Blanton} M., {Hirata} C.~M.,
  {Brinkmann} J., 2006{\natexlab{a}}, \mnras, 372, 758,
  \eprint{astro-ph/0605476}

\bibitem[{{Mandelbaum} {et~al}\mbox{.}(2006{\natexlab{b}}){Mandelbaum},
  {Seljak}, {Kauffmann}, {Hirata}, \& {Brinkmann}}]{Mandelbaum06a}
{Mandelbaum} R., {Seljak} U., {Kauffmann} G., {Hirata} C.~M., {Brinkmann} J.,
  2006{\natexlab{b}}, \mnras, 368, 715, \eprint{astro-ph/0511164}

\bibitem[{{Mandelbaum} {et~al}\mbox{.}(2013){Mandelbaum}, {Slosar}, {Baldauf},
  {Seljak}, {Hirata}, {Nakajima}, {Reyes}, \& {Smith}}]{Mandelbaum13}
{Mandelbaum} R., {Slosar} A., {Baldauf} T., {Seljak} U., {Hirata} C.~M.,
  {Nakajima} R., {Reyes} R., {Smith} R.~E., 2013, \mnras, 432, 1544,
  \eprint{arXiv:1207.1120}

\bibitem[{{Mandelbaum} {et~al}\mbox{.}(2005{\natexlab{b}}){Mandelbaum},
  {Tasitsiomi}, {Seljak}, {Kravtsov}, \& {Wechsler}}]{Mandelbaum05a}
{Mandelbaum} R., {Tasitsiomi} A., {Seljak} U., {Kravtsov} A.~V., {Wechsler}
  R.~H., 2005{\natexlab{b}}, \mnras, 362, 1451, \eprint{astro-ph/0410711}

\bibitem[{{Marian} {et~al}\mbox{.}(2010){Marian}, {Smith}, \&
  {Bernstein}}]{Marian10}
{Marian} L., {Smith} R.~E., {Bernstein} G.~M., 2010, \apj, 709, 286,
  \eprint{arXiv:0912.0261}

\bibitem[{{Merson} {et~al}\mbox{.}(2013){Merson}, {Baugh}, {Helly},
  {Gonzalez-Perez}, {Cole}, {Bielby}, {Norberg}, {Frenk}, {Benson}, {Bower},
  {Lacey}, \& {Lagos}}]{Merson13}
{Merson} A.~I. {et~al.}, 2013, \mnras, 429, 556, \eprint{arXiv:1206.4049}

\bibitem[{{Moster} {et~al}\mbox{.}(2013){Moster}, {Naab}, \&
  {White}}]{Moster13}
{Moster} B.~P., {Naab} T., {White} S.~D.~M., 2013, \mnras, 428, 3121,
  \eprint{arXiv:1205.5807}

\bibitem[{{Moster} {et~al}\mbox{.}(2010){Moster}, {Somerville}, {Maulbetsch},
  {van den Bosch}, {Macci{\`o}}, {Naab}, \& {Oser}}]{Moster10}
{Moster} B.~P., {Somerville} R.~S., {Maulbetsch} C., {van den Bosch} F.~C.,
  {Macci{\`o}} A.~V., {Naab} T., {Oser} L., 2010, \apj, 710, 903,
  \eprint{arXiv:0903.4682}

\bibitem[{{Munari} {et~al}\mbox{.}(2013){Munari}, {Biviano}, {Borgani},
  {Murante}, \& {Fabjan}}]{Munari13}
{Munari} E., {Biviano} A., {Borgani} S., {Murante} G., {Fabjan} D., 2013,
  \mnras, 430, 2638, \eprint{arXiv:1301.1682}

\bibitem[{{Murray} {et~al}\mbox{.}(2013){Murray}, {Power}, \& {Robotham}}]{HMF}
{Murray} S.~G., {Power} C., {Robotham} A.~S.~G., 2013, Astronomy and Computing,
  3, 23, \eprint{arXiv:1306.6721}

\bibitem[{{Nakajima} {et~al}\mbox{.}(2012){Nakajima}, {Mandelbaum}, {Seljak},
  {Cohn}, {Reyes}, \& {Cool}}]{Nakajima12}
{Nakajima} R., {Mandelbaum} R., {Seljak} U., {Cohn} J.~D., {Reyes} R., {Cool}
  R., 2012, \mnras, 420, 3240, \eprint{arXiv:1107.1395}

\bibitem[{{Navarro} {et~al}\mbox{.}(1996){Navarro}, {Frenk}, \&
  {White}}]{NFW96}
{Navarro} J.~F., {Frenk} C.~S., {White} S.~D.~M., 1996, \apj, 462, 563,
  \eprint{arXiv:astro-ph/9508025}

\bibitem[{{Navarro} {et~al}\mbox{.}(1997){Navarro}, {Frenk}, \&
  {White}}]{NFW97}
{Navarro} J.~F., {Frenk} C.~S., {White} S.~D.~M., 1997, \apj, 490, 493,
  \eprint{arXiv:astro-ph/9611107}

\bibitem[{{Oliva-Altamirano} {et~al}\mbox{.}(2014){Oliva-Altamirano}, {Brough},
  {Lidman}, {Couch}, {Hopkins}, {Colless}, {Taylor}, {Robotham},
  {Gunawardhana}, {Ponman}, {Baldry}, {Bauer}, {Bland-Hawthorn}, {Cluver},
  {Cameron}, {Conselice}, {Driver}, {Edge}, {Graham}, {van Kampen},
  {Lara-L{\'o}pez}, {Liske}, {L{\'o}pez-S{\'a}nchez}, {Loveday}, {Mahajan},
  {Peacock}, {Phillipps}, {Pimbblet}, \& {Sharp}}]{Oliva14}
{Oliva-Altamirano} P. {et~al.}, 2014, \mnras, 440, 762,
  \eprint{arXiv:1402.4139}

\bibitem[{{Parker} {et~al}\mbox{.}(2005){Parker}, {Hudson}, {Carlberg}, \&
  {Hoekstra}}]{Parker05}
{Parker} L.~C., {Hudson} M.~J., {Carlberg} R.~G., {Hoekstra} H., 2005, \apj,
  634, 806, \eprint{astro-ph/0508328}

\bibitem[{{Reddick} {et~al}\mbox{.}(2013){Reddick}, {Wechsler}, {Tinker}, \&
  {Behroozi}}]{Reddick13}
{Reddick} R.~M., {Wechsler} R.~H., {Tinker} J.~L., {Behroozi} P.~S., 2013,
  \apj, 771, 30, \eprint{arXiv:1207.2160}

\bibitem[{{Reyes} {et~al}\mbox{.}(2012){Reyes}, {Mandelbaum}, {Gunn},
  {Nakajima}, {Seljak}, \& {Hirata}}]{Reyes12}
{Reyes} R., {Mandelbaum} R., {Gunn} J.~E., {Nakajima} R., {Seljak} U., {Hirata}
  C.~M., 2012, \mnras, 425, 2610

\bibitem[{{Robotham} {et~al}\mbox{.}(2011){Robotham}, {Norberg}, {Driver},
  {Baldry}, {Bamford}, {Hopkins}, {Liske}, {Loveday}, {Merson}, {Peacock},
  {Brough}, {Cameron}, {Conselice}, {Croom}, {Frenk}, {Gunawardhana}, {Hill},
  {Jones}, {Kelvin}, {Kuijken}, {Nichol}, {Parkinson}, {Pimbblet}, {Phillipps},
  {Popescu}, {Prescott}, {Sharp}, {Sutherland}, {Taylor}, {Thomas}, {Tuffs},
  {van Kampen}, \& {Wijesinghe}}]{G3C}
{Robotham} A.~S.~G. {et~al.}, 2011, \mnras, 416, 2640, \eprint{arXiv:1106.1994}

\bibitem[{{Schneider} {et~al}\mbox{.}(2013){Schneider}, {Cole}, {Frenk},
  {Kelvin}, {Mandelbaum}, {Norberg}, {Bland-Hawthorn}, {Brough}, {Driver},
  {Hopkins}, {Liske}, {Loveday}, \& {Robotham}}]{Schneider13}
{Schneider} M.~D. {et~al.}, 2013, \mnras, 433, 2727, \eprint{arXiv:1306.4963}

\bibitem[{{Schneider} {et~al}\mbox{.}(2012){Schneider}, {Frenk}, \&
  {Cole}}]{Schneider12}
{Schneider} M.~D., {Frenk} C.~S., {Cole} S., 2012, \jcap, 5, 30,
  \eprint{arXiv:1111.5616}

\bibitem[{{Schneider}(2005)}]{Schneider05}
{Schneider} P., 2005, arXiv:astro-ph/0509252, \eprint{astro-ph/0509252}

\bibitem[{{Schneider} \& {Rix}(1997)}]{Schneider97}
{Schneider} P., {Rix} H.-W., 1997, \apj, 474, 25, \eprint{astro-ph/9601190}

\bibitem[{{Sheldon} {et~al}\mbox{.}(2009){Sheldon}, {Johnston}, {Scranton},
  {Koester}, {McKay}, {Oyaizu}, {Cunha}, {Lima}, {Lin}, {Frieman}, {Wechsler},
  {Annis}, {Mandelbaum}, {Bahcall}, \& {Fukugita}}]{Sheldon09}
{Sheldon} E.~S. {et~al.}, 2009, \apj, 703, 2217, \eprint{arXiv:0709.1153}

\bibitem[{{Sheth} {et~al}\mbox{.}(2001){Sheth}, {Mo}, \& {Tormen}}]{Sheth2001}
{Sheth} R.~K., {Mo} H.~J., {Tormen} G., 2001, \mnras, 323, 1,
  \eprint{astro-ph/9907024}

\bibitem[{{Skibba} {et~al}\mbox{.}(2011){Skibba}, {van den Bosch}, {Yang},
  {More}, {Mo}, \& {Fontanot}}]{Skibba11}
{Skibba} R.~A., {van den Bosch} F.~C., {Yang} X., {More} S., {Mo} H.,
  {Fontanot} F., 2011, \mnras, 410, 417, \eprint{arXiv:1001.4533}

\bibitem[{{Springel} {et~al}\mbox{.}(2005){Springel}, {White}, {Jenkins},
  {Frenk}, {Yoshida}, {Gao}, {Navarro}, {Thacker}, {Croton}, {Helly},
  {Peacock}, {Cole}, {Thomas}, {Couchman}, {Evrard}, {Colberg}, \&
  {Pearce}}]{Millennium}
{Springel} V. {et~al.}, 2005, \nat, 435, 629, \eprint{astro-ph/0504097}

\bibitem[{{Taylor} {et~al}\mbox{.}(2011){Taylor}, {Hopkins}, {Baldry}, {Brown},
  {Driver}, {Kelvin}, {Hill}, {Robotham}, {Bland-Hawthorn}, {Jones}, {Sharp},
  {Thomas}, {Liske}, {Loveday}, {Norberg}, {Peacock}, {Bamford}, {Brough},
  {Colless}, {Cameron}, {Conselice}, {Croom}, {Frenk}, {Gunawardhana},
  {Kuijken}, {Nichol}, {Parkinson}, {Phillipps}, {Pimbblet}, {Popescu},
  {Prescott}, {Sutherland}, {Tuffs}, {van Kampen}, \& {Wijesinghe}}]{GAMASM}
{Taylor} E.~N. {et~al.}, 2011, \mnras, 418, 1587, \eprint{arXiv:1108.0635}

\bibitem[{{Tinker} {et~al}\mbox{.}(2008){Tinker}, {Kravtsov}, {Klypin},
  {Abazajian}, {Warren}, {Yepes}, {Gottl{\"o}ber}, \& {Holz}}]{Tinker2008}
{Tinker} J., {Kravtsov} A.~V., {Klypin} A., {Abazajian} K., {Warren} M.,
  {Yepes} G., {Gottl{\"o}ber} S., {Holz} D.~E., 2008, \apj, 688, 709,
  \eprint{arXiv:0803.2706}

\bibitem[{{Velander} {et~al}\mbox{.}(2013){Velander}, {van Uitert}, {Hoekstra},
  {Coupon}, {Erben}, {Heymans}, {Hildebrandt}, {Kitching}, {Mellier}, {Miller},
  {Van Waerbeke}, {Bonnett}, {Fu}, {Giodini}, {Hudson}, {Kuijken}, {Rowe},
  {Schrabback}, \& {Semboloni}}]{CFHTVelander}
{Velander} M. {et~al.}, 2013, arXiv:1304.4265, \eprint{arXiv:1304.4265}

\bibitem[{{Wang} {et~al}\mbox{.}(2013{\natexlab{a}}){Wang}, {De Lucia}, \&
  {Weinmann}}]{WangL13}
{Wang} L., {De Lucia} G., {Weinmann} S.~M., 2013{\natexlab{a}}, \mnras, 431,
  600, \eprint{arXiv:1211.4308}

\bibitem[{{Wang} {et~al}\mbox{.}(2013{\natexlab{b}}){Wang}, {Farrah}, {Oliver},
  {Amblard}, {B{\'e}thermin}, {Bock}, {Conley}, {Cooray}, {Halpern}, {Heinis},
  {Ibar}, {Ilbert}, {Ivison}, {Marsden}, {Roseboom}, {Rowan-Robinson},
  {Schulz}, {Smith}, {Viero}, \& {Zemcov}}]{WangLY13}
{Wang} L. {et~al.}, 2013{\natexlab{b}}, \mnras, 431, 648,
  \eprint{arXiv:1203.5828}

\bibitem[{{Wang} \& {Jing}(2010)}]{WangL10}
{Wang} L., {Jing} Y.~P., 2010, \mnras, 402, 1796, \eprint{arXiv:0911.1864}

\bibitem[{{Wang} {et~al}\mbox{.}(2006){Wang}, {Li}, {Kauffmann}, \& {De
  Lucia}}]{WangL06}
{Wang} L., {Li} C., {Kauffmann} G., {De Lucia} G., 2006, \mnras, 371, 537,
  \eprint{astro-ph/0603546}

\bibitem[{{Wang} {et~al}\mbox{.}(2014){Wang}, {Sales}, {Henriques}, \&
  {White}}]{Wenting}
{Wang} W., {Sales} L.~V., {Henriques} B.~M.~B., {White} S.~D.~M., 2014, \mnras,
  442, 1363, \eprint{arXiv:1403.2409}

\bibitem[{Wilks(1938)}]{Wilks}
Wilks S., 1938, Annals Math.Statist., 9, 60

\bibitem[{{Wright} \& {Brainerd}(2000)}]{NFWlens}
{Wright} C.~O., {Brainerd} T.~G., 2000, \apj, 534, 34

\bibitem[{{Yang} {et~al}\mbox{.}(2003){Yang}, {Mo}, \& {van den
  Bosch}}]{Yang03}
{Yang} X., {Mo} H.~J., {van den Bosch} F.~C., 2003, \mnras, 339, 1057,
  \eprint{astro-ph/0207019}

\bibitem[{{Yang} {et~al}\mbox{.}(2008){Yang}, {Mo}, \& {van den
  Bosch}}]{Yang08}
{Yang} X., {Mo} H.~J., {van den Bosch} F.~C., 2008, \apj, 676, 248,
  \eprint{arXiv:0710.5096}

\bibitem[{{Yang} {et~al}\mbox{.}(2009){Yang}, {Mo}, \& {van den
  Bosch}}]{Yang09}
{Yang} X., {Mo} H.~J., {van den Bosch} F.~C., 2009, \apj, 695, 900,
  \eprint{arXiv:0808.0539}

\bibitem[{{Zheng} {et~al}\mbox{.}(2007){Zheng}, {Coil}, \& {Zehavi}}]{Zheng07}
{Zheng} Z., {Coil} A.~L., {Zehavi} I., 2007, \apj, 667, 760,
  \eprint{astro-ph/0703457}

\bibitem[{{Zitrin} {et~al}\mbox{.}(2012){Zitrin}, {Bartelmann}, {Umetsu},
  {Oguri}, \& {Broadhurst}}]{Zitrin12}
{Zitrin} A., {Bartelmann} M., {Umetsu} K., {Oguri} M., {Broadhurst} T., 2012,
  \mnras, 426, 2944, \eprint{arXiv:1208.1766}

\end{thebibliography}

\appendix
\section{Ellipticity measurements}\label{app_shape}
The shapes of the source galaxies in this work are measured using the re-Gaussianisation technique by \citet{Mandelbaum05,Reyes12}, which we briefly describe here. The convolution of a galaxy image with a point spread function (PSF)
has two major effects that we seek to remove.  Firstly, since the PSF
is generally close to round, it circularises the apparent galaxy
shape; this is known as PSF dilution, and the correction for it can be
a factor of $\sim 2$ for typical-sized galaxies.  Secondly, since the
PSF has some small ellipticity (the PSF anisotropy), that ellipticity
is imprinted coherently into the shapes of all galaxies.  If
uncorrected then this gives rise to a coherent additive systematic error
in galaxy shapes and inferred lensing shears.  The goal of a PSF
correction method is to allow one to infer galaxy shapes by
correcting for both of these effects, and thereby infer the lensing
shear.

Historically, the earliest methods of PSF correction were based on
correcting the second moments of the observed galaxy image using the
second moments of the PSF to derive the correction factor (e.g.,
\citealt{KSB95}).  The method that we use here, re-Gaussianization, is
a moments-based technique that corrects for non-Gaussianity of the PSF
(provided that it is small, as for ground-based PSFs) and for kurtosis
of the galaxy profile.  In brief, the correction proceeds in two
steps.  In the first step, the PSF is split into a Gaussian image,
$G(x)$, plus a residual, $\epsilon(x)$, so that the observed image can
be written as
\begin{equation}
I = (G + \epsilon) \otimes f = G\otimes f + \epsilon \otimes f,
\end{equation}
where $f$ is the pre-seeing galaxy profile, and all quantities are
implicitly functions of position (but we have suppressed the
argument).  We make a simple elliptical Gaussian approximation, $f'$ to
$f$, and use that to derive a new image $I'$ defined as
\begin{equation}
I' = I - \epsilon\otimes f' \simeq G\otimes f.
\end{equation}
While our approximation to $f$ is a simple one, we rely on the fact
that the residual from a Gaussian PSF ($\epsilon$) is quite small.
Our new image, $I'$, can be interpreted as an image of what the galaxy
would have looked like if it had been convolved with a simple Gaussian
PSF.  We can therefore carry out the second step of our PSF correction
process using a moments-based method that corrects for the lowest
order of non-Gaussianity in the galaxy profiles, but assumes a
Gaussian PSF \citep{BJ02} in order to estimate a per-object galaxy
ellipticity.  For more details on this entire process, see
\cite{Hirata03}.  The re-Gaussianization method has been tested
extensively on real and simulated data \citep{Mandelbaum05, Reyes12,
  Mandelbaum12, Mandelbaum13} with calibration that is well-controlled
at the percent level.

\section{Further discussion of systematics}
We expand in this section several discussions on systematics related to various datacuts introduced in Section~\ref{sec_cut}. These include our choice of the redshift cut, the influence of unmodelled lenses mentioned in the virial cut and in Section~\ref{sec_sys}, and the effect of the multiplicity cut, which was discussed in Sections~\ref{sec_ML} and \ref{sec_SMHM}. 

\subsection{Suppressing foreground contamination}\label{app_boost}
\begin{figure}
\myplot{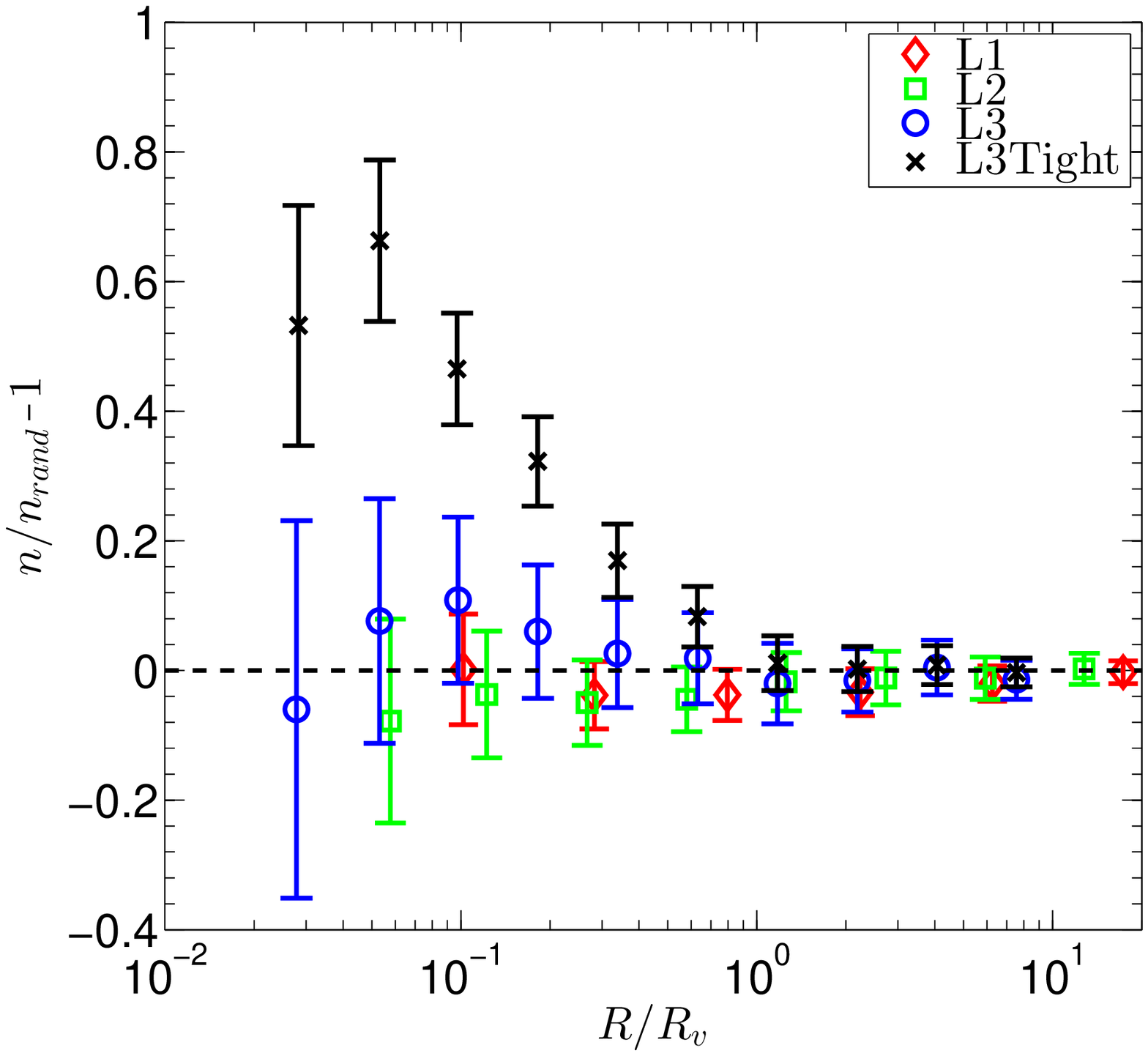}
\caption{Contamination in the background source sample from foreground
  galaxies, for lenses in three luminosity mass bins. The L1, L2
  and L3 bins have mean mass
  $5\times10^{12}$, $1.5\times10^{13}$ and $8\times10^{13}\msunh$, and mean
  redshifts of $0.12$, $0.18$ and $0.25$, and a minimum redshift separation
  of $\Delta z=0.3$. For
  comparison we also show the L3 bin with $\Delta z=0.01$
  (L3Tight). Error-bars mark the $2\sigma$ uncertainty estimated from
  1000 mock samples. 
}\label{fig_contamination}
\end{figure}

To quantify the amount of contamination in the source sample from
foreground group member galaxies, we extract the correlation function of background
galaxies with foreground lenses. Specifically, we count the average
number density, $n$, of background galaxies around foreground lenses,
and compare them with the average number density of random
galaxies around lenses, $n_{\rm rand}$. The random galaxies are generated by
randomizing the position of background galaxies inside the survey
region, so that the random sample will have the same size and follow
the same redshift distribution as the real sample. The contamination
level, or the lens-source correlation function, can then be estimated
from $\xi=n/n_{\rm rand}-1$. In Fig.~\ref{fig_contamination} we show the
estimated correlation for our adopted minimum redshift separation of
$\Delta z=0.3$, 
around different mass haloes. For comparison, the correlation for the
most massive bin with $\Delta z=0.01$ is also shown. While it is
obvious that the contamination is large with a small $\Delta z$, it
can be mostly eliminated with our redshift cut. 

\subsection{Influence of the two halo term}\label{app_2halo}
We have ignored the contribution of the two halo term \citep[see, e.g.][]{
Mandelbaum05a,Johnston07,Hayashi08} on large scales
throughout this paper. Since this term arises from the correlated
distribution of haloes, the missing contribution comes from
unmodelled haloes. By adding a two halo term to
the mass model in our Monte Carlo shear map simulation, and fitting the
simulated map with our standard procedure, we have estimated that the
bias introduced by completely missing the two halo term is a $\sim 3$
per cent overestimate in mass.Since this procedure double-counts the two-halo
term if we include it for every halo at every separation, the
influence from the two-halo term is already overestimated. Hence we
ignore it throughout this paper. 

\subsection{Absolute multiplicity dependence of the mass-observable relations}\label{app_Nabs}
\begin{figure*}
\myplot{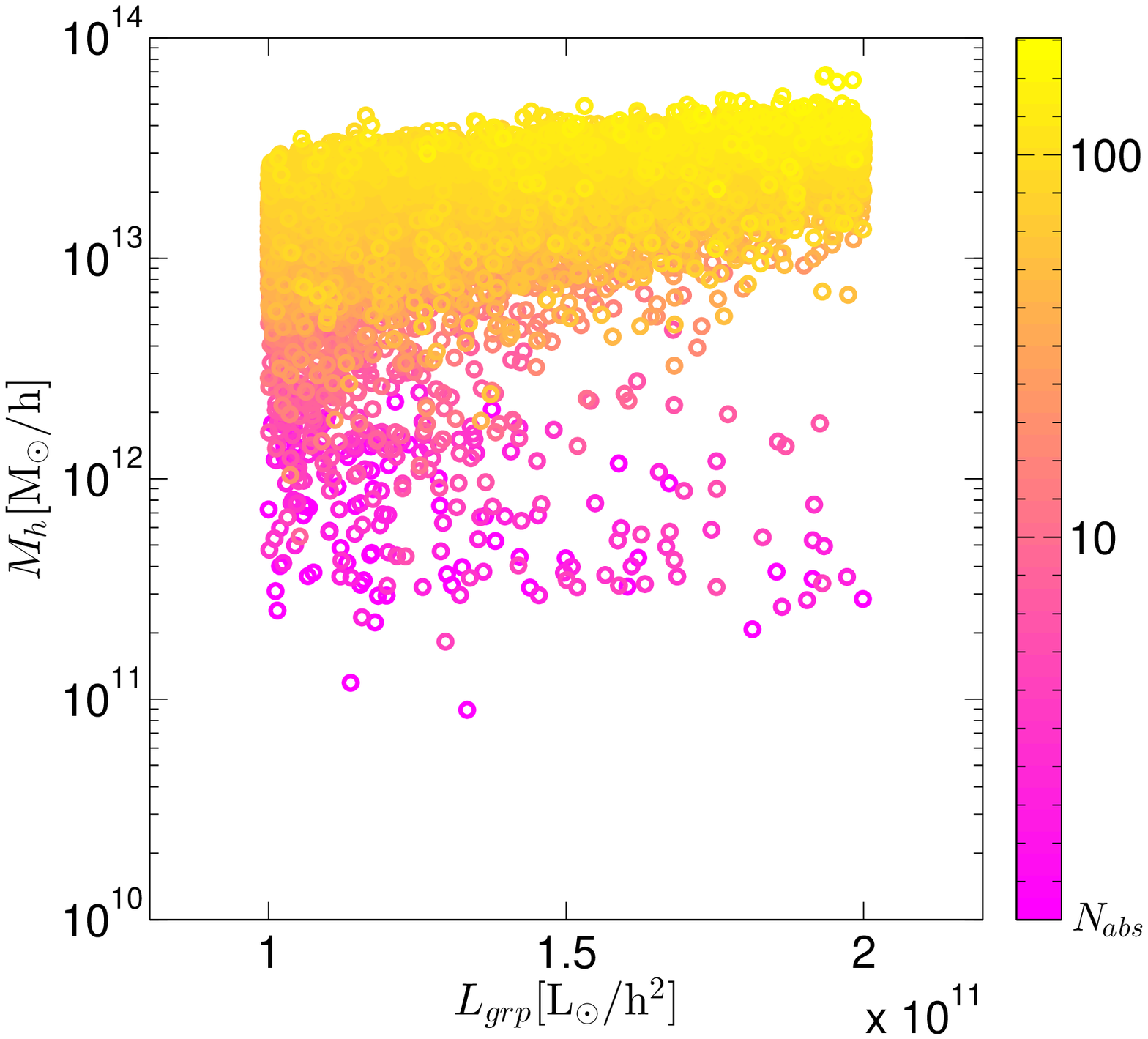}\myplot{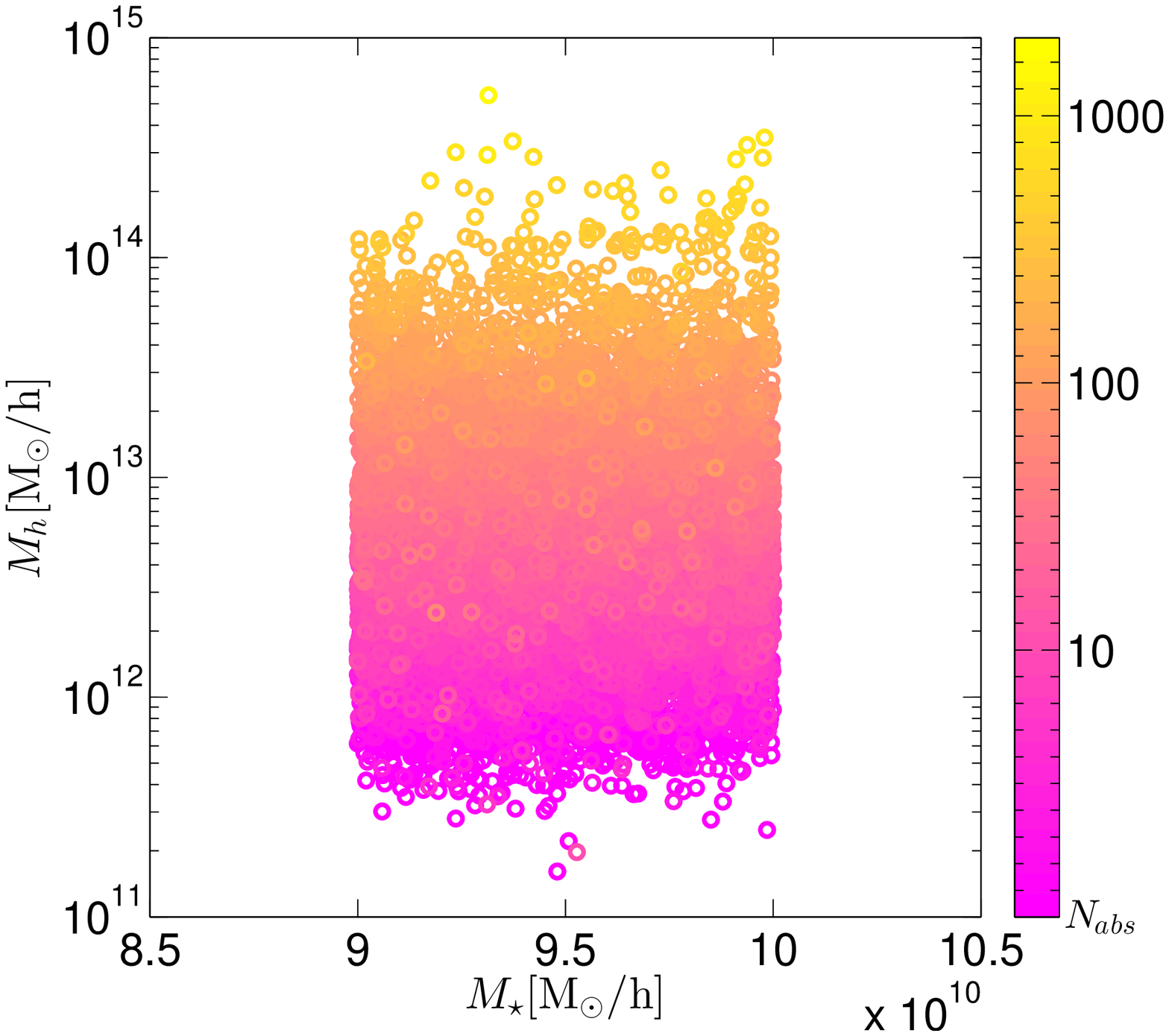}
\caption{The role of group absolute multiplicity in the
  mass-observable relations. In the left panel, we plot the halo
  mass-luminosity relation for all haloes with $10^{11}
  \lsunhh < L_{\rm grp} < 2\times 10^{11} \lsunhh$, colour-coded by
  $N_{\rm abs}$. The right panel is similar but for the halo mass-central
  galaxy stellar mass relation of systems with central galaxy stellar
  masses in the range $9\times 10^{10} \msunhh <M_\star
  <10^{11} \msunhh$.} \label{fig_Nabs}
\end{figure*}

The group selection function of our current catalogue is described by the multiplicity volume $V_N$ (see Equation~\ref{eq_VolMult} and \ref{eq_VolMultLim}), or equivalently the absolute multiplicity. To understand better how this selection function affects our measurements of mass-observable relations, in this section we explore explicitly the dependence of these relations on the absolute multiplicity.
We have selected, from the \cite{Bower06} model in the
Millennium database, two samples of haloes within a narrow range of either
luminosity or central galaxy stellar mass. In Fig.~\ref{fig_Nabs}, We plot the halo mass of these objects as a function of their luminosity or central stellar mass, and color-code them with absolute multiplicity. Here the absolute
multiplicity $N_{\rm abs}$ is defined to be the 
number of galaxies in the halo with an absolute $r$-band magnitude
$r<-13$. The scatter of these mass observable relations at
fixed observable value is clearly not stochastic, but correlates
strongly with $N_{\rm abs}$. As a result, the measured mass-observable relations in a multiplicity-limited group sample will generally be higher than those in a volume-limited sample, since groups with lower absolute multiplicities are more likely to be missing in the sample, consistent with what we see in Fig.~\ref{fig_MMstar_comp} of Section~\ref{sec_SMHM}.

\section{Stellar mass-halo mass relations}\label{app_HOD}
We convert the fitted average central galaxy stellar mass-halo mass
relations in the literature to the following form where possible, and
list the parameters in Table~\ref{table_hod}, along with their adopted
HOD dispersion: 
\begin{equation}\label{eq_hod}
M_\star=\frac{A M_h}{\left[\left(\frac{M_{h}}{M_0}\right)^{\alpha}+\left(\frac{M_{h}}{M_0}\right)^{\beta}\right]^\gamma}.
\end{equation}
The halo mass $M_h$ is defined to have an average density of 200 times
the background matter density. This functional form, especially with
$\gamma=1$ as in \cite{Yang03}, or similar functions to represent two
power-laws with a smooth transition, has been frequently
used to fit the observed galaxy stellar mass distribution to a
modelled halo mass distribution
\citep[e.g.][]{WangL06,Yang08,WangL10,Moster10,Moster13,Guo10,
  Behroozi10,Behroozi13,WangLY13}.
The different relations largely agree at the low mass end, where there are
good constraints, and differ significantly at the high mass end. All the listed relations adopt a Chabrier IMF. These relations are compared with our measurements in Section~\ref{sec_SMHM}.

\begin{table*}
\caption{HOD model parameters for the central galaxy stellar mass-halo
  mass distribution, with the average relation of the form given by
  Equation~\eqref{eq_hod}, and halo mass defined to have an average
  density of $200$ times the background matter
  density. The final column lists the assumed dispersion in the
  stellar mass at fixed halo mass.}\label{table_hod} 
\begin{center}
\begin{tabular}{ccccccc}
\hline
\hline Model & $A$ & $\log(M_0/\msunh)$ & $\alpha$& $\beta$ & $\gamma$ & $\sigma_{\log(M_\star)}$\\
\hline WangL13 \citep{WangL13} & 0.0387 & 11.67 & -1.56 & 0.66 & 1 & 0.17 \\
\hline WangLY13 \citep{WangLY13} & 0.0372 & 11.70 & -1.16 & 0.71 & 1 & 0.22 \\
\hline Moster13 \citep{Moster13} & 0.0370 & 11.58 & -1.38 & 0.61 & 1 & 0.1 \\
\hline Guo10 \citep{Guo10} & 0.0690 & 11.40 & -0.926 & 0.261 & 2.44 & 0 \\
\hline
\end{tabular}
\end{center}
\end{table*}

\section{Stacked group density profiles}\label{app_stack}
As a sanity check, in this section we show the stacked density profiles of our group sample, and compare them with the predictions from our MLWL fits. 

Following \citet{Mandelbaum06b}, we adopt the following estimator for
the average comoving surface overdensity profile around haloes of
similar mass stacked in comoving cooordinates:  
\begin{equation}\label{eq_sig}
\langle\DS(r)_{\rm cmv}\rangle=\frac{\sum_i{w_i \chi_{t,i}\Sigma_{\rm crit,i}a^2_{\ell ,i}}}{2\res\sum_i{w_i}}.
\end{equation}
Here, $\Sigma_{\rm cmv}(r)=\bar{\rho}\int_{l.o.s}\delta_{m,{\rm cmv}}(r) dl$ is the comoving
overdensity of matter integrated along the line of sight, and
$\langle\DS_{\rm cmv}(r)\rangle=\langle \Sigma_{\rm cmv}(<r) \rangle - \langle \Sigma_{\rm cmv}(r) \rangle$ is the
difference between the average surface overdensity within a radius $r$
and that at $r$. The subscript $i$ runs over all the lens-source pairs
in the sample. $a_{\ell }$ is the scale factor at the lens
redshift, $\chi_{t}$ is the tangential ellipticity of the source
galaxy with respect to the lens, and the weighting function is chosen to
be 
\begin{equation}
w_i=\frac{1}{(\Sigma_{\rm crit,i}a^2)^{2}(\sigma_{\chi_i}^2+\sigma_{SN}^2)}.
\end{equation}
With this weighting scheme, the responsivity is calculated using \citep{BJ02}
\begin{equation}
\res=\frac{\sum_i{w_i[1-(1-f_i)\sigma_{SN}^2-{f_i^2 \chi_{i}^2}/{2}}]}
{\sum_i{w_i}},
\end{equation}
where $f_i={\sigma_{SN}^2}/({\sigma_{SN}^2+\sigma_{\chi_i}^2})$.
The responsivity is almost independent of radius. 

Ignoring the error on shear responsivity, the covariance of the
estimated surface density at radii $r_{I}$ and $r_{J}$ can be written as 
\begin{multline}\label{eq_cov}
C(\DS_I,\DS_J)=\\
\frac{\sum_{i\in I,j\in J}{w_i\Sigma_{\rm crit,i}a^2_{\ell,i}}{w_j\Sigma_{\rm crit,j}a^2_{\ell,j}}C(\chi_{t,i},\chi_{t,j})}{4\res^2\sum_{i\in I}{w_i}\sum_{j\in J}{w_j}}.
\end{multline}
Suppose the ellipticities of different galaxies are independent, then
correlations of tangential ellipticity only exist if the two
lens-source pairs are constructed from the same source galaxy, i.e. 
\begin{multline}
\mathrm{C}(\chi_{t,i},\chi_{t,j})=\\
\left\{ \begin{array}{rl}
 \cos(2\phi_{ij})(\sigma_{SN}^2+\sigma_{\chi_i}^2) &\mbox{common source galaxy} \\
  0 &\mbox{ otherwise,}
\end{array} \right.
\end{multline}
where $\phi_{ij}$ is the angle subtended by lens-source pair $ij$.
The equations above fully account for the correlated error
introduced by multiple use of the same source galaxy in the
stacking. These estimated errors give consistent results
with chunked bootstrap measurements. \citet{Jeong09} have
also derived a continuous version of the covariance matrix for stacked lensing that
accounts for cosmic variance.

\begin{figure}
\includegraphics[width=9.5cm]{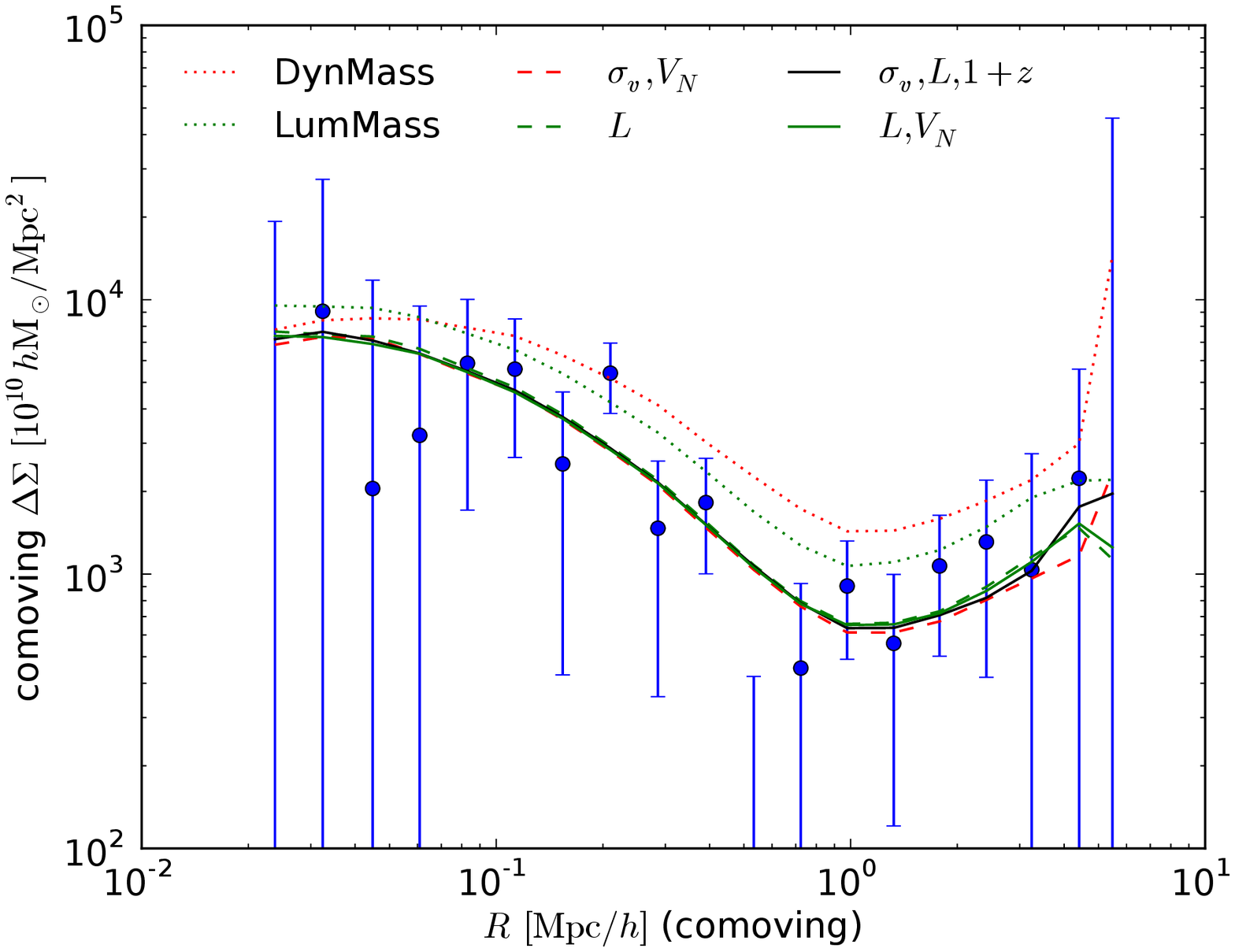}
\caption{Stacked surface density profile for all the groups used in this
  work. Points with errorbars are the stacked
  profiles. Different lines are the predicted surface density profiles from
  various mass estimates, stacked exactly the same way as for the
  data. As labelled in the figure, the dynamical and luminosity masses
  are the standard G$^3$Cv5 calibrated mass estimators, while other lines
  are power-law combinations of the listed observables.}\label{fig_profile} 
\end{figure}

\begin{figure*}
\includegraphics[width=\textwidth]{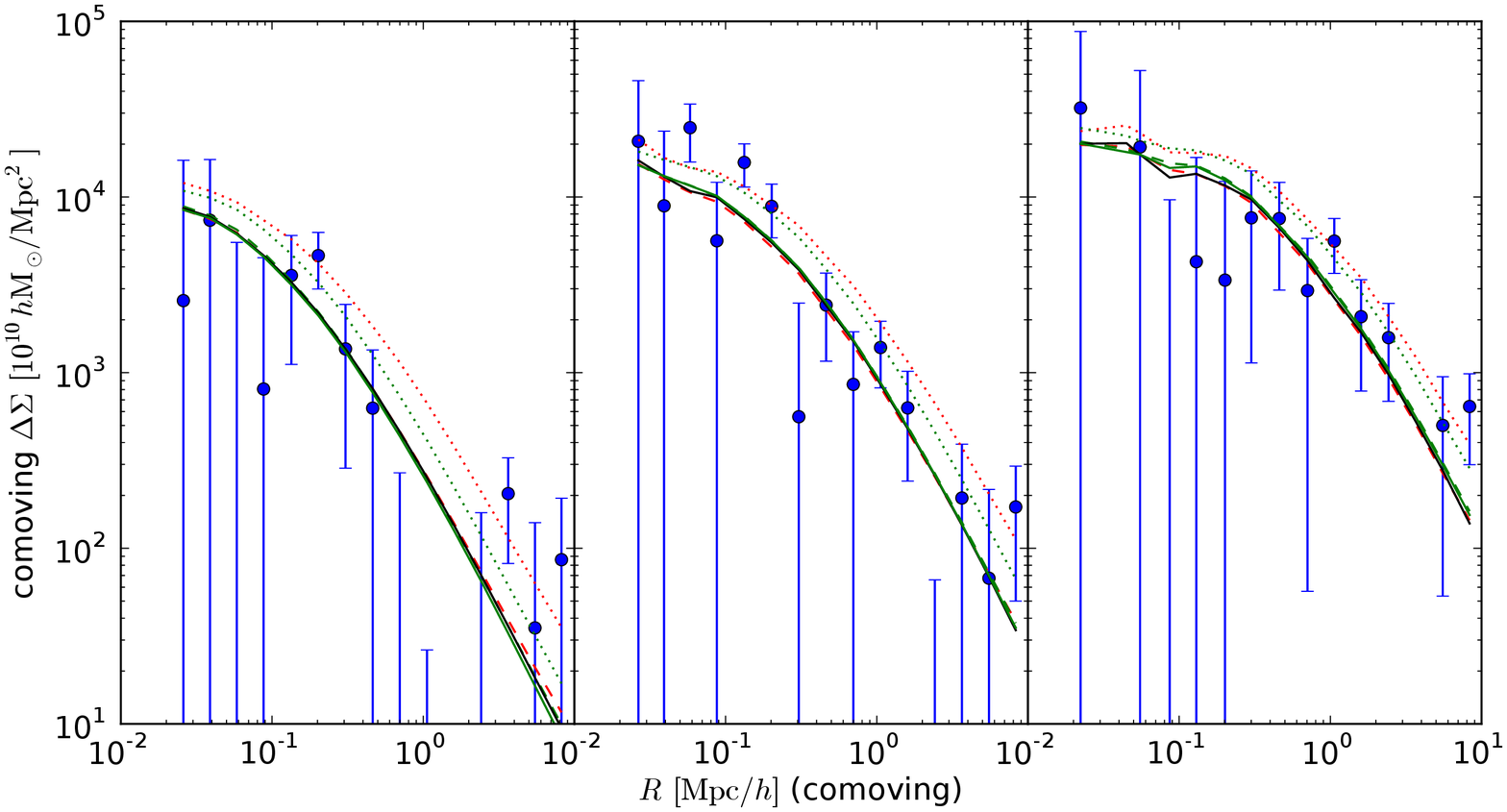}
\caption{Stacked surface density profile for groups with different
  luminosities. From left to right, groups are selected by
  luminosities in the range of $(0.1-1)\times10^{11}\lsunhh$,
  $(1-5)\times10^{11}\lsunhh$ and $(5-50)\times10^{11}\lsunhh$.
  Line styles are the same as in
  Fig.~\ref{fig_profile}.  Note that no radial cut has been applied in producing this plot.}\label{fig_profile2} 
\end{figure*}

Fig.~\ref{fig_profile} shows the stacked surface density profile of
groups, with the same data cuts as applied in the likelihood
analysis. Since we have halo mass estimates for each individual group
from Section~\ref{sec_estimator}, we can stack the inferred projected
NFW profiles in exactly the same way as we stack the data. This gives
predicted stacked profiles that are directly comparable with the
measured profiles, free from any averaging ambiguities, under the
assumption that the predicted mass is taken as the real mass of each
group. No systematic corrections are applied in the mass estimates
during stacking, to make a fair comparison with the measured profiles
for which no correction is made either. It can be seen that our newly
calibrated mass estimators lead to model stacked profiles that agree
very well with each other, as well as with the measurement, while the
G$^3$Cv5 estimates overpredict the measured profile. The rise in the
profile at large radius is caused by our virial cut, $R<2 R_{200b}$,
for each group, which implies that the smaller haloes cannot contribute
at large radii and are hence unable to dilute the average stacked
surface density profile here.

In Fig.~\ref{fig_profile2}, we show the stacked surface density
profiles split into 
three luminosity bins. This time no radial cut is applied. The
measured and modelled profiles are in good agreement in all cases.
Good agreement is also observed between the data and our estimators.

\end{document}